\pdfoutput=1
\newcommand*{\ATLASLATEXPATH}{./}
\documentclass[cernpreprint, USenglish, texlive=2017]{\ATLASLATEXPATH atlasdoc}

\usepackage{\ATLASLATEXPATH atlaspackage}
\usepackage{\ATLASLATEXPATH atlasbiblatex}

\usepackage{\ATLASLATEXPATH atlasphysics}

\graphicspath{{./}}

\usepackage{EXOT-2016-36-PAPER-defs}

\usepackage{tabularx}
\usepackage{rotating}
\usepackage{multirow}
\usepackage{colortbl}

\usepackage{amssymb}  
\usepackage{bm}

\AtlasTitle{Search for lepton-flavor-violating decays of the $Z$ boson into a $\tau$ lepton and a light lepton with the ATLAS detector}

\AtlasAbstract{%
Direct searches for lepton flavor violation in decays of the $Z$ boson with the ATLAS detector at the LHC are presented. Decays of the $Z$ boson into an electron or muon and a hadronically decaying $\tau$ lepton are considered.
The searches are based on a data sample of proton--proton collisions collected by the ATLAS detector in 2015 and 2016, corresponding to an integrated luminosity of \SI{36.1}{\ifb} at a center-of-mass energy of $\sqrt{s} = \SI{13}{\TeV}$. No statistically significant excess of events above the expected background is observed, and upper limits on the branching ratios of lepton-flavor-violating decays are set at the 95\%\ confidence level:  
$\BR (Z\to e\tau) < 5.8 \times 10^{-5}$ and $\BR (Z\to \mu\tau) < 2.4 \times 10^{-5}$. This is the first limit on $\BR (Z\to e\tau)$ with ATLAS data. The upper limit on $\BR (Z\to \mu\tau)$ is combined with a previous ATLAS result based on \SI{20.3}{\ifb} of proton--proton collision data at a center-of-mass energy of $\sqrt{s} = \SI{8}{\TeV}$ and the combined upper limit at 95\%\ confidence level is $\BR (Z\to \mu\tau) < 1.3 \times 10^{-5}$.
}

\author{The ATLAS Collaboration}

\AtlasRefCode{EXOT-2016-36}

\PreprintIdNumber{CERN-EP-2018-052}

\AtlasJournal{PRD}
\AtlasJournalRef{Phys. Rev. D 98, 092010}
\AtlasDOI{10.1103/PhysRevD.98.092010}

\hypersetup{pdftitle={ATLAS document},pdfauthor={The ATLAS Collaboration}}

\begin{document}

\maketitle


\section{Introduction}
\label{sec:intro}

One of the main goals of the physics program of the Large Hadron Collider (LHC) at CERN is to discover
physics beyond the Standard Model (SM).   
The observation of lepton flavor
violation in decays of the $Z$ boson into a pair of leptons of different flavors would give a
clear indication for new physics.
These decays can  
occur within the SM only via neutrino oscillations and would have a rate too small to be detected \cite{Illana:1999ww}.  
This paper presents searches by the ATLAS Collaboration for the decays of the $Z$ boson into a $\tau$ lepton and an electron or a muon, hereafter referred to as a light lepton or $\ell$. Only final states with a hadronically decaying $\tau$ lepton are considered.

Lepton-flavor-violating (LFV) $Z$ boson decays are predicted by models with heavy neutrinos \cite{Illana:2000ic}, extended gauge models \cite{Kuo:1985jt} and supersymmetry \cite{Gabbiani:1988pp}. The most stringent bounds on such decays with a $\tau$ lepton in the final state are set by the LEP experiments: $\BR (Z\to e\tau) < 9.8 \times 10^{-6}$ \cite{LEPLFVZetau} and $\BR (Z\to \mu\tau) < 1.2 \times 10^{-5}$ \cite{LEPLFVZmutau} at 95\%\ confidence level (CL). The ATLAS experiment has set the upper limit $\BR (Z\to \mu\tau) < 1.7 \times 10^{-5}$ at 95\% CL \cite{HIGG-2015-09} by analyzing \SI{20.3}{\ifb} of proton--proton collisions at a center-of-mass energy of \SI{8}{\TeV}.   
There are no previously published limits
on $\BR (Z\to e\tau)$ with ATLAS data. Regarding the LFV $Z\to e\mu$ decays, at the LHC the CMS experiment set the upper bound at $\BR (Z\to e\mu) < 7.3\times 10^{-7}$ at 95\% CL \cite{CMSLFVZemu}. The ATLAS experiment obtained a similar result with the upper bound set at $\BR (Z\to e\mu) < 7.5\times 10^{-7}$ at 95\% CL \cite{EXOT-2013-02}.

The searches for LFV $Z$ decays presented in this paper use a data sample of proton--proton collisions collected at a center-of-mass energy of $\sqrt{s} = \SI{13}{\TeV}$ with the ATLAS detector at the LHC. These data correspond to an integrated luminosity of \SI{36.1}{\ifb}. The signal model used assumes unpolarized $\tau$ leptons.
Events are classified using neural networks, and the output distribution is used in a template fit to data to extract the $Z$ boson lepton-flavor-violating branching ratios, or otherwise set upper limits on these values. The major backgrounds to the search are reducible backgrounds such as $W$+jets, top-quark pair production and \Zll, and the irreducible background $Z\to\tau\tau\to\ell+\mathrm{hadrons}+3\nu$. Reducible backgrounds from events with a quark- or gluon-initiated jet misidentified as a hadronically decaying $\tau$ lepton, so-called ``fakes'', are estimated via 
a data-driven  
method.
The reducible backgrounds from \Zll, where one light lepton fakes a hadronic $\tau$ lepton decay signature, are estimated using simulation. An event selection specifically designed to reduce the contribution from this background is applied. The 
shape of the template for the
irreducible background from $Z\to\tau\tau$ is estimated via simulations and its magnitude  
is determined in the fit to data.

The results of the search for the LFV $Z\to \mu\tau$ decays presented in this paper are combined with the previous ATLAS results based on \SI{8}{\TeV} data.

This paper is structured as follows. Section~\ref{sec:detector} briefly describes the ATLAS detector and the reconstruction of the detected particles. Section~\ref{sec:samples} details the data sample and the simulations used in the analysis. Section~\ref{sec:event-selection} describes the event selection and classification criteria. Section~\ref{sec:bkg-estimation-paper} discusses the methodology used to estimate the yield of events from background sources, and Section~\ref{sec:syst-paper} lists the experimental and theoretical systematic uncertainties affecting the analysis. The statistical interpretation of the observed data and the results are presented in Section~\ref{sec:result}. The combination of the result in the $Z\to \mu\tau$ channel with the previous ATLAS result from \SI{8}{\TeV} data is also presented. Finally, Section~\ref{sec:summary} summarizes the analysis. 

\section{The ATLAS detector and object reconstruction}
\label{sec:detector}

The ATLAS detector\footnote{ATLAS uses a right-handed coordinate system with its origin at the nominal interaction point (IP) in the center of the detector
and the $z$-axis along the beam pipe. The $x$-axis points from the IP to the center of the LHC ring, and the $y$-axis points
upward. Cylindrical coordinates $(r, \phi)$ are used in the transverse plane, with $\phi$ as the azimuthal angle around the beam pipe.
The pseudorapidity is defined in terms of the polar angle $\theta$ as $\eta = -\ln\tan(\theta/2)$. 
The transverse momentum and the transverse energy are defined as $\pT = p \sin\theta$ and $\ET = E \sin\theta$, respectively. The distance $\Delta R$ in $\eta$--$\phi$ space is defined as $\Delta R = \sqrt{(\Delta\eta)^2 + (\Delta\phi)^2}$.
}~\cite{PERF-2007-01} at the LHC is a multipurpose particle detector with a forward-backward symmetric
cylindrical geometry and a nearly $4\pi$ coverage in solid angle. It consists of an inner tracking detector, electromagnetic and hadronic calorimeters, and a muon spectrometer. The inner detector (ID),
immersed in a 2 T axial magnetic field provided by a thin superconducting solenoid, includes silicon pixel and microstrip detectors, which provide precision tracking in the pseudorapidity range $|\eta|<2.5$, and a transition-radiation tracker providing additional tracking and information for electron identification for $|\eta|<2.0$. For the $\sqrt{s}=13$ TeV data-taking period, the ID was upgraded with a silicon-pixel insertable B-layer \cite{ATLAS-TDR-19}  
that provides
additional tracking information  
closer to the interaction point.
The solenoid is surrounded by sampling calorimeters: a lead/liquid-argon (LAr) electromagnetic calorimeter covering the region $|\eta|<3.2$, a hadronic calorimeter with a steel/scintillator-tile barrel section for $|\eta|<1.7$ and two copper/LAr endcaps for $1.5<|\eta|<3.2$. The forward region is covered by additional LAr calorimeters with coarser granularity up to $|\eta|=4.9$. The calorimeter is surrounded by the muon spectrometer, which is based on three large superconducting toroid magnets each containing eight coils. Precise momentum measurements for muons with pseudorapidity up to $|\eta|=2.7$ are provided by three layers of tracking chambers. The muon spectrometer also includes separate trigger chambers covering $|\eta|<2.4$. A two-level trigger system \cite{TRIG-2016-01} was used during the $\sqrt{s}=13$ data-taking period. The first-level trigger (L1) is implemented in hardware and uses a subset of the detector information. This is followed by a software-based level which runs algorithms similar to the offline reconstruction software, reducing
the event rate to approximately 1 kHz on average from the maximum L1 rate of 100 kHz.

Electron candidates are reconstructed from energy deposits in the electromagnetic calorimeter which are matched to a charged-particle track measured in the inner detector. These candidates are required to satisfy ``medium'' likelihood-based identification criteria \cite{ATLAS-CONF-2016-024}, to have a transverse momentum $\pT > \SI{30}{\GeV}$  and to be in the acceptance region $|\eta| < 2.47$ of the inner detector. Candidates in the transition region $1.37 < |\eta| < 1.52$ between the barrel and endcap calorimeters are excluded. 

Muon candidates are reconstructed from track segments in the muon spectrometer which are matched to tracks reconstructed in the inner detector 
which satisfy 
$|\eta| < 2.5$. The matched tracks are re-fitted using the complete track information from both detector subsystems. Muon candidates are required to satisfy ``medium'' identification criteria \cite{PERF-2015-10} and to have a $\pT > \SI{30}{\GeV}$. 

Isolation criteria are applied to both the electrons and muons using calorimeter- and track-based information. The calorimeter-based variables are corrected for the average energy contributions from additional proton--proton interactions in the same and neighboring bunch crossings (``pileup'') \cite{ATLAS-CONF-2016-024}. A standard working point is used, where a 90\% efficiency is obtained for leptons with $\pt = \SI{25}{~GeV}$, rising to 99\% efficiency at $\pt = \SI{60}{~GeV}$ in $Z\to\ell\ell$ events. Events with electrons or muons that fail the isolation criteria are used to estimate background contributions, as described in Section~\ref{sec:bkg-estimation-paper}.

Topological clusters of energy deposits in the calorimeter are reconstructed into jets with the anti-$k_t$ algorithm \cite{Cacciari:2008gp} and radius parameter $R = 0.4$ using the \textsc{FastJet} software package \cite{FastJet}.
Energy deposits from pileup are subtracted using an average pileup energy density and the jet area.
Jets are then calibrated as described in Ref.~\cite{PERF-2016-04}. Jet candidates are required to have $\pT > \SI{20}{\GeV}$ and $|\eta| < 2.5$. To further reduce the effect of pileup, a jet vertex tagger (JVT) algorithm is used for jets with $\pT < \SI{60}{\GeV}$ and $|\eta| < 2.4$. The JVT algorithm employs a multivariate technique based on jet energy, vertexing, and tracking variables in order to determine the likelihood that jets originate from or are heavily contaminated by pileup \cite{PERF-2016-06}. 

In order to identify jets containing $b$-hadrons ($b$-jets), a multivariate algorithm is used that depends on the presence of tracks with a large impact parameter with respect to the primary vertex \cite{ATL-PHYS-PUB-2015-026}, on the presence of displaced secondary vertices, and on the reconstructed flight paths of $b$- and $c$-hadrons associated with the jet \cite{ATL-PHYS-PUB-2016-012}. Using this algorithm, jets are $b$-tagged if they satisfy criteria of a standard working point, which provides a $b$-jet efficiency of 77\% and a light-jet rejection rate of 134 in simulated $\ttbar$ events

Hadronic $\tau$ lepton decays result in a neutrino and a set of visible decay products (\tauhadvis), typically one or three charged pions and up to two neutral pions \cite{Patrignani:2016xqp}. The reconstruction of the visible decay products \cite{PERF-2013-06} is seeded by jets. Selected \tauhadvis candidates are required to have $\pT > \SI{20}{\GeV}$, $|\eta| < 2.5$ excluding $1.37 < |\eta| < 1.52$, one (1-prong) or three (3-prong) associated tracks with $\pT>\SI{1}{\GeV}$, and an electric charge of $\pm 1$. A boosted decision tree (BDT) identification procedure that is based on calorimetric shower shapes and tracking information is used to discriminate 
$\tau$ lepton decays from jet backgrounds \cite{ATL-PHYS-PUB-2015-045,ATLAS-CONF-2017-029}.  All events used in this analysis  
must
have a \tauhadvis candidate that passes the ``loose'' identification working point. 
For events in the signal region, the \tauhadvis candidate must satisfy the ``tight'' identification criterion.
Selected events that are not in the signal region
are used to estimate backgrounds (Section~\ref{sec:bkg-estimation-paper}). 
The combined reconstruction and identification efficiencies for ``loose'' and ``tight'' criteria are 60\% (50\%) and 45\% (30\%) for 1-prong (3-prong) hadronic $\tau$ lepton decays, 
and are 
independent of the \tauhadvis $\pT$ and the number of pileup interactions. To reduce the number of muons misidentified as \tauhadvis, a \tauhadvis candidate is excluded if it is within $\Delta R = 0.2$ of a reconstructed muon with $\pT > \SI{2}{\GeV}$. An additional BDT, denoted hereafter by eBDT, is used to reduce the number of electrons misidentified as \tauhadvis, providing 85\% (95\%) efficiency for 1-prong (3-prong) hadronic $\tau$ lepton decays. The leading-$\pT$ candidate is selected as the \tauhadvis candidate, while any other candidates are considered to be jets.

To avoid potential ambiguities among objects, light lepton and \tauhadvis candidates are required to be separated from each other and from jets in the following order: (a) jets within $\Delta R = 0.2$ of selected \tauhadvis candidates are excluded; (b) jets within $\Delta R = 0.4$ of an electron or a muon are excluded; (c) any \tauhadvis within $\Delta R = 0.2$ of an electron or a muon is excluded; and (d) electrons within $\Delta R = 0.2$ of a muon are excluded.

The missing transverse momentum, with magnitude \MET, is calculated as the negative vectorial sum of the transverse momenta of all fully reconstructed and calibrated (``hard'') physics objects and inner-detector tracks that originate from the hard-scattering vertex but are not matched to a reconstructed object (``soft term'') \cite{Aaboud:2018tkc}. The soft term is an important contribution for improving both the \MET scale and its resolution.

\section{Data and simulated event samples}
\label{sec:samples}

This search analyzes proton--proton collisions recorded by the ATLAS detector at the LHC during 2015 and 2016 at a center-of-mass energy of $\sqrt{s} = \SI{13}{\TeV}$. The data correspond to a total integrated luminosity of \SI{36.1}{\ifb} after requiring that all relevant components of the ATLAS detector 
were in good working condition during data collection.  The uncertainty in the combined 2015 and 2016 integrated luminosity is 2.1\%. It was estimated following a methodology similar to the one described in Ref.~\cite{DAPR-2013-01}.
The events considered for the $e\tau$ ($\mu\tau$) channel were selected by single-lepton triggers which require the presence of at least 
one electron (muon) candidate with transverse momentum above $\SI{24}{\GeV}$ ($\SI{20}{\GeV}$) in 2015 data and $\SI{26}{\GeV}$ ($\SI{26}{\GeV}$) in 2016 data.
These triggers apply isolation criteria for electron (muon) candidates with \pT below $\SI{60}{\GeV}$ ($\SI{40}{\GeV}$ in 2015 and $\SI{50}{\GeV}$ in 2016).
These isolation requirements are looser than the ones applied offline in the light-lepton selections used in this analysis.

Simulated Monte Carlo (MC) samples are used to predict the $Z/\gamma^* \to \ell\tau$ signal and
the background contributions from $Z/\gamma^*$+jets, $W$+jets, \ttbar, single top-quark, Higgs boson and diboson ($WW$, $WZ$
and $ZZ$) production.

Signal samples were simulated using $\textsc{Pythia}$ 8.186 \cite{PYTHIA8} with the NNPDF2.3 parton distribution function (PDF) set \cite{Ball:2012cx} and a set of tuned parameters called the A14 tune \cite{ATL-PHYS-PUB-2014-021}. 
The lepton-flavor-violating $Z/\gamma^*$ decay was modeled assuming unpolarized $\tau$ leptons in the final state.
In order to use the same production cross section for both signal and the main background, $Z/\gamma^*\to\tau\tau$, event weights computed as a function of the true boson transverse momentum are applied to the signal events to match the more accurate modeling of the $Z/\gamma^*$ production in the $Z/\gamma^*\to\tau\tau$ simulation described in the following. After this reweighting procedure, the signal events, together with the $Z/\gamma^*\to\tau\tau$ events, are normalized to the $Z/\gamma^*$ production cross section determined from data in the template fit described in Section~\ref{sec:result}. Therefore, 
the analysis is independent of the theoretical uncertainty in the $Z/\gamma^*$ production cross section.
The SM value of this cross section is $\SI{2.1}{nb}$, calculated at NNLO accuracy \cite{Melnikov:2006kv}.

The production of $Z/\gamma^*\to\tau\tau$ events 
was simulated with $\textsc{Sherpa}$ 2.2.1 \cite{Sherpa}. The NNPDF 3.0 NNLO PDF set \cite{Ball:2014uwa} was used for both the matrix element calculation and the dedicated parton-shower tuning developed by the authors of $\textsc{Sherpa}$. The event generation utilized Comix \cite{Gleisberg:2008fv} and OpenLoops \cite{Cascioli:2011va} for the matrix element calculation, which was then matched to the $\textsc{Sherpa}$ parton shower using the ME+PS@NLO prescription \cite{Hoeche:2012yf}. The matrix elements were calculated for up to two additional partons at NLO and for three and four partons at LO in QCD. 
As stated above, the normalization of this background process, together with the signal events, is determined in a fit to data.

The $Z/\gamma^*\to\mu\mu,ee$ events were simulated with $\textsc{Powheg-Box}$ \cite{Alioli:2010xd,Nason:2004rx,Frixione:2007vw} using the CT10 PDF set \cite{Lai:2010vv} and the AZNLO tune \cite{STDM-2012-23}, and interfaced to $\textsc{Pythia}$ 8.186. 
The normalization of the $Z/\gamma^*\to\mu\mu,ee$ events is determined from data in a dedicated region enhanced in $Z\to\mu\mu$ events (Section~\ref{sec:bkg-estimation-paper}) as a function of the reconstructed transverse momentum of the $Z/\gamma^*$ boson. 

The other simulated processes account for only a small fraction (less than 0.3\%) of the background events. Samples of $W(\to\tau\nu)$ + jets events were simulated with $\textsc{Sherpa}$ 2.2.1. Events with a top-quark pair or a  
single
top quark produced via electroweak t-channel, s-channel and $Wt$-channel processes were simulated with $\textsc{Powheg-Box}$ using the CT10 PDF set. The parton shower, fragmentation and underlying event were simulated using $\textsc{Pythia}$ 6.428 \cite{PYTHIA6} 
with the Perugia 2012 tune \cite{Skands:2010ak}. EvtGen \cite{EvtGen} was used to decay bottom and charm hadrons. Diboson processes were simulated with $\textsc{Sherpa}$ 2.1.0 with the CT10 PDF set. Higgs boson events, $H\to WW,\tau\tau,\ell\ell$, produced via gluon--gluon fusion and vector-boson fusion were simulated with $\textsc{Powheg-Box}$. 

Simulated minimum-bias events were overlaid on all simulated samples to include the effect of  
pileup. These minimum-bias events were generated with $\textsc{Pythia}$ 8.186, using the A2 tune \cite{ATL-PHYS-PUB-2012-003} and the MSTW2008LO PDF set \cite{Martin:2009iq}.
Each simulated event was processed using the $\textsc{Geant}$-based ATLAS detector simulation \cite{GEANT4,SOFT-2010-01} and the same event reconstruction algorithms  
used for the data. Reconstruction and identification efficiencies, as well as energy calibrations for all selected objects in simulated 
events, are corrected to match those measured in data.

\section{Event selection and classification}
\label{sec:event-selection}

Of the events satisfying the trigger and the quality criteria described in Section~\ref{sec:samples}, the events selected in this analysis are required to contain exactly one isolated electron or muon that is geometrically matched to the object that fired the trigger, and no additional light leptons. These events must also contain at least one \tauhadvis candidate that passes the tight identification. 
The isolated light lepton and the \tauhadvis candidate are required to have opposite charge, 
$q_\ell q_{\tauhadvis} = -1$. 
Events with one or more $b$-tagged jets are removed to reject background events with a top-quark pair or a singly produced top quark. To reduce the $Z\to\ell\ell$ background, events with 1-prong \tauhadvis candidates 
that satisfy $|\eta(\tauhadvis)| > 2.2$ for the $e\tau$ channel or  
$|\eta(\tauhadvis)| < 0.1$ for the $\mu\tau$ channel are rejected. These regions of the detector are excluded because they are insufficiently instrumented and therefore affected by higher $\ell\to\tau$ misreconstruction and misidentification rates.
The selection described here, denoted hereafter to as preselection, defines the sample of events used for the training of the neural network. 

Further kinematic selections are applied to define the sample of events in the ``signal region'' (SR) which are used in the final template fit. Orthogonal sets of events in the so-called ``calibration regions'' (CR) are defined by inverting some of the preselection or SR selection requirements and used to estimate background contributions in the SR, as described in Section~\ref{sec:bkg-estimation-paper}.

\begin{figure}[tbp]
    \centering
    \includegraphics[width=0.45\textwidth]{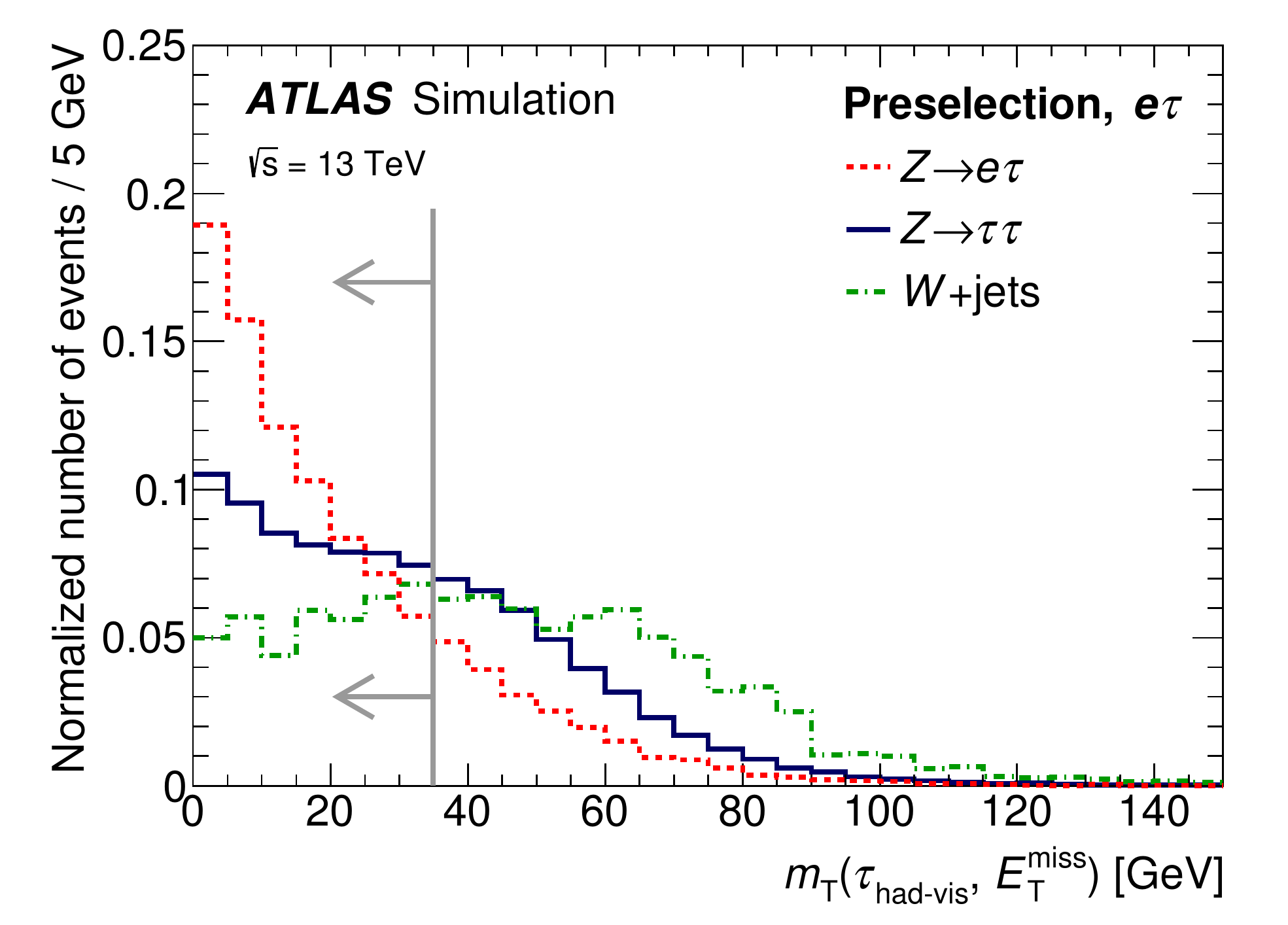}
    \includegraphics[width=0.45\textwidth]{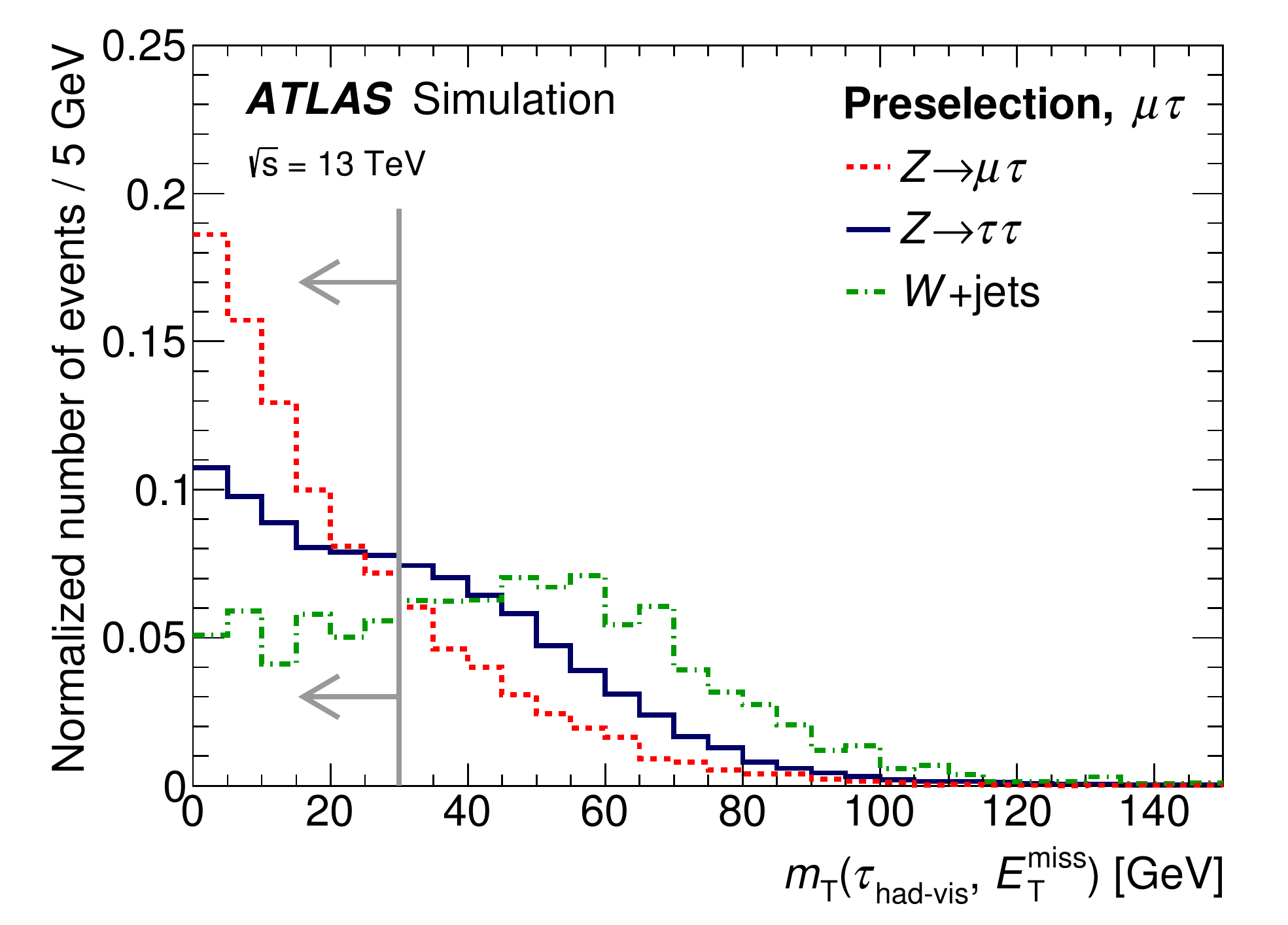}
  \caption{Expected distributions of $\mT(\tauhadvis,\MET)$ in $Z/\gamma^*\to\tau\tau$, $W(\to\ell\nu/\tau\nu)$+jets and signal events in the $e\tau$ (left) and $\mu\tau$ (right) channels after 
preselection requirements. The $Z/\gamma^*\to\tau\tau$ and $W(\to\ell\nu/\tau\nu)$+jets distributions also include the contributions to fakes from the corresponding processes as predicted by MC simulations. All distributions are normalized to unity.} 
  \label{fig:preselection-tau-mT}
\end{figure}

\begin{figure}[tbp]
    \centering
    \includegraphics[width=0.45\textwidth]{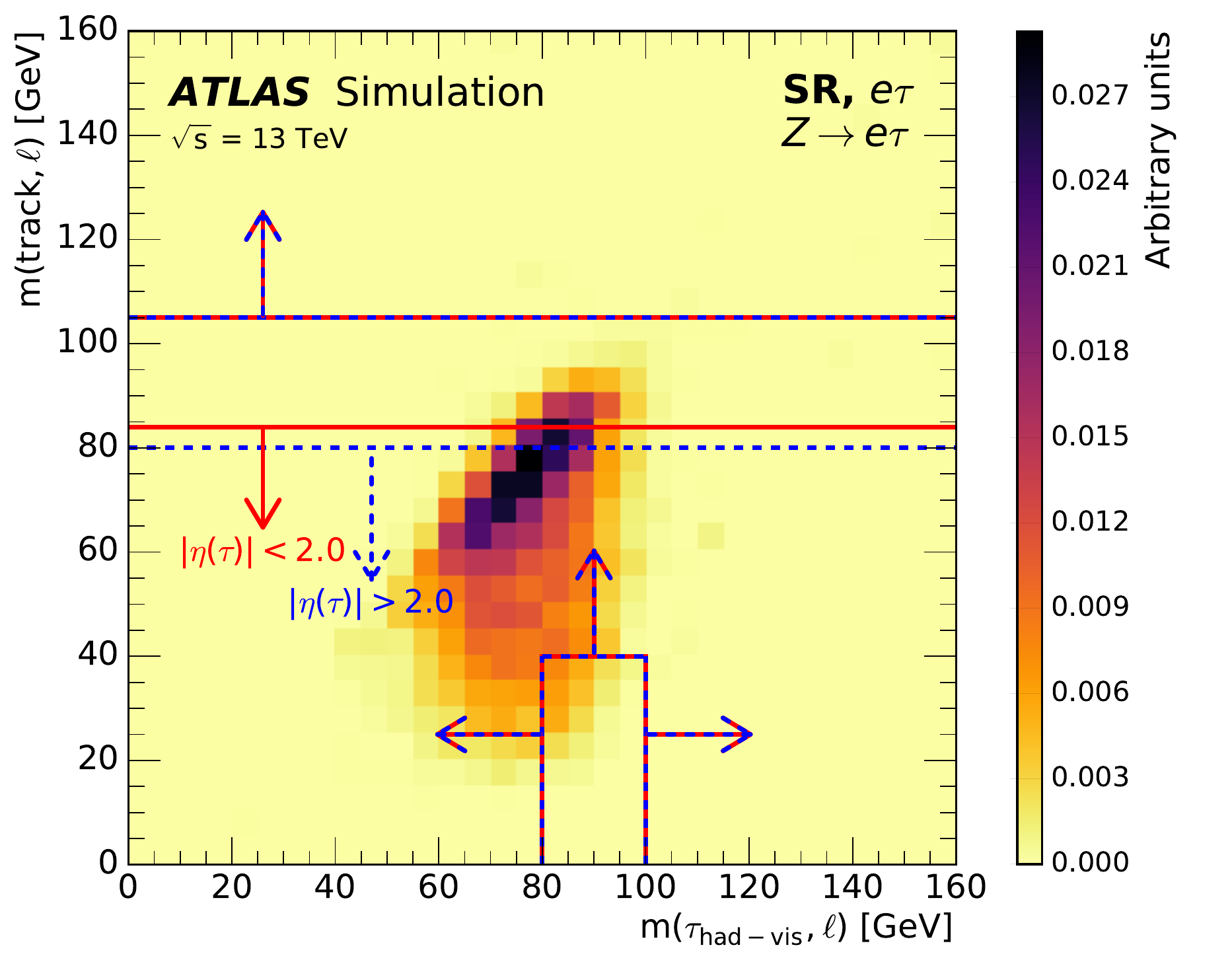}
    \includegraphics[width=0.45\textwidth]{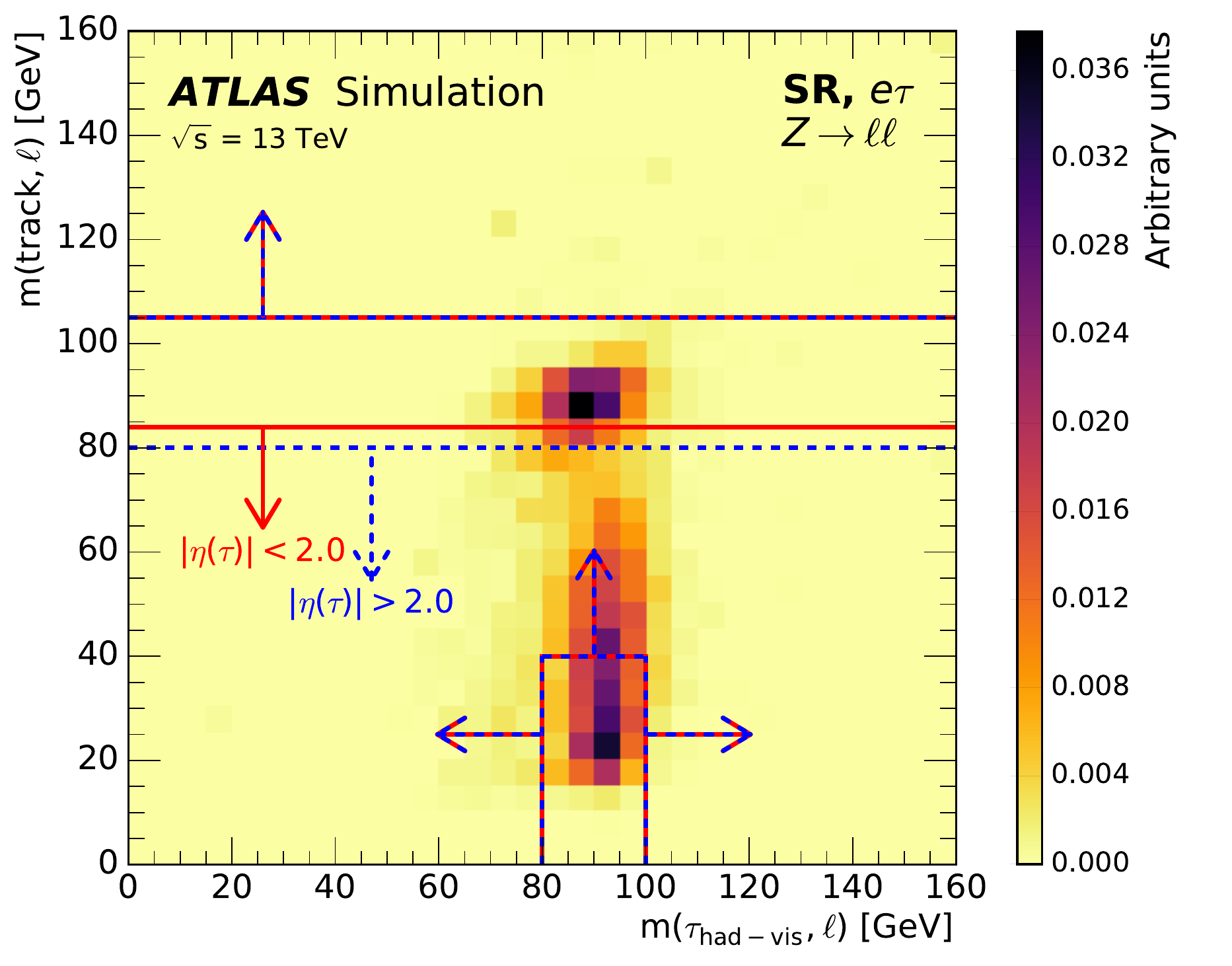} \\
    \includegraphics[width=0.45\textwidth]{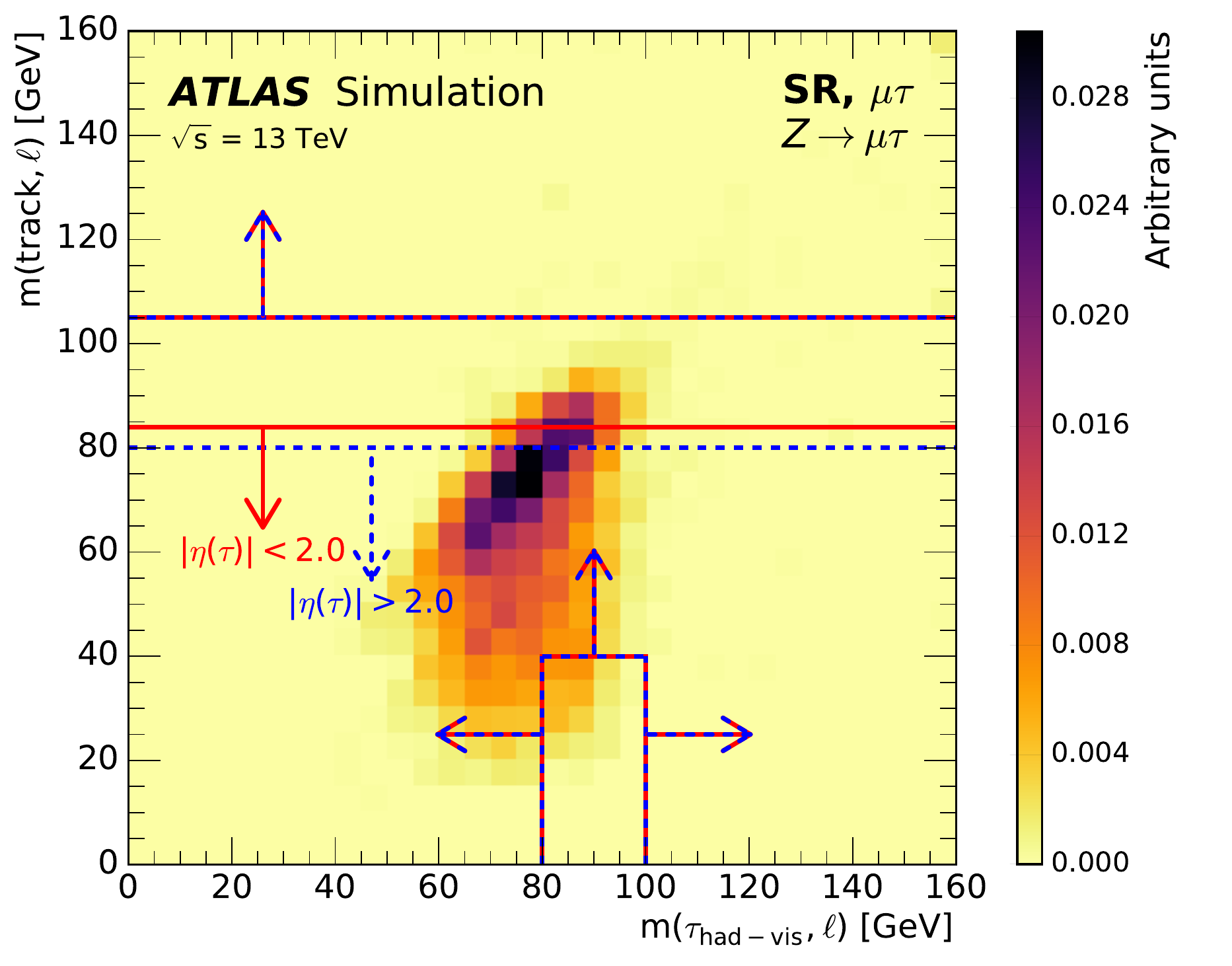}
    \includegraphics[width=0.45\textwidth]{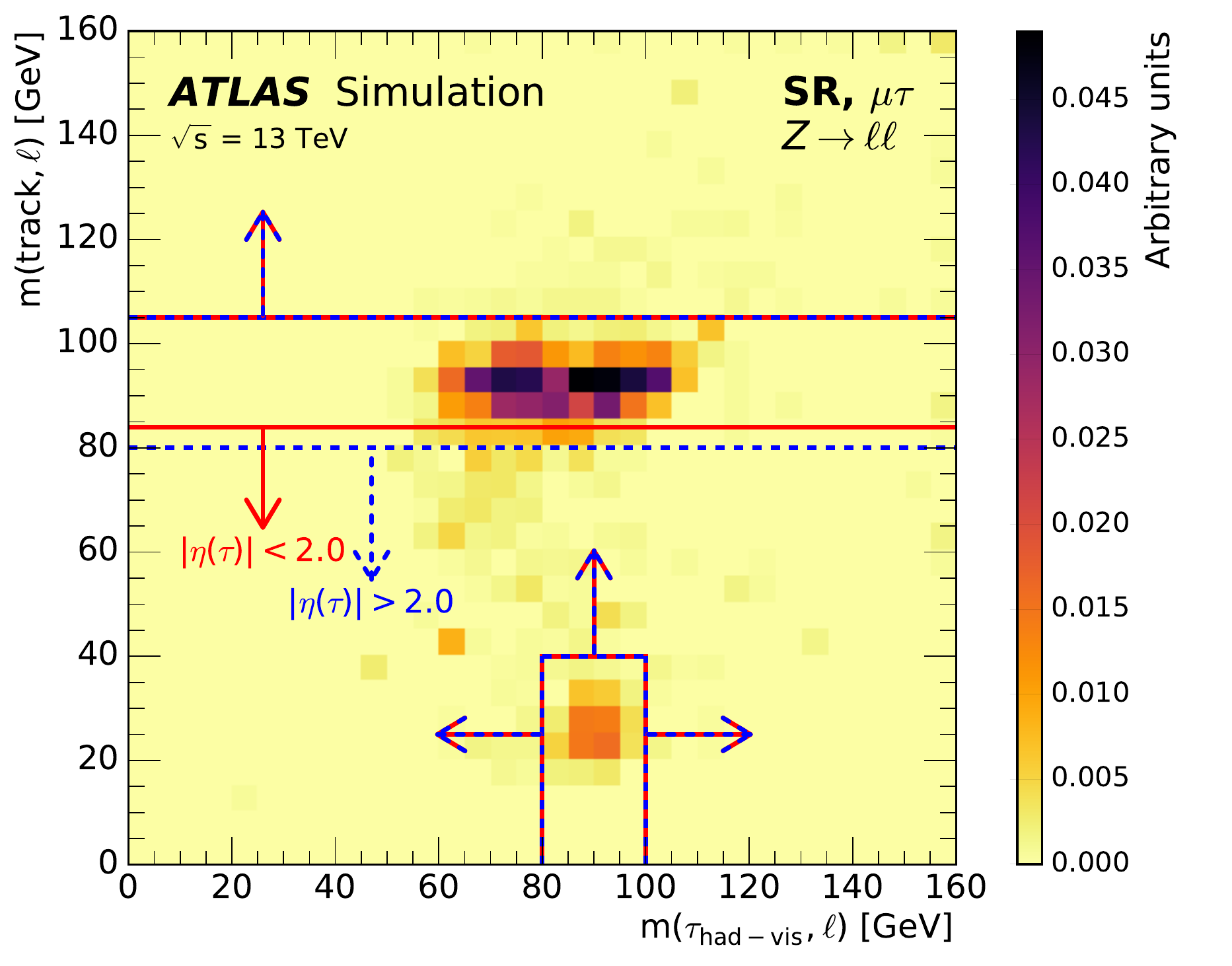}
  \caption{Expected distributions of $m(\mathrm{track},\ell)$ versus $m(\tauhadvis,\ell)$ in signal (left) and $Z\to\ell\ell$ (right) events with 1-prong \tauhadvis candidates in the $e\tau$ (top) and $\mu\tau$ (bottom) channels after the SR selection except for the cuts on these two variables (Table~\ref{tab:regions-overview-paper}).} 
  \label{fig:SR-track}
\end{figure}

Events accepted in the SR must satisfy the preselection and the following selections. The transverse mass, 
\begin{equation*}
    \mT(\tauhadvis, \MET) \equiv \sqrt{2\pT(\tauhadvis) \MET \left[1-\cos\left(\Delta\phi(\tauhadvis, \MET)\right) \right]},
\end{equation*}
is required to be smaller than 35(30) GeV in the $e\tau$($\mu\tau$) channel. Signal events are expected to have the missing transverse momentum from the neutrino in a direction close to the \tauhadvis candidate, resulting in small $\mT(\tauhadvis, \MET)$ values. The $W(\to\ell\nu/\tau\nu)$+jets events and some of the $Z/\gamma^*\to\tau\tau$ events have instead higher $\mT(\tauhadvis, \MET)$ values. This selection allows the definition of a CR that is dominated by $W$+jets events, which are the major contribution to fakes. The selection is illustrated in Figure~\ref{fig:preselection-tau-mT}.
In events with a 1-prong \tauhadvis candidate, an additional selection is applied to further reduce the $Z\to\ell\ell$ background. 
In most of these events, the momentum of the track matched to the 1-prong \tauhadvis candidate corresponds to the original momentum of the light lepton misidentified as \tauhadvis, while the energy deposited in the calorimeter and used to estimate the energy of the \tauhadvis originates from radiation (light-lepton bremsstrahlung) or other sources. Therefore, events in which the invariant mass of the \tauhadvis track and the light lepton ($m(\mathrm{track},\ell)$) is compatible with the $Z$ boson mass are rejected. In particular, events with a 1-prong \tauhadvis candidate are accepted when $m(\mathrm{track},\ell)<84$ GeV or $m(\mathrm{track},\ell)>105$ GeV if $|\eta(\tauhadvis)|<2.0$, and when $m(\mathrm{track},\ell)<80$ GeV or $m(\mathrm{track},\ell)>105$ GeV if $|\eta(\tauhadvis)|>2.0$. A wider range in $m(\mathrm{track},\ell)$ is rejected at high $|\eta(\tauhadvis)|$ because of the smaller signal contribution and the higher $Z\to\ell\ell$ background rate. Moreover, events in which the invariant mass of the 1-prong \tauhadvis candidate and the light lepton satisfies $\SI{80}{\GeV}<m(\tauhadvis,\ell)<\SI{100}{\GeV}$ are required to have $m(\mathrm{track},\ell)>40$ GeV. These selections have been optimized in the $m(\tauhadvis, \ell)-m(\mathrm{track}, \ell)$ plane to specifically reject the $Z\to\ell\ell$ background events at a small acceptance loss for signal. The impact of these selections is illustrated in Figure~\ref{fig:SR-track}. 

The signal selection efficiency in the SR is 3.2\% (3.5\%) for the $e\tau$ ($\mu\tau$) channel. The $Z\to\tau\tau$, $Z\to\ell\ell$ and $W\to\ell\nu$+jets background selection efficiencies are, including their contributions to fakes as predicted by MC simulations, $7.1\times10^{-3}$ ($6.5\times10^{-3}$), $9.4\times10^{-5}$ ($5.9\times10^{-5}$) and $2.6\times10^{-5}$ ($2.9\times10^{-5}$) respectively.
A summary of the event selection criteria is given in Table~\ref{tab:regions-overview-paper}.

\begin{table}[tbp]
\caption{Overview of the event selection. More details are given in Sections \ref{sec:detector} and \ref{sec:event-selection}.}
    \begin{center}
        \begin{tabular}{@{}rl@{}}
            \toprule
            Preselection & one isolated tight light lepton with $\pt>30$ GeV matched to a lepton selected at trigger level\\
                                 & leading \tauhadvis with $\pt>20$ GeV,  $N^{\mathrm{tracks}}_{\tau}=1$ or $3$ and passing tight identification \\
                                 & if $N^{\mathrm{tracks}}_{\tau}=1$: $0.0(0.1)<|\eta_{\tau}|<1.37$  or $1.52<|\eta_{\tau}|<2.2(2.5)$ in $e\tau$($\mu\tau$) events \\
                                 & if $N^{\mathrm{tracks}}_{\tau}=3$: $0.0<|\eta_{\tau}|<1.37$  or $1.52<|\eta_{\tau}|<2.5$ \\
                                 & $q_\ell \times q_{\tau} = -1$ \\
                                 &  no $b$-jet, no additional light lepton \\
            \midrule
            Signal Region &$m_{\mathrm{T}}(\tauhadvis, \MET) < 35 (30)$ GeV in $e\tau$ ($\mu\tau$) events \\
                                   & if $N^{\mathrm{tracks}}_{\tau}=1$ and $|\eta_{\tau}|<2.0$: $m(\mathrm{track},\ell)<84$ GeV or $m(\mathrm{track},\ell)>105$ GeV\\
                                   & if $N^{\mathrm{tracks}}_{\tau}=1$ and $|\eta_{\tau}|>2.0$: $m(\mathrm{track},\ell)<80$ GeV or $m(\mathrm{track},\ell)>105$ GeV\\
                                   & if $N^{\mathrm{tracks}}_{\tau}=1$ and $80<m(\tauhadvis,\ell)<100$ GeV: $m(\mathrm{track},\ell)>40$ GeV\\
            \bottomrule
        \end{tabular}
    \end{center}
   \label{tab:regions-overview-paper}
\end{table}

Events accepted in the SR are classified using neural networks (NNs) trained to discriminate \Zlfv signal from \Ztt, \Zll and \Wln+jets background events. 
The classification is based on event kinematic properties that are extracted by the NN from the reconstructed momenta of the selected particles, as well as from other event variables. The NN  
achieves good performance
using low-level  
variables, such as the particle momentum components, due to the network's capability to build non-linear relations between input variables. 

Three types of NN classifiers, ``Z'', ``Zll'' and ``W'', are trained to distinguish signal from \Ztt, \Zll and \Wln backgrounds, respectively. These classifiers are trained separately in the $e\tau$ and $\mu\tau$ channels because of the different detector acceptances, but  
combine 1-prong and 3-prong \tauhadvis candidates. 
Simulated events passing the preselection (Table~\ref{tab:regions-overview-paper}) are used to train, optimize and validate the classifiers. 
In order to increase the size of the available training samples 
for \Zlfv and \Ztt processes with a true hadronic $\tau$ lepton decay, 
all events with a \tauhadvis candidate that passes the loose   
identification are used.
Moreover, in the events used for the Zll classifiers, the misreconstructed \tauhadvis is required to be either a true muon or electron. With these requirements, about 40\,000 signal events, 200\,000 \Ztt events and 80\,000 \Wln events are used for training in each channel. For \Zll, about 30\,000 events are used in the $e\tau$ channel and only 5000 events in the $\mu\tau$ channel. The limited number of $Z\to\mu\mu$ events is due to the low $\mu\to\tau$ misreconstruction rate, and leads to poor classification power for the Zll NN in the $\mu\tau$ channel. However, the $Z\to\mu\mu$ 
background is effectively reduced by the selection on $m(\mathrm{track},\ell)$ and $m(\tauhadvis,\ell)$ described earlier.

The input variables common to all the classifiers are: the light lepton, \tauhadvis and \met momentum components, assuming vanishing masses; the collinear mass $m_{\mathrm{coll}}$, defined as the invariant mass of the $\ell$--$\tauhadvis$--$\nu$ system, where $\nu$ is the neutrino from the $\tau$ decay, which is assumed to have a momentum that is equal in the transverse plane to the measured \met and collinear in $\eta$ with the \tauhadvis candidate; and $\Delta\alpha$ \cite{Davidson:2012wn}:
\begin{equation}
    \Delta\alpha = \frac{1}{2}\frac{m_Z^2 - m_\tau^2}{p(\tauhadvis)\cdot p(\ell)} - \frac{\pT(\ell)}{\pT(\tauhadvis)}, \label{eq:dalpha}
\end{equation}
where $p(\tauhadvis)$ and $p(\ell)$ are the four-momenta of the \tauhadvis and the light-lepton candidates respectively, and the rest masses $m_Z$ and $m_\tau$ take on values reported by the Particle Data Group \cite{Patrignani:2016xqp}. The variable $\Delta\alpha$ helps to discriminate signal events, expected to be around $\Delta\alpha=0$, from \Ztt events,
where $\Delta\alpha$ is negative 
due to the presence of 
additional neutrinos. Even though not specifically targeted by this variable, \Zll and \Wln events tend to be at vanishing and positive values of $\Delta\alpha$, respectively, as shown later in 
Figures~\ref{fig:postfit_SR_el_NNInputs_1P}--\ref{fig:postfit_SR_mu_NNInputs_3P}.
The invariant mass $m(\ell,\tauhadvis)$ is also used in the Zll classifier. In the limit of very large training statistics, the light lepton, \tauhadvis and \met momentum components would be sufficient for the NN to learn the full event kinematics. However, with the available training samples, the high-level variables $m_{\mathrm{coll}}$, $\Delta\alpha$ and $m(\ell,\tauhadvis)$ were found to be able to improve the NN classification power and were therefore included among the NN inputs.
 
The NN inputs are preprocessed to harmonize their magnitudes and to remove known symmetries as is required for optimal training. The preprocessing consists of the following steps:
\begin{enumerate}
\item Boost: after computing $m_{\mathrm{coll}}$, $\Delta\alpha$ and $p^\mathrm{tot}=p(\ell)+p(\tauhadvis)+\met$ in the lab frame, the light lepton, \tauhadvis and \met momenta are boosted to the frame in which their total momentum vanishes. The longitudinal component of the three-momentum of \met is zero in the lab frame.
\item Rotation: the light lepton, \tauhadvis and \met momenta are first rotated so that the 
three-momentum
of the light lepton is along the positive $z$-axis. A second rotation about the $z$-axis is applied so that the \tauhadvis momentum has a vanishing component on the $y$-axis.
\item ``Feature scaling'': each input variable is scaled by subtracting its mean and by dividing by its standard deviation, where the mean and the standard deviation are computed on the set of signal and background events used in the training of each classifier.
\end{enumerate}
The boost and the rotation are used to remove the degeneracy among apparently different events which are instead equivalent under Lorentz transformation.
``Feature scaling'' is needed because the network works best with input variables of the same magnitude. 
The same preprocessing procedure, with the same mean and standard deviation values, is applied to all the events on which the classifiers are evaluated. After preprocessing, six of the twelve components of the light lepton, \tauhadvis and \met momenta are either vanishing or redundant, and therefore not included in the network inputs. The resulting lists of input variables are given in Table~\ref{tab:NNinputs1}. The transverse component, $\pt^\mathrm{tot}$, of the total momentum $p^\mathrm{tot}$ in the lab frame is also included as otherwise this information would be lost after the preprocessing.
\begin{table}[tbp]
    \caption{Input variables for the NN classifiers.
The first six quantities are in the boosted and rotated frame described in the
text; the last four are in the laboratory frame.}
    \begin{center}
        \begin{tabular}{lcccc}
            \toprule
            \textbf{Variable} & \textbf{Description}  & \textbf{Z NN} & \textbf{Zll NN} & \textbf{W NN} \\
            \midrule
            $\hat{E}^{\mathrm{lep}}$ & light-lepton energy & $\checkmark$ & $\checkmark$ & $\checkmark$ \\
            $\hat{p}^{\tauhadvis}_{x}$ & \tauhadvis $p_{x}$ & $\checkmark$ & $\checkmark$ & $\checkmark$ \\
            $\hat{p}^{\tauhadvis}_{z}$ & \tauhadvis $p_{z}$ & $\checkmark$ & $\checkmark$ & $\checkmark$ \\
            $\hat{E}^{\tauhadvis}$ & \tauhadvis energy & $\checkmark$ & $\checkmark$ & $\checkmark$ \\
            $\hat{p}^{\mathrm{miss}}_{z}$ & \met component along $z$-axis & $\checkmark$ & $\checkmark$ & $\checkmark$ \\
            $\hat{E}^{\mathrm{miss}}$ &   magnitude of \met & $\checkmark$ & $\checkmark$ & $\checkmark$ \\
            \midrule
            $\pt^\mathrm{tot}$ & transverse component of total momentum & $\checkmark$ & $\checkmark$ & $\checkmark$ \\
            $m_{\mathrm{coll}}$ & collinear mass & $\checkmark$ & $\checkmark$ & $\checkmark$ \\
            $\Delta\alpha$ & see Eq. (\ref{eq:dalpha}) \cite{Davidson:2012wn} & $\checkmark$ & $\checkmark$ & $\checkmark$ \\
            $m(\ell,\tauhadvis)$ &  invariant mass of light lepton and \tauhadvis& & $\checkmark$ &  \\
            \bottomrule
        \end{tabular}
    \end{center}
    \label{tab:NNinputs1}
\end{table}
The distributions of some of the NN input variables are shown in Section~\ref{sec:result}.

The NN classifiers are sequential models optimized for binary classification. They are based on the \textsc{Keras} 1.1.1 \cite{keras} and \textsc{Tensorflow} 0.11 \cite{tensorflow} packages, using a standard implementation for binary classifiers having two hidden dense layers with 16 nodes each.

In order to obtain a single discriminating variable, the outputs of the classifiers evaluated in each event are combined in the following way.
In events with 3-prong \tauhadvis candidates, where no further rejection is needed against the \Zll events, the Z and W classifiers are combined as the distance in the two-dimensional plane from the point with highest NN outputs, where the NN outputs can range within $[0,1]$:
\begin{equation*}
    \mathrm{combined~output (3P)} = 1 - \sqrt{(1-\mathrm{output_W})^{2}+(1-\mathrm{output_Z})^{2}}/\sqrt{2}.
\end{equation*}
In a similar fashion, for events with 1-prong \tauhadvis candidates,  the Z, W and Zll classifiers are combined as:
\begin{equation*}
    \mathrm{combined~output (1P)} = 1 - \sqrt{(1-\mathrm{output_W})^{2}+(1-\mathrm{output_Z})^{2}+(1-\mathrm{output_{Zll}})^{2}}/\sqrt{3}.
\end{equation*}
The chosen procedure to combine the individual NN outputs reduces the dimensionality of the classifiers while maintaining the correlations among these classifiers for each event. The binned distributions of these combined classifiers for the events selected in the SR are used in the final template fit, as discussed in Section \ref{sec:result}.

\section{Background estimation}
\label{sec:bkg-estimation-paper}

Background processes are categorized according to the origin of the \tauhadvis candidate, which can be a true $\tau$ lepton, or a misidentified light lepton, or a misidentified quark- or gluon-initiated jet.
Different techniques are used to estimate these background contributions in the SR, as well as to model their expected combined NN output distributions, which are used in the template fit to data (Section~\ref{sec:result}). As described in the following, the shapes of all components are determined prior to the fit, as are the normalizations for
all but the \Ztt and fake components, which are determined in the fit.

Backgrounds from processes with a true hadronically decaying $\tau$ lepton are estimated from simulation. The $Z\to\tau\tau$ decays are the dominant source of these events. As detailed in Section~\ref{sec:samples}, they are modeled via simulation but their total yield in the SR is left unconstrained in the template fit to data in order to remove the theoretical systematic uncertainties in the $Z$ production cross section.

Processes where the $\tauhadvis$ candidate is a misidentified light lepton are also estimated from simulation. These are mostly $Z\to\ell\ell$ events. The simulated 
rate for misidentifying electrons as 1-prong $\tauhadvis$ candidates 
is corrected using data \cite{ATLAS-CONF-2017-029}. 
Due to the lack of dedicated measurements of the 
rates of misidentifying electrons as 3-prong $\tauhadvis$ candidates and muons as 1-prong $\tauhadvis$ candidates,
conservative uncertainties are assigned which have negligible impact on the precision of the measured $\BR (Z\to\ell\tau)$.

The normalization of the $Z\to\ell\ell$ events is determined from data with a sample of events with an opposite-charge muon pair with $\SI{81}{\GeV}<m_{\mu\mu}<\SI{101}{\GeV}$. 
The preselection requirements on the leading muon, the absence of $b$-tagged jets and the veto on additional light leptons are imposed. 
A  
correction factor derived as the relative difference between the predicted and observed numbers of $Z\to\mu\mu$ events is applied to both the $Z\to ee$ and $Z\to\mu\mu$ yields in the SR. This correction is applied as a function of the reconstructed transverse momentum of the $Z/\gamma^*$ boson.
In the $Z\to\mu\mu$-enhanced region, the $Z/\gamma^*$ boson momentum is computed as the vector sum of the muon pair, while in the SR it is the vector sum of the misidentified $\tauhadvis$ candidate and the remaining light lepton.
The uncertainty in this correction is statistical only. Differences between the electron and muon acceptances are covered by the systematic uncertainties in the electron and muon selections, which are accounted for in the $Z\to\ell\ell$ predictions in the SR.

Events where the $\tauhadvis$ candidate originates from a quark- or gluon-initiated jet are estimated from data. This contribution is referred to as ``fakes'' and is dominated by $W$+jets and multi-jet processes. A data-driven  
fake-factor
technique is used to estimate this contribution. It uses events in the so-called ``fail sideband'', which is the set of events passing all but one of the SR selection requirements: the \tauhadvis candidate is required to fail the tight identification requirement. This is a set of events orthogonal to the ones selected in the SR and enhanced with fakes. The yield of these events is corrected by the fake factor, which is the transfer factor needed to scale the fail sideband sample to the amount of background expected in the signal region, which requires an identified \tauhadvis candidate. This factor is process-specific as it depends on the fractions of quark- and gluon-initiated jets that are misidentified as $\tauhadvis$ candidates. It also depends on properties of the $\tauhadvis$ candidate. To capture these effects, different fake factors are measured in samples of events dominated by different processes and different \tauhadvis kinematic properties.

Fake factors $F_\mathrm{W}$, $F_\mathrm{T}$, $F_\mathrm{Zll}$, and $F_\mathrm{QCD}$ are measured in four data samples of events dominated by $W$ + jets (``CRW''), \ttbar and single-top (``CRT''), $Z\to\ell\ell$ + jets (``CRZll''), and multi-jet (``CRQ'') events, respectively. 
The selections that define these ``calibration regions'' (CR) are similar to the SR selection but define orthogonal samples dominated by the target source of background. These selections are detailed in Table~\ref{tab:CRs-paper} together with the expected purities in each CR for the target process as estimated from simulation. For CRQ the purity is estimated as the number of events in data, after subtracting the contribution from other processes estimated from simulation, divided by the total number of events.

\begin{table}[tbp]
    \caption{Calibration regions used to derive fake factors. Differences from the SR selection (Table~\ref{tab:regions-overview-paper}) are listed together with the purities for the target processes as expected from simulation.}
    \begin{center}
    \begin{tabular}{llcc}
        \toprule
        \textbf{Region} &  \textbf{Change relative to SR selection} & \multicolumn{2}{c}{\textbf{Purity [\%]}}\\
               &   & $e\tau$ & $\mu\tau$ \\
        \midrule
        CRZll  & Two same-flavor opposite-charge light leptons with $81<m_{\ell\ell}<101$ GeV & 98 & 98\\
        CRW & $\mT(\ell, \MET)>40$ GeV and $\mT(\tauhadvis,\MET)>35 (30)$ GeV in $e\tau$ ($\mu\tau$) events & 84 & 85\\ 
        CRT  &$N_{b\text{-jets}}\ge2$ & 98 & 98\\
        CRQ & Inverted light-lepton isolation & 75 & 37\\
        \bottomrule
    \end{tabular}
    \end{center}
    \label{tab:CRs-paper}
\end{table}

In each CR, $F_i$ ($i=\text{W, T, Zll, QCD}$) is measured in data as the ratio of the number of events where the \tauhadvis candidate passes the tight identification to the number of events where the \tauhadvis candidate fails in bins of the \tauhadvis \pt. Contributions from background processes that are not the target process of the CR or from events where the \tauhadvis candidate does not originate from a jet are subtracted from data using simulation. The four $F_i$ are combined into a weighted average $F = \sum_i R_i F_i$, where $R_i$ is the fraction of events from fakes in the SR as predicted by simulation for each process. For multi-jet events, this fraction is defined as $R_\mathrm{QCD} = 1-R_\mathrm{W}-R_\mathrm{Zll}-R_\mathrm{T}$.
Fake factors are measured separately for \tauhadvis candidates with one and with three associated tracks. For 1-prong candidates, they are estimated in two-dimensional bins of \tauhadvis \pt and \tauhadvis track \pt, since the associated track momentum is used in the selection of these candidates, while for 3-prong
candidates 
they are estimated only in bins of \tauhadvis \pt. The  
choice of bin boundaries
is optimized to capture the statistically significant variations of the fake factors as a function of the \tauhadvis properties, while retaining enough events per bin. An additional binning as a function of \tauhadvis $|\eta|$ was  
found to be unnecessary.  
The measured fake factors are shown in Table \ref{table:fake-factors}. For events with low \tauhadvis \pt and high \tauhadvis track \pt, the fake factors are large and  
have large statistical uncertainties because there are few events in the calibration regions. However, these fake factors are applied only to a small fraction of events in the sidebands.

\begin{table}[htp]
\caption{The fake factors binned in \tauhadvis \pt and \tauhadvis track \pt for 1-prong, and \tauhadvis \pt for 3-prong events as determined in the SR.}
\begin{center}
\begin{tabular}{r|ccc||ccc}
\toprule
\textbf{1-prong} & \multicolumn{3}{c||}{\textbf{$e\tau$ events}} & \multicolumn{3}{c}{\textbf{$\mu\tau$ events}} \\
 \tauhadvis $\pt$ & 20--30 GeV & 30--40 GeV & $>40$ GeV & 20--30 GeV & 30--40 GeV & $>40$ GeV \\
\midrule
\tauhadvis track $\pt$ \\
1--15 GeV  & 0.29$\pm$0.02 & 0.32$\pm$0.04 & 0.29$\pm$0.04 & 0.35$\pm$0.06 & 0.32$\pm$0.04 & 0.28$\pm$0.05 \\
15--20 GeV & 0.54$\pm$0.06 & 0.46$\pm$0.07 & 0.33$\pm$0.11 & 0.54$\pm$0.07 & 0.40$\pm$0.09 & 0.30$\pm$0.11 \\
20--60 GeV & 1.34$\pm$0.18 & 0.80$\pm$0.15 & 0.52$\pm$0.08 & 1.3$\pm$0.2 & 0.78$\pm$0.14 & 0.52$\pm$0.07 \\
$>60$ GeV  & 1.0$\pm$1.0 & $2.6^{+5.3}_{-2.6}$ & 0.67$\pm$0.19 & 0.5$\pm$0.4 & 0.8$\pm$0.7 & 0.7$\pm$0.4 \\
\midrule
\textbf{3-prong}  & \multicolumn{3}{c||}{\textbf{$e\tau$ events}} & \multicolumn{3}{c}{\textbf{$\mu\tau$ events}} \\
 \tauhadvis $\pt$ & 20--30 GeV & 30--40 GeV & $>40$ GeV & 20--30 GeV & 30--40 GeV & $>40$ GeV \\
\midrule
  & 0.21$\pm$0.01 & 0.22$\pm$0.02 & 0.20$\pm$0.01 & 0.20$\pm$0.03 & 0.24$\pm$0.06 & 0.19$\pm$0.02 \\
\bottomrule
\end{tabular}
\end{center}
\label{table:fake-factors}
\end{table}%

The number of events from fakes in the SR is:
\begin{equation*}
N_{\mathrm{SR}}^{\mathrm{fake}} = 
\sum_k
\Big(N_{\mathrm{SR},\mathrm{data}}^{\mathrm{fail}} - N_{\mathrm{SR},\mathrm{MC},\mathrm{not}~\mathrm{jet}\rightarrow\tau}^{\mathrm{fail}} \Big)_k \times F_k,
\end{equation*}
where $F_k$ is the fake factor corresponding to the \pt (and track \pt for 1-prong \tauhadvis) bin $k$,
$N_{\mathrm{SR},\mathrm{data}}^{\mathrm{fail}}$ is the number of data events in the fail sideband in bin $k$, and $N_{\mathrm{SR},\mathrm{MC},\mathrm{not}~\mathrm{jet}\rightarrow\tau}^{\mathrm{fail}}$ is the number of events in the fail sideband in bin $k$ for which the \tauhadvis candidate did not originate from a jet as predicted by simulation.

The sources of uncertainty in the estimate of the fake background 
are the statistical uncertainties in the $F$ measurements in each bin, the statistical uncertainties of the data in the fail sideband and the uncertainty in $R_i$. All  
statistical uncertainties are 
treated as independent.
The uncertainty in $R_i$ is estimated by varying the estimated $R_{\mathrm{W}}$ by 50\%, although this has a negligible impact on the sensitivity.


The simulation and the data-driven techniques used to model the signal and background processes were validated in samples enriched with fakes and \Ztt events. Both the predicted NN input and output distributions are in agreement with data.  

\section{Systematic uncertainties}
\label{sec:syst-paper}

Systematic uncertainties affecting the estimations of signal and background contributions arise from the theoretical predictions and the detector modeling used in simulation, the luminosity measurement, and the data-driven background estimations. 

The theoretical uncertainties in the production cross section affect only the predictions of the simulated $W$+jets, top, diboson and Higgs boson events with a true hadronically decaying $\tau$ lepton,
since the $Z\to\tau\tau$ and signal yields are determined in the template fit to data.
These constitute a small fraction of the background events in the SR, and a conservative uncertainty in their production cross sections was assigned with negligible impact on the final results. As described in Section~\ref{sec:bkg-estimation-paper}, $Z\to\ell\ell$ events are normalized to data using $Z\to\mu\mu$ events, so 
the theoretical uncertainty in the $Z\to\ell\ell$ normalization is 
irrelevant. The statistical uncertainty of 0.1\% in this normalization correction is included as a systematic uncertainty. 

Uncertainties arising from the simulation of the detector and pileup conditions in the reconstruction of \tauhadvis candidates, muons, electrons, jets (including $b$-tagging) and \met are evaluated.
Sources of uncertainty in the \tauhadvis candidate include the reconstruction and identification efficiencies and the energy calibration. 
These are applied only to 
\tauhadvis candidates from hadronically decaying $\tau$ leptons. For misidentified \tauhadvis candidates originating from an electron or a muon,  
systematic uncertainties in the misidentification rates are assigned
using a data-driven method, 
as detailed in Section~\ref{sec:bkg-estimation-paper}. For the simulation of electron and muon candidates, uncertainties in the trigger, reconstruction, identification and isolation efficiencies are accounted for. 
The effect of uncertainties in the light-lepton momentum scale and resolution is also evaluated. 
For jets, uncertainties in the jet momentum scale and resolution, as well as in the $b$-tagging (in)efficiencies are accounted for. All experimental uncertainties are propagated to the \met calculation. In addition, uncertainties in the energy scale and resolution of the \met soft term are considered.

The 2.1\% uncertainty in the measured luminosity (Section~\ref{sec:samples}) is only considered for the simulated $W$+jets, top, diboson and Higgs boson contributions, whose normalizations are based purely on simulation, without any data-driven estimate. 

Data-driven techniques are used to estimate the background contributions from events with a \tauhadvis candidate originating from either a light lepton or a quark- or gluon-initiated jet. The systematic uncertainties in these methods are described in Section~\ref{sec:bkg-estimation-paper}.

\begin{figure}[tbp]
    \centering
    \includegraphics[width=0.45\textwidth]{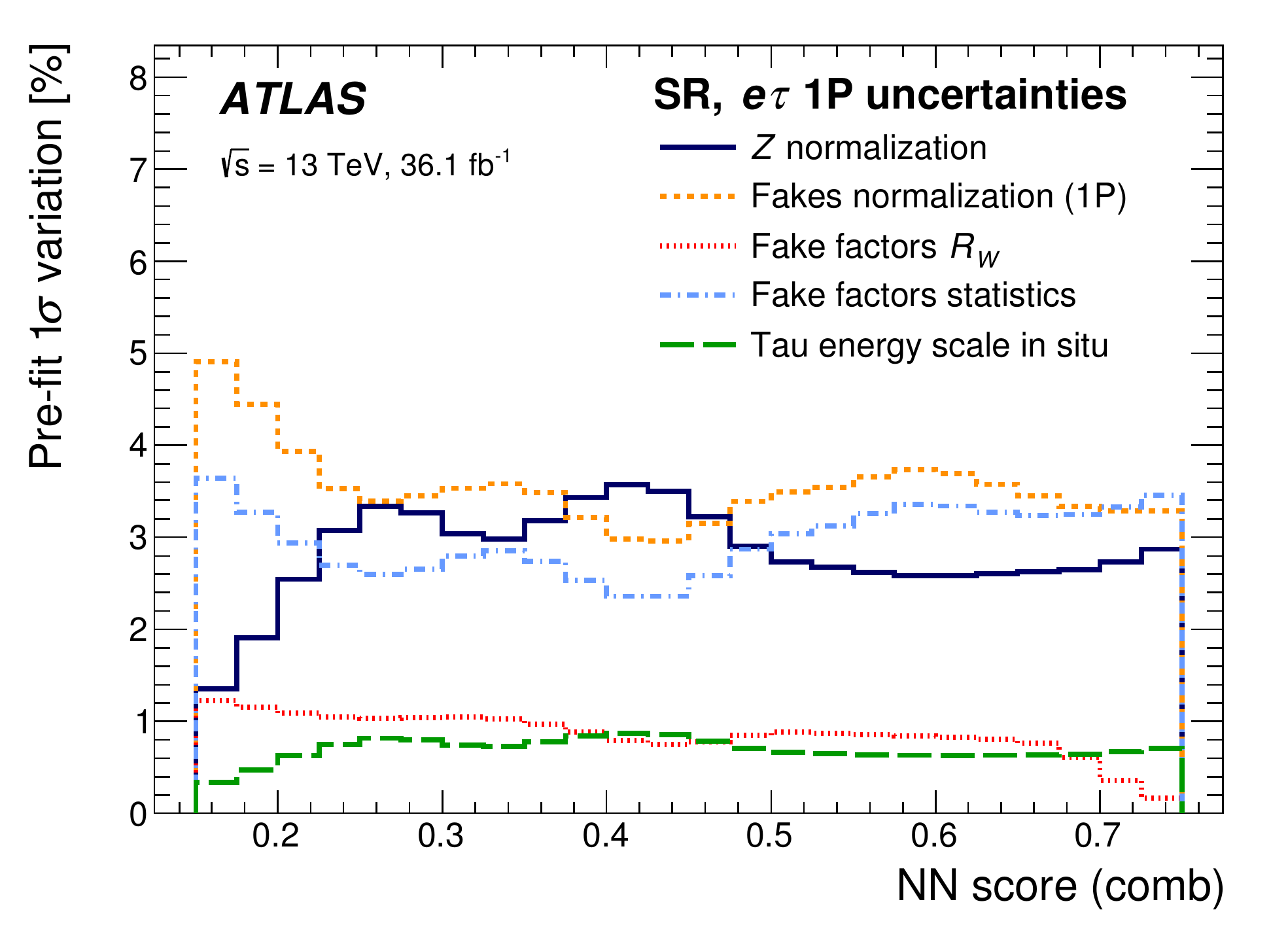}
    \includegraphics[width=0.45\textwidth]{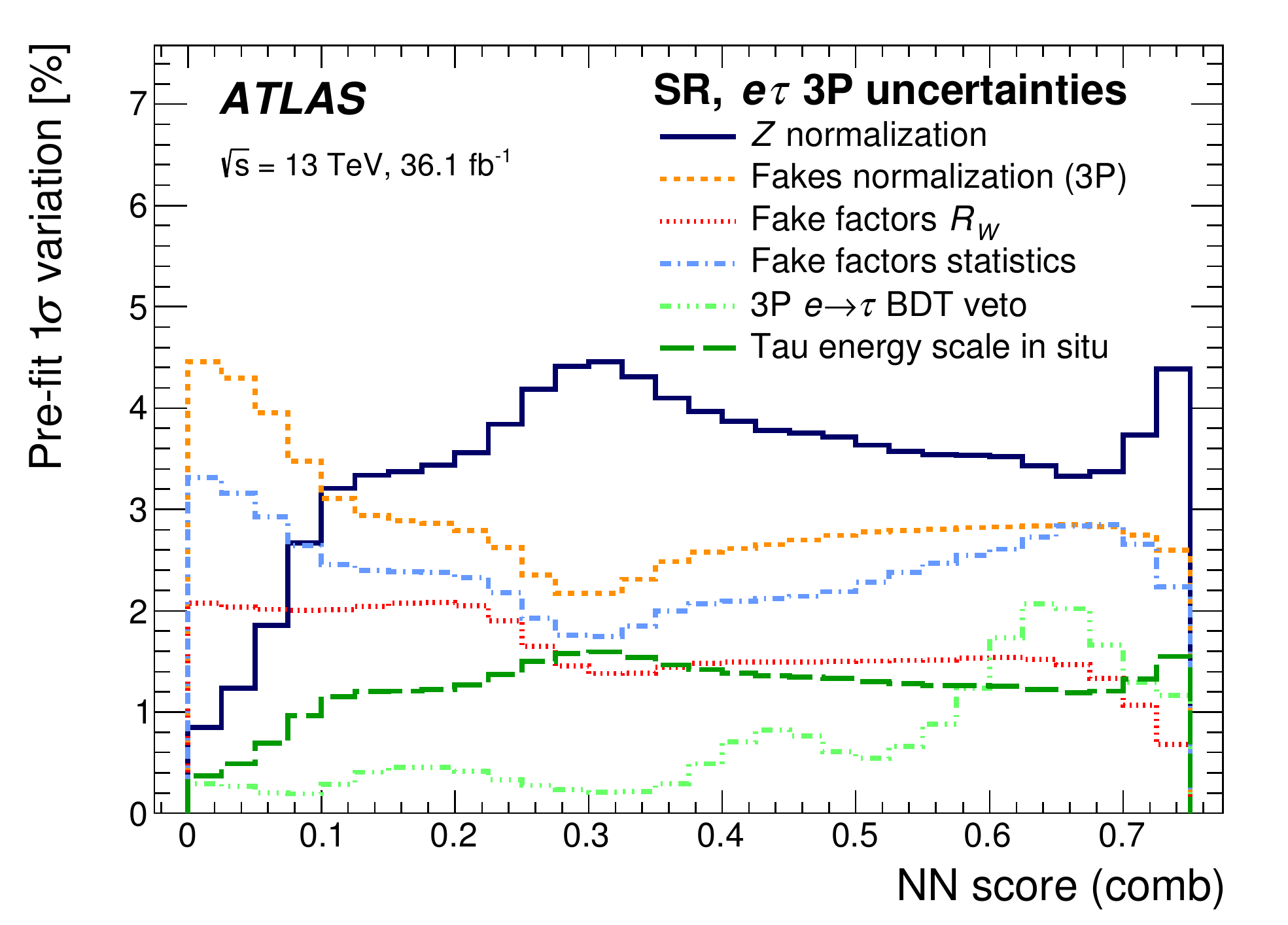} \\
    \includegraphics[width=0.45\textwidth]{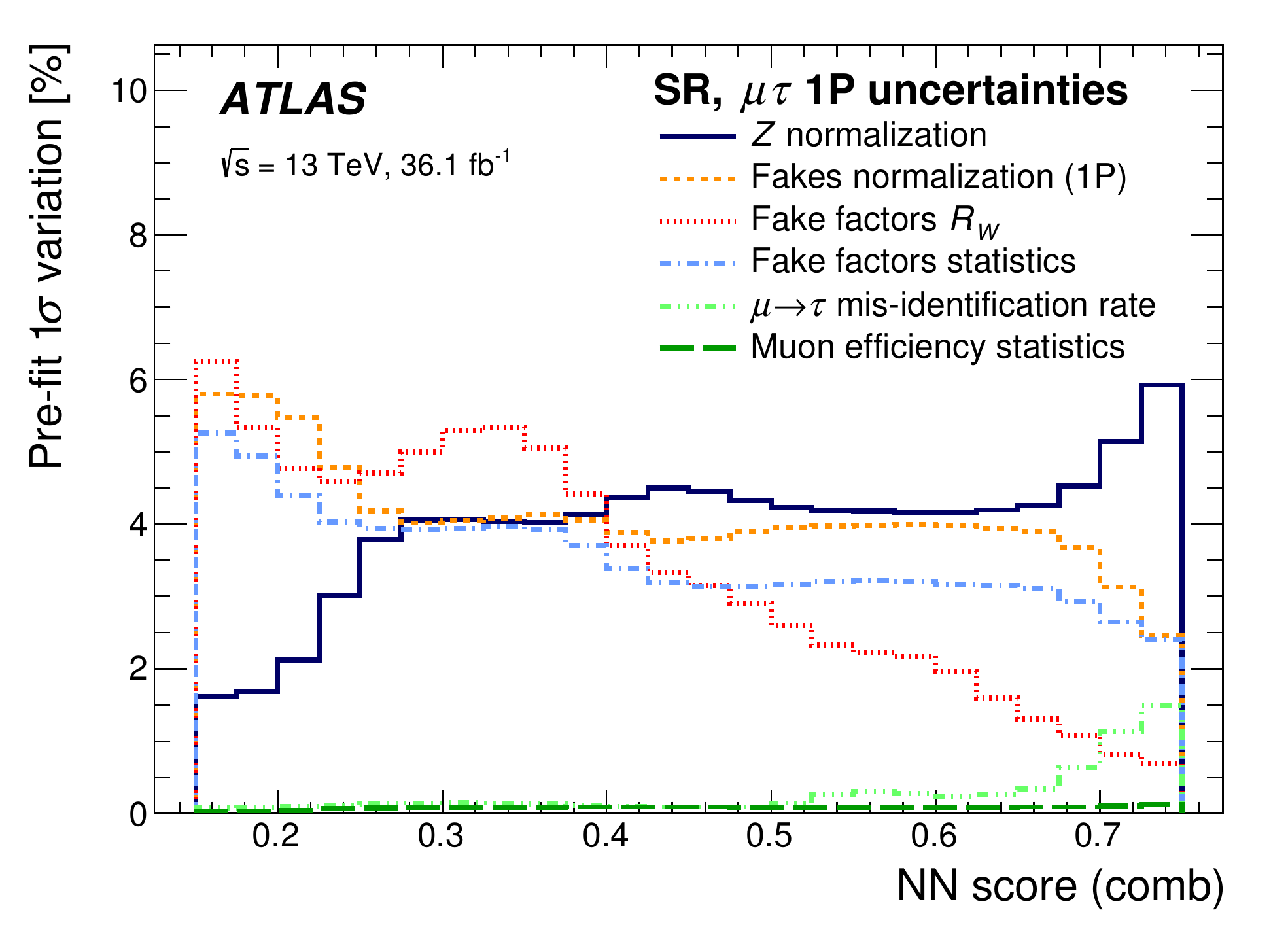}
    \includegraphics[width=0.45\textwidth]{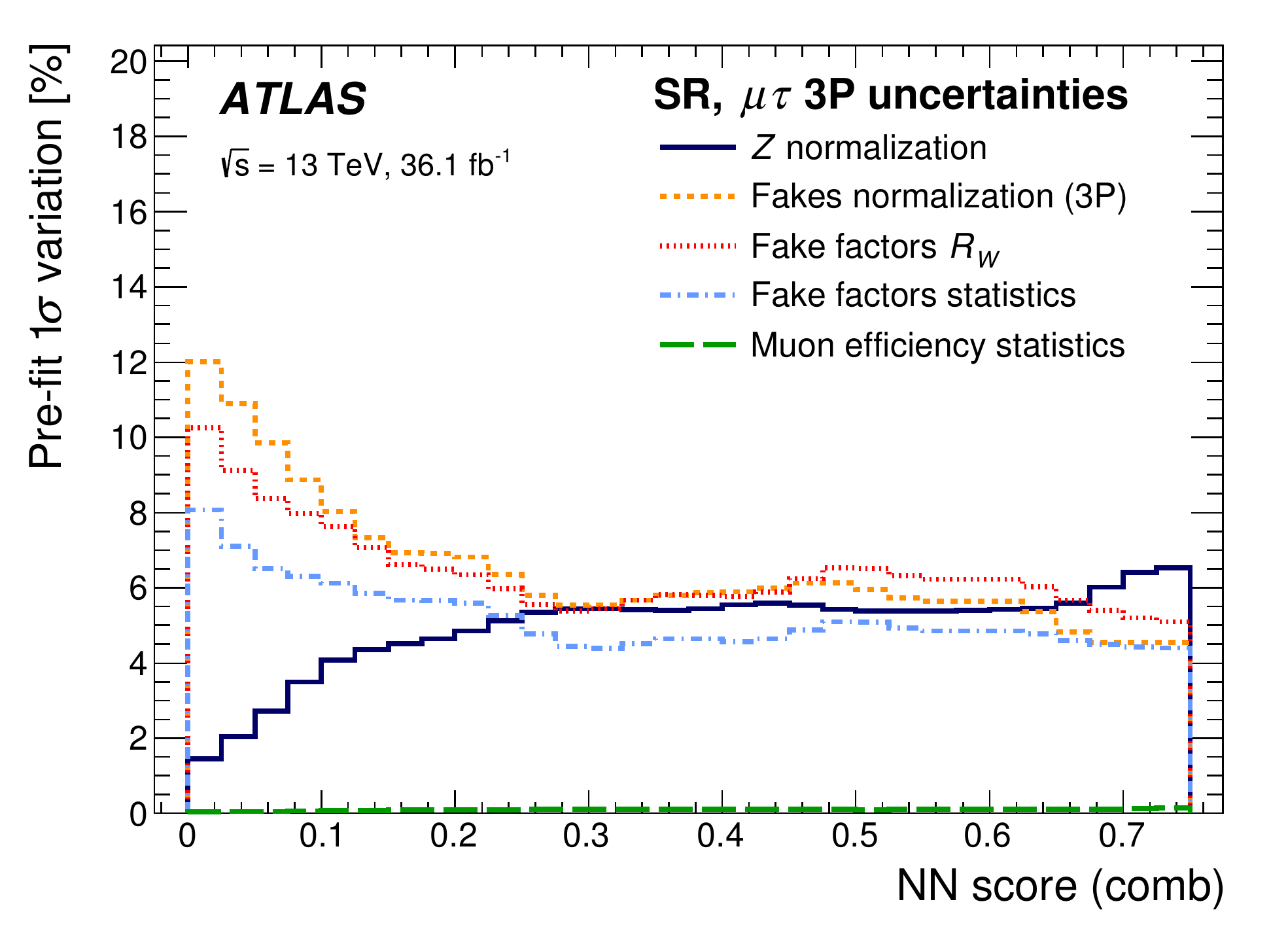}
  \caption{Expected uncertainties in the total background predictions in the SR as a function of the combined NN output for the dominant systematic uncertainties in $e\tau$ (top) and $\mu\tau$ (bottom) channels with 1-prong (left) and 3-prong (right) $\tauhadvis$ candidates. The uncertainties in the normalizations of the $Z$ and fake components are based on the expected statistical power of the fit described in Section~\ref{sec:result}. ``Muon efficiency statistics" refers to the statistical uncertainty of the corrections applied to the simulated muon reconstruction efficiency~\cite{PERF-2015-10}. ``Tau energy scale in situ" refers to the uncertainty of the corrections applied to the energy of the \tauhadvis candidate based on measurements with \Ztt data \cite{ATLAS-CONF-2017-029}.} 
  \label{fig:prefit-syst-impacts-vs-NN}
\end{figure}

To illustrate the sizes of the systematic uncertainties, Figure~\ref{fig:prefit-syst-impacts-vs-NN} shows the relative uncertainties of the total background predictions as a function of the combined NN output for the dominant systematic uncertainties. The uncertainties in the 
normalizations of the $Z$ and fake components, estimated from the expected statistical power of the fit described in Section~\ref{sec:result}, and the  
statistical uncertainty in the fake factor 
are the largest sources of systematic uncertainty, contributing on average between 3\% and 6\%. The systematic uncertainty in $R_{\mathrm{W}}$ is also relevant and ranges between 1\% and 6\% 
over the different final states. All other systematic uncertainties affect the total background prediction by less than 
one percent.

\section{Results and statistical interpretation}
\label{sec:result}

A binned maximum-likelihood fit to data, performed with the statistical analysis packages RooFit \cite{RooFit}, RooStats \cite{RooStats} and HistFitter \cite{HistFitter}, is used to compare the observed binned distributions of the combined NN classifiers in the SR with the model, and to extract evidence of signal events. The parameter of interest 
in such fit is the signal strength modifier $\mu_\text{sig}$, which quantifies the size of the LFV decay branching fraction $\BR (Z\to\ell\tau)$.

Fits are performed independently for the $e\tau$ and $\mu\tau$ channels, and in each fit events with a 1-prong $\tau_\text{had-vis}$ candidate and those with a 3-prong candidate are considered separately. In the fits of events with 1-prong \tauhadvis candidates, because of the way the NN classifiers are combined, only
a few background-like events 
have an NN output value below 0.15; these are excluded.
Independent templates, estimated as described in previous sections, are used for signal, $Z\to\tau\tau$, fakes, $Z\to\ell\ell$, top events, and $W(\to\tau\nu)$+jets events. The small contributions from Higgs boson and diboson events are summed into a single template, referred to as ``Other''. 

The likelihood is the product of Poisson probability density functions describing the observed number of events in each bin. It also includes Gaussian, Poisson and log-normal distributions to constrain the nuisance parameters associated with the systematic, statistical and theoretical uncertainties in the predicted number of events, respectively.
In addition to the parameter of interest and the nuisance parameters, three normalization parameters are included: $\mu(Z)$ determines the normalizations of the $Z\to\tau\tau$ and signal events while $\mu(\mathrm{fakes\_1P})$ and $\mu(\mathrm{fakes\_3P})$ control the normalization of the fake component in events with a 1-prong or a 3-prong \tauhadvis candidate, respectively.
These parameters are fit independently in the $e\tau$ and $\mu\tau$ channels. Within the same channel, the same $\mu(Z)$ is used to fit events with 1-prong and 3-prong \tauhadvis candidates, while $\mu(\mathrm{fakes\_1P})$ and $\mu(\mathrm{fakes\_3P})$ are used to fit independently the corresponding contributions from fakes.
By fitting the overall normalizations of the $Z\to\tau\tau$ and signal event yields, the $\mu(Z)$ parameter accounts for uncertainties in these contributions due to theoretical uncertainties on the $Z$ production cross section as well as experimental uncertainties in the measurement of the integrated luminosity and in the acceptance times efficiency of the $\ell\tau$ final state (uncertainties due to trigger, reconstruction, isolation and identification efficiencies). Therefore, $\mu(Z)$ ensures that the same Z production cross section and the same $\ell\tau$ acceptance efficiencies are used in the predictions of the signal and the $Z\to\tau\tau$ background contribution. The normalization with $\mu(Z)=1$ corresponds to the $Z$ production cross section of 2.1 nb, the SM value calculated at NNLO accuracy.
The normalization parameters $\mu(\mathrm{fakes\_1P})$ and $\mu(\mathrm{fakes\_3P})$ account for the systematic uncertainties in the overall normalizations of the fake contributions, so that only the systematic uncertainties in the template shape are implemented as nuisance parameters.
The fitted values of these parameters are sensitive to the yields of events with low NN outputs, which are dominated by contributions from $Z\to\tau\tau$ and fakes. 
Fitting these normalization parameters reduces the systematic uncertainties in the predictions of the $Z\to\tau\tau$ and fake backgrounds in the bins at high NN output, which are sensitive to the $Z\to\ell\tau$ signal.

\begin{figure}[tbp]
  \centering
    \includegraphics[width=0.45\textwidth]{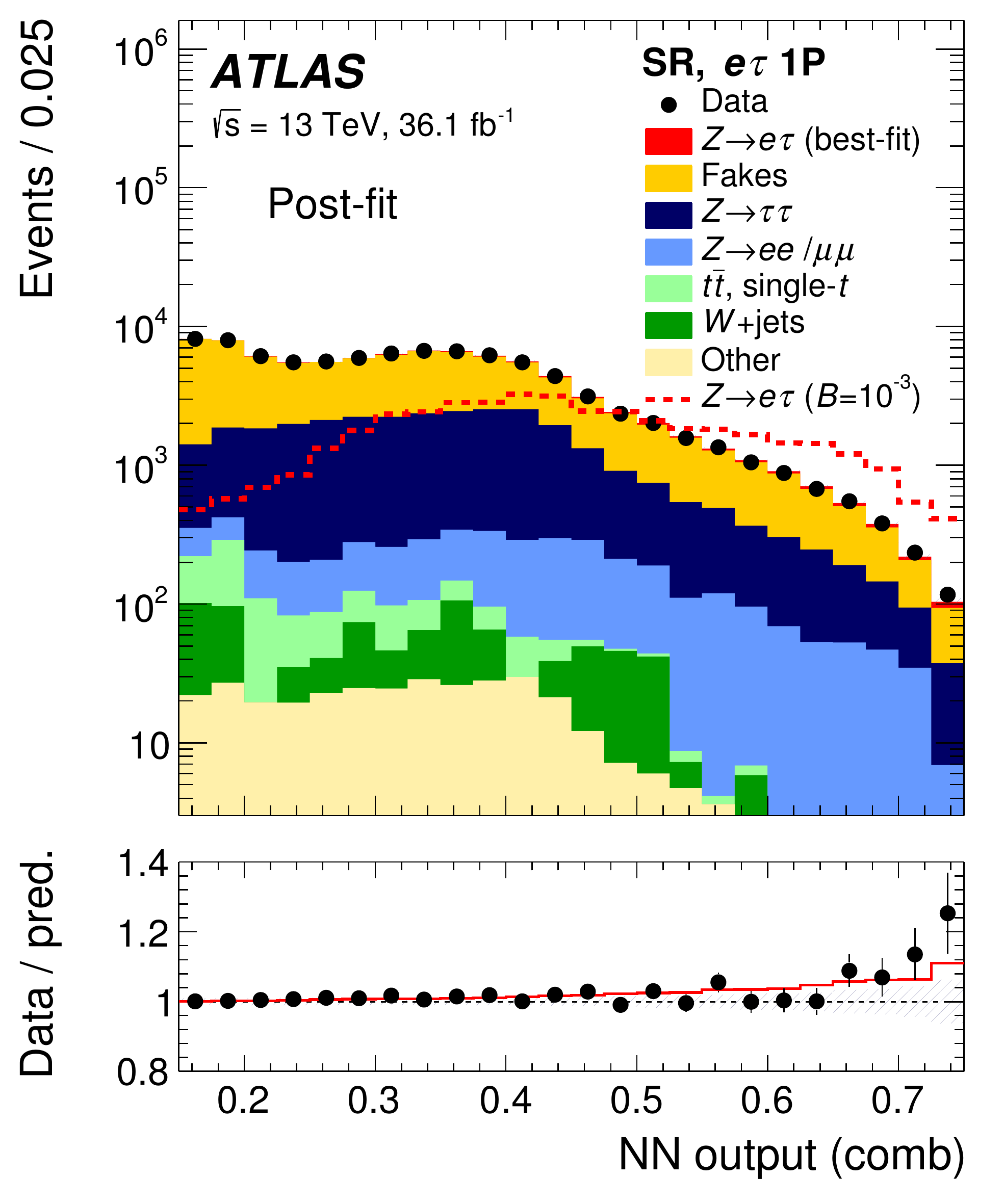}
    \includegraphics[width=0.45\textwidth]{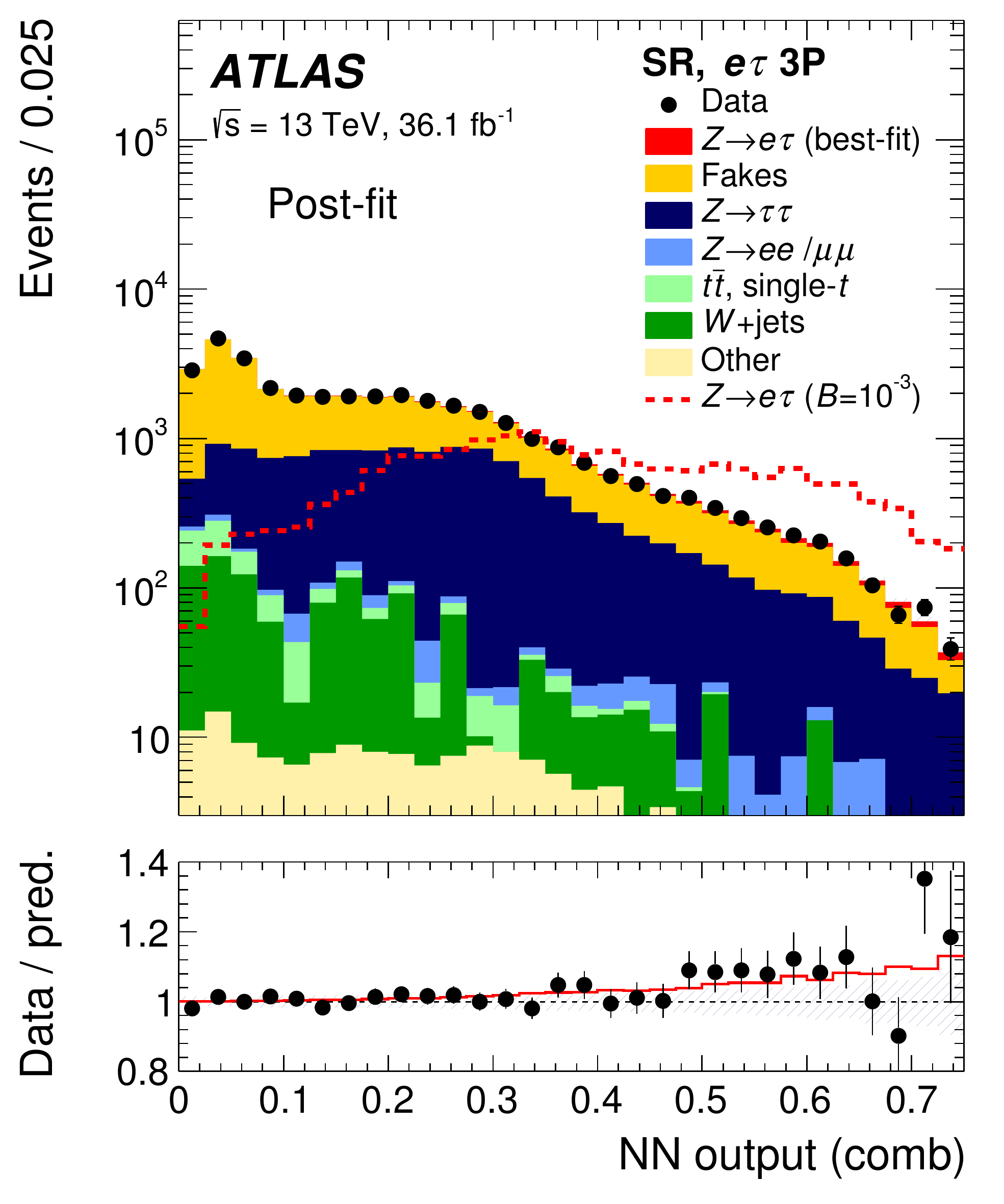} \\
    \includegraphics[width=0.45\textwidth]{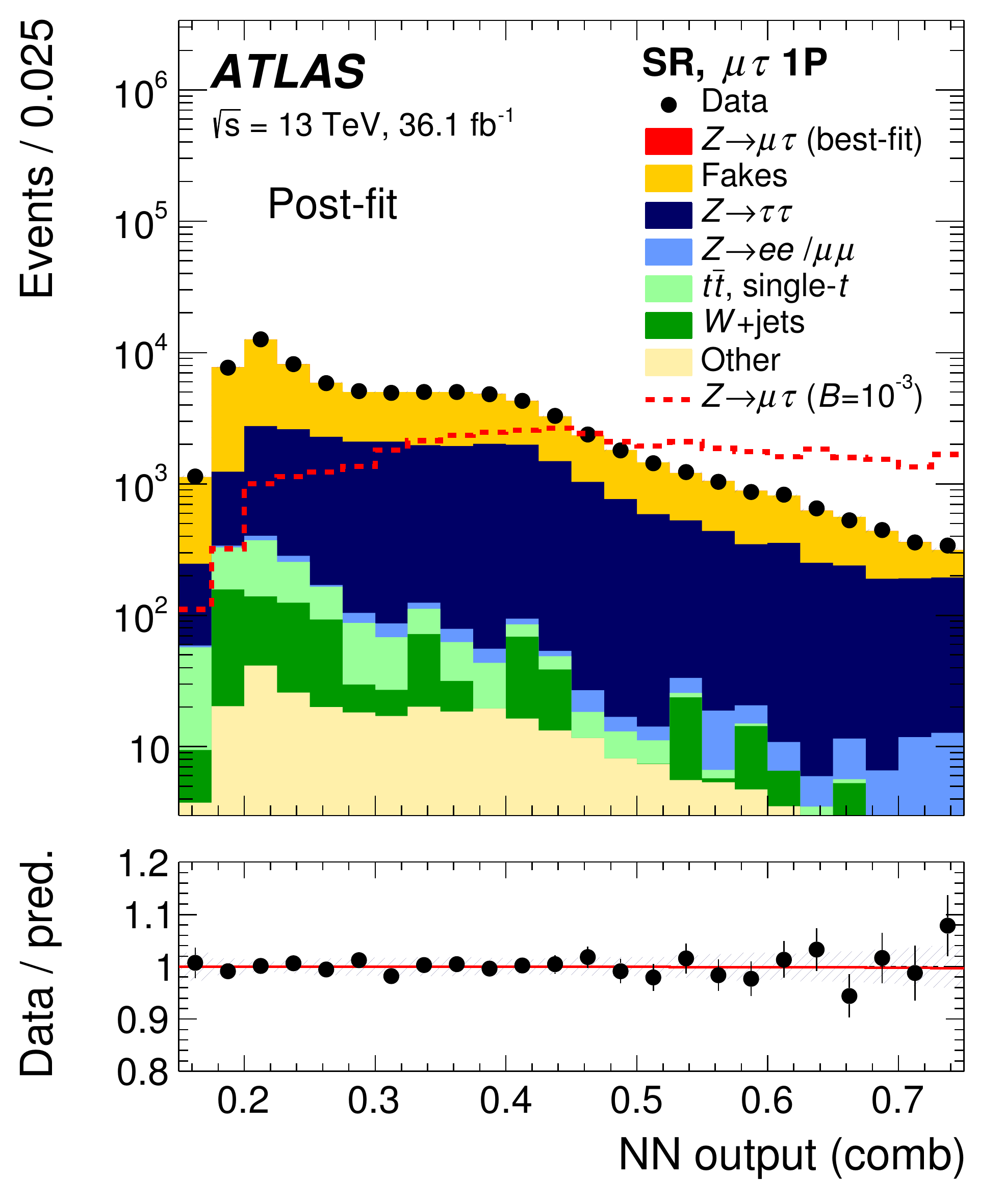}
    \includegraphics[width=0.45\textwidth]{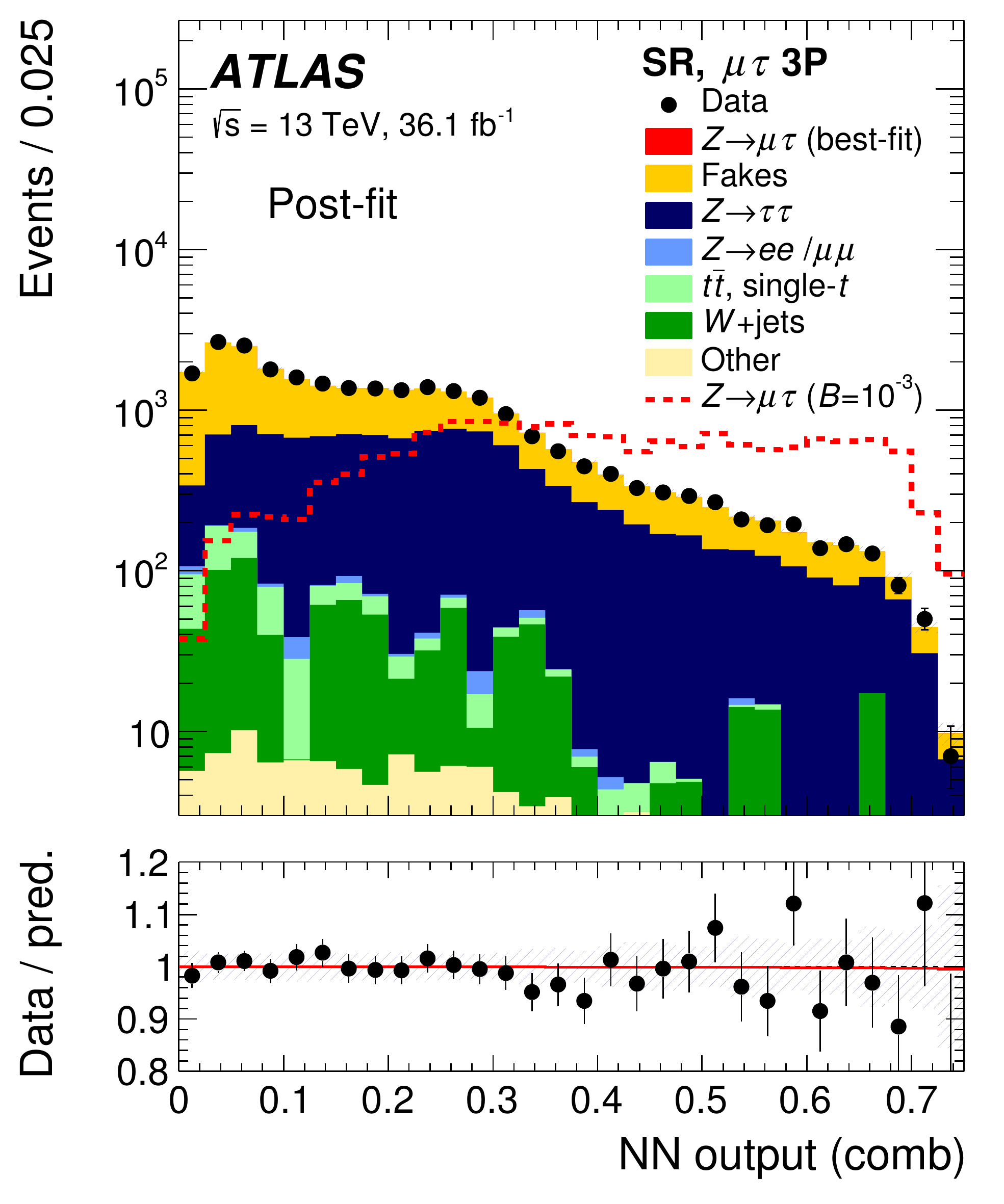}
  \caption{Observed and expected post-fit distributions of the combined NN output in SR for the $e\tau$ (top) and $\mu\tau$ (bottom) channels, for 1-prong (left) and 3-prong (right) \tauhadvis candidates. The filled histogram stacked on top of the backgrounds represents the signal normalized to the best-fit $\BR(Z\to\ell\tau)$. The overlaid dashed line represents the expected distribution for the signal normalized to $\BR(Z\to\ell\tau)=10^{-3}$. In the panels below each plot, the ratios of the observed data (dots) and the post-fit background plus signal (solid line) to the post-fit background are shown. The hatched error bands represent the combined statistical and systematic uncertainties. The first and last bins include underflow and overflow events, respectively.} 
\label{fig:post-fit-SR}
\end{figure}

Table~\ref{tab:yields_unblindNNfit_combined} reports the total observed number of events and post-fit event yields in the SR after a fit to data. The observed and post-fit expected distributions of the combined NN output are shown in Figure~\ref{fig:post-fit-SR}. As reported in Table~\ref{tab:post-fit-mu}, the best-fit values for $\mu(Z)$, $\mu(\text{fakes\_1P})$ and $\mu(\text{fakes\_3P})$ are consistent between the $e\tau$ and $\mu\tau$ channels, while the best-fit value for $\BR (Z\to\ell\tau)$ is consistent with zero in the $\mu\tau$ channel, $\BR (Z\to\mu\tau)=(-0.1^{+1.2}_{-1.2})\times10^{-5}$, and slightly fluctuating to positive values in the $e\tau$ channel, $\BR (Z\to e\tau)=(3.3^{+1.5}_{-1.4})\times10^{-5}$.

\begin{figure}[tbp]
  \centering
  \subfloat[$\hat{E}^{\mathrm{lep}}$]{
    \includegraphics[width=0.32\textwidth]{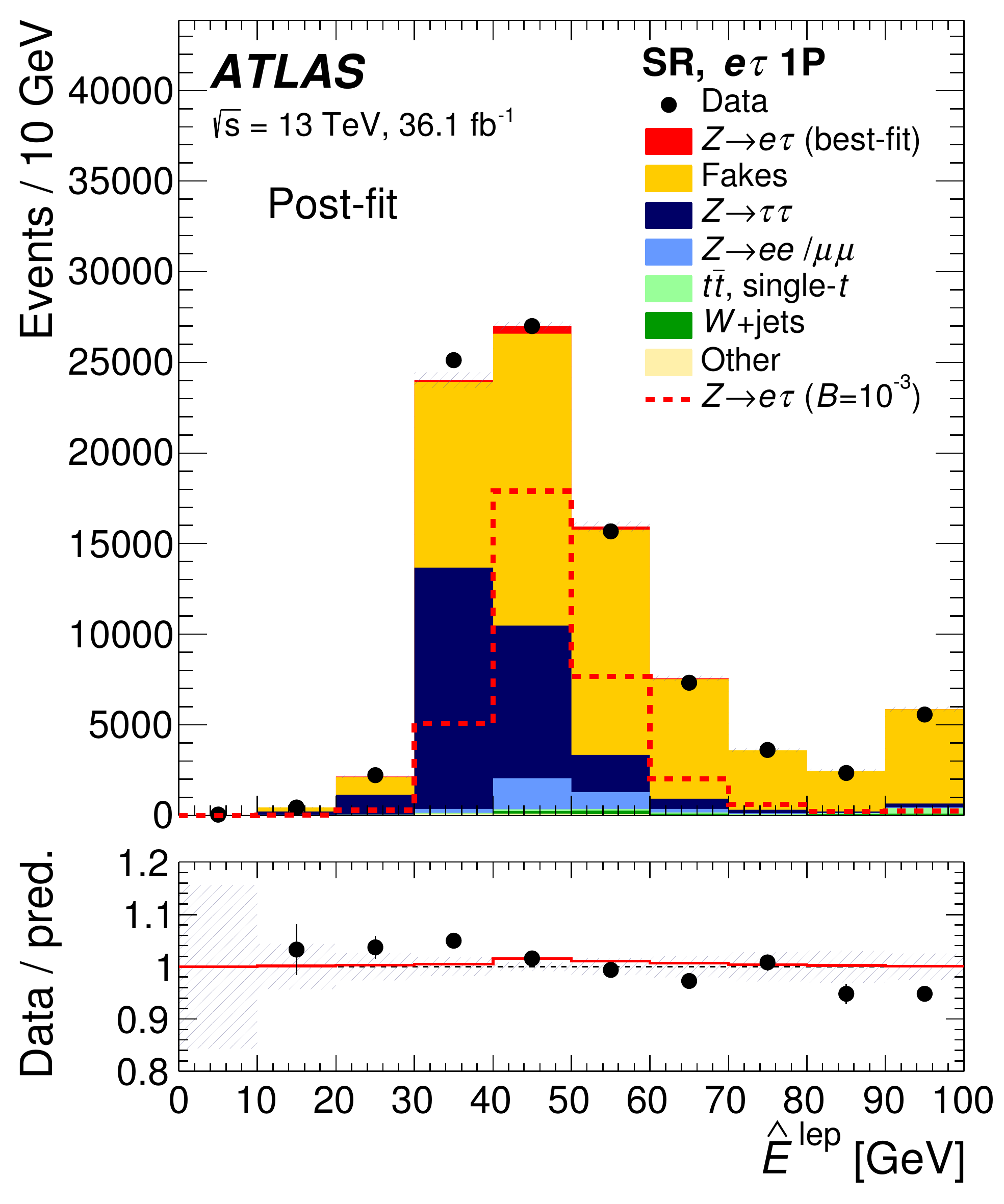}
    \label{}
  }
  \hfill
  \subfloat[$\hat{E}^{\tauh}$]{
    \includegraphics[width=0.32\textwidth]{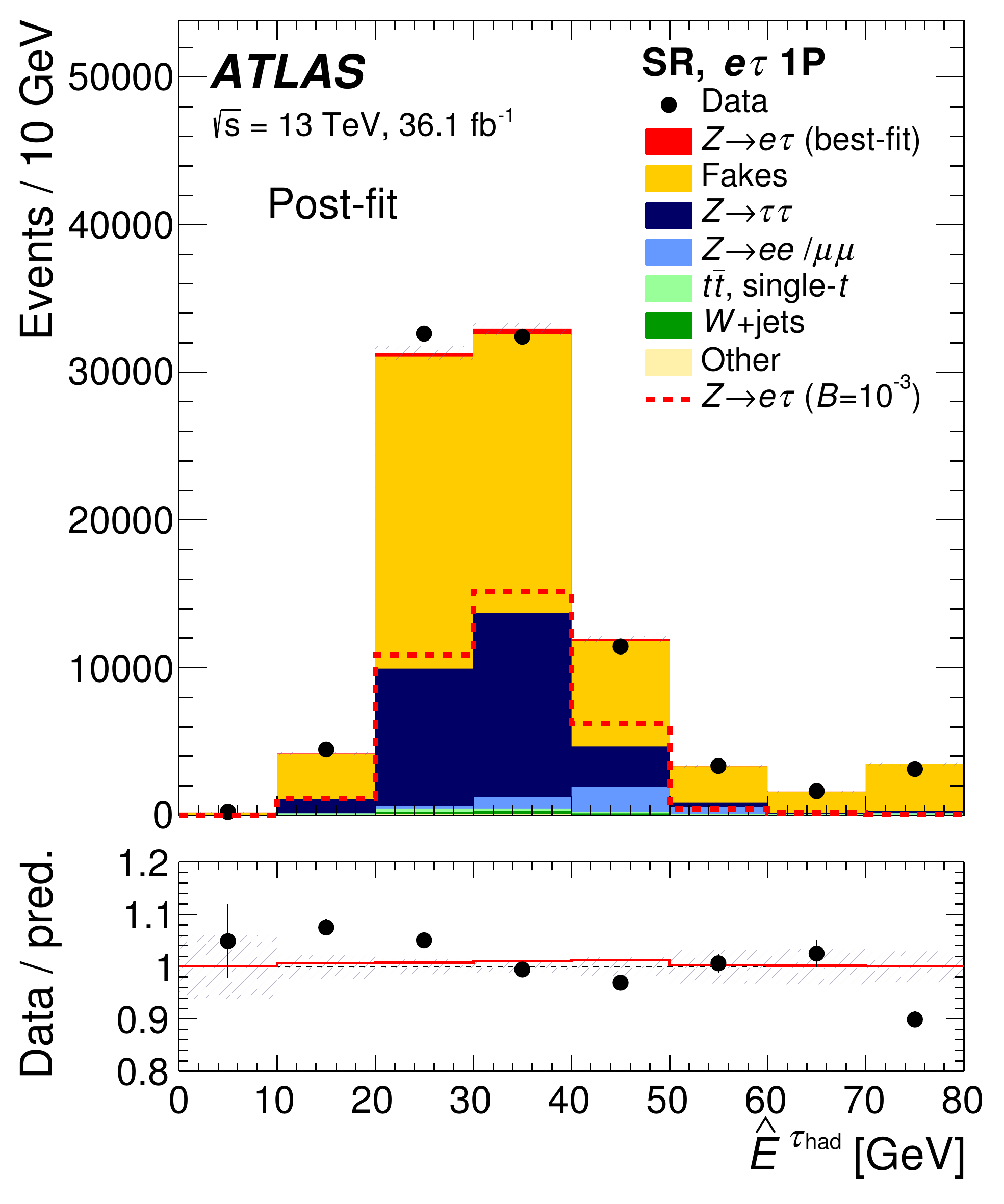}
    \label{}
  }
  \hfill
    \subfloat[$\hat{E}^{\mathrm{miss}}$]{
    \includegraphics[width=0.32\textwidth]{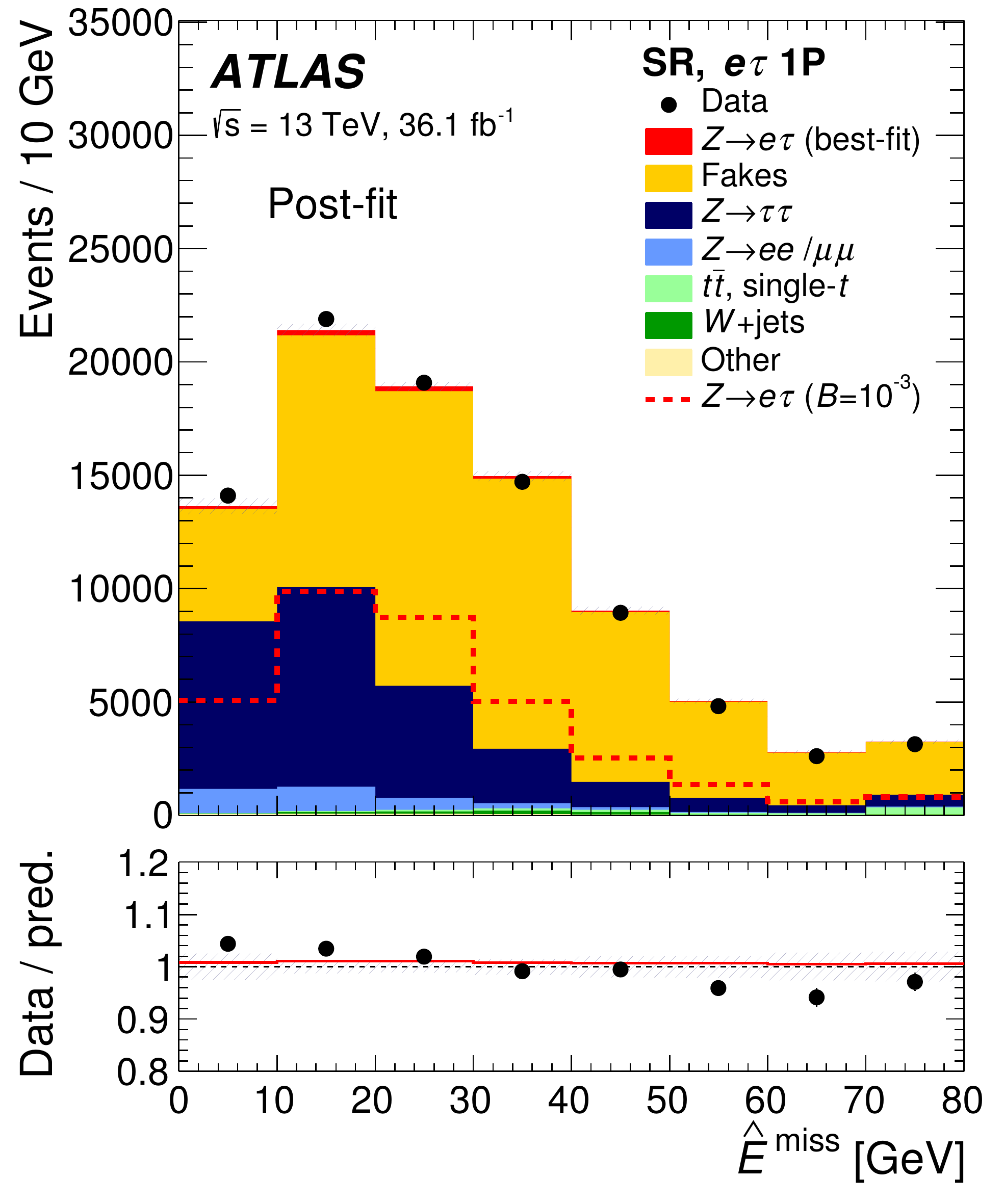}
    \label{}
  }
  \hfill
  \subfloat[$\Delta\alpha$]{
    \includegraphics[width=0.32\textwidth]{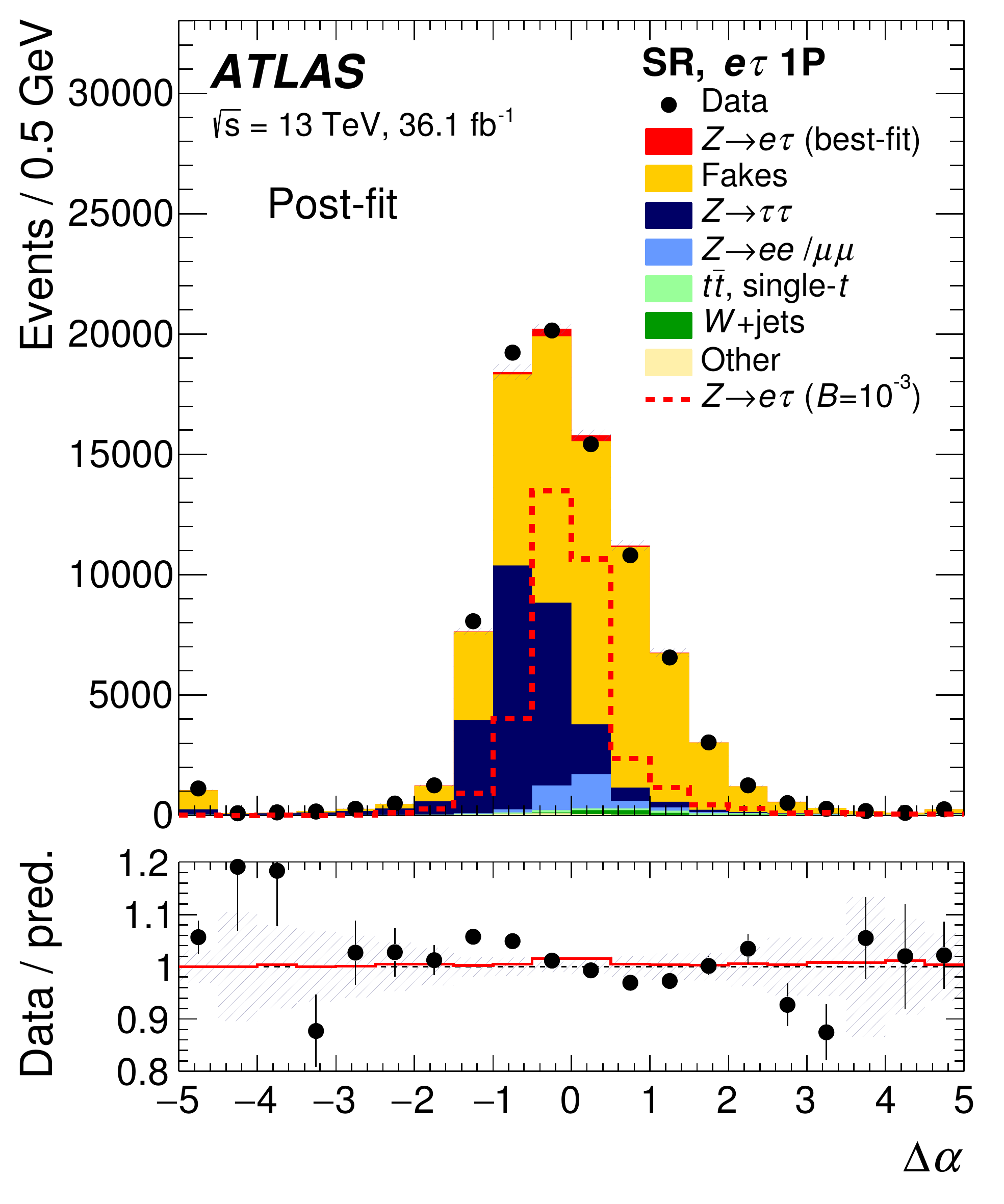}
    \label{}
  }
  \hfill
  \subfloat[$m_{\mathrm{coll}}$]{
    \includegraphics[width=0.32\textwidth]{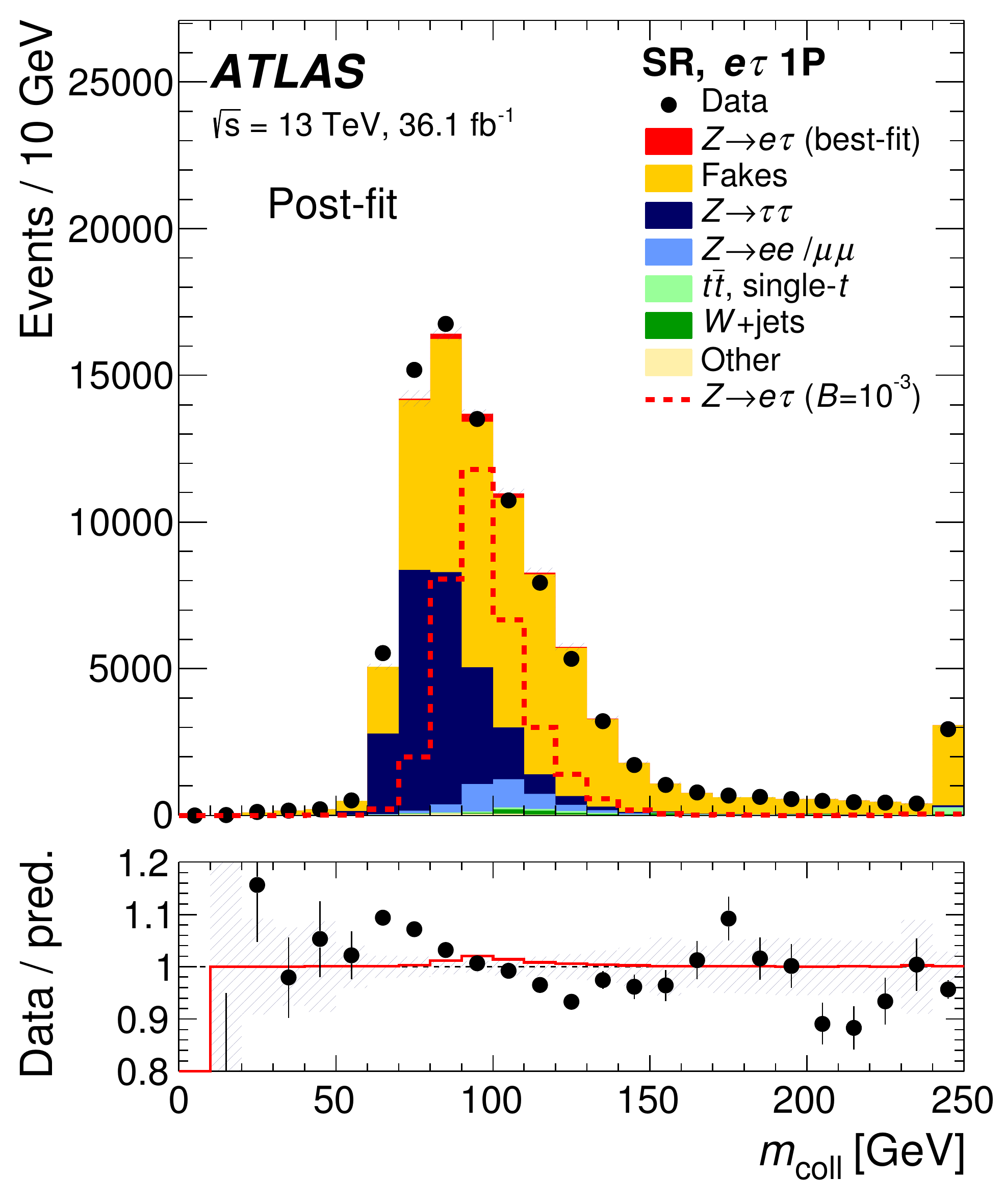}
    \label{}
  }
    \hfill
  \subfloat[$\pt^\mathrm{tot}$]{
    \includegraphics[width=0.32\textwidth]{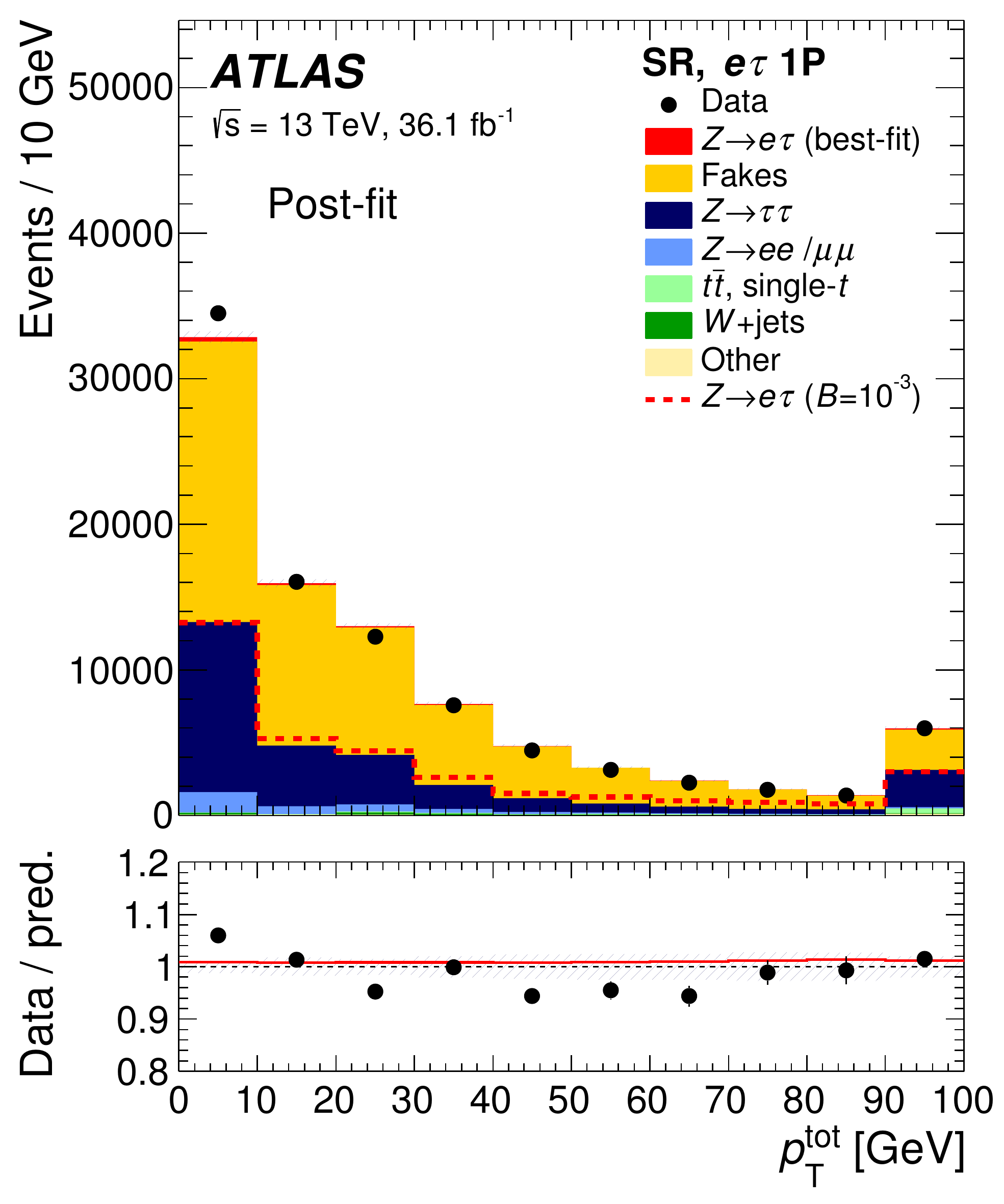}
    \label{}
  }
  \caption{Observed and expected post-fit distributions of unscaled NN inputs in SR for the $e\tau$ channel with 1-prong \tauhadvis candidates. The fit is based on profiling on the combined NN classifier, but not directly on these variables. The filled histogram stacked on top of the backgrounds represents the signal normalized to the best-fit $\BR(Z\to\ell\tau)$. The overlaid dashed line represents the expected distribution for the signal normalized to $\BR(Z\to\ell\tau)=10^{-3}$. In the panels below each plot, the ratios of the observed data (dots) and the post-fit background plus signal (solid line) to the post-fit background are shown. The hatched error bands represent the combined statistical and systematic uncertainties. The first and last bins include underflow and overflow events, respectively.} 
\label{fig:postfit_SR_el_NNInputs_1P}
\end{figure}

\begin{figure}[tbp]
  \centering
  \subfloat[$\hat{E}^{\mathrm{lep}}$]{
    \includegraphics[width=0.32\textwidth]{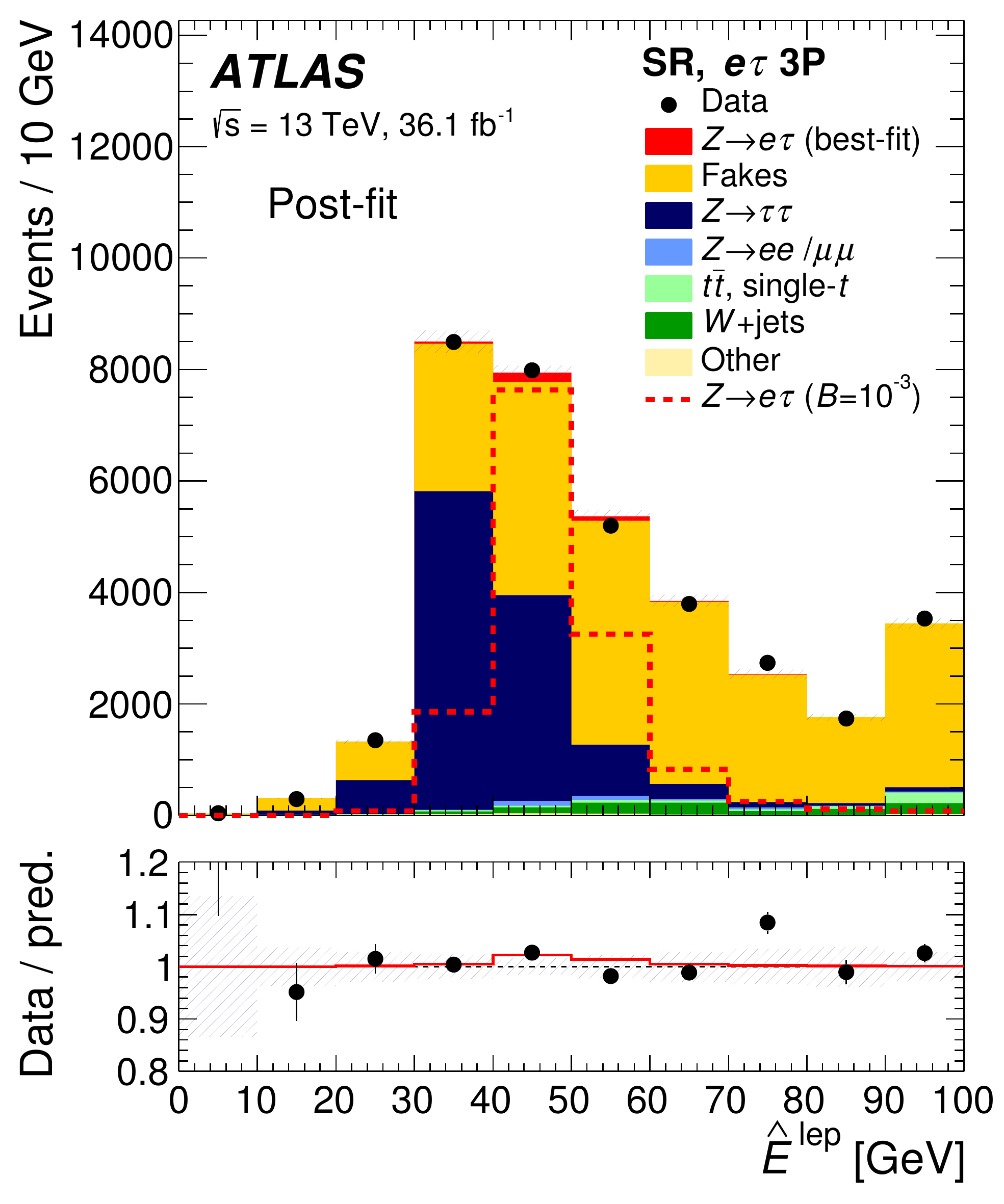}
    \label{}
  }
  \hfill
  \subfloat[$\hat{E}^{\tauh}$]{
    \includegraphics[width=0.32\textwidth]{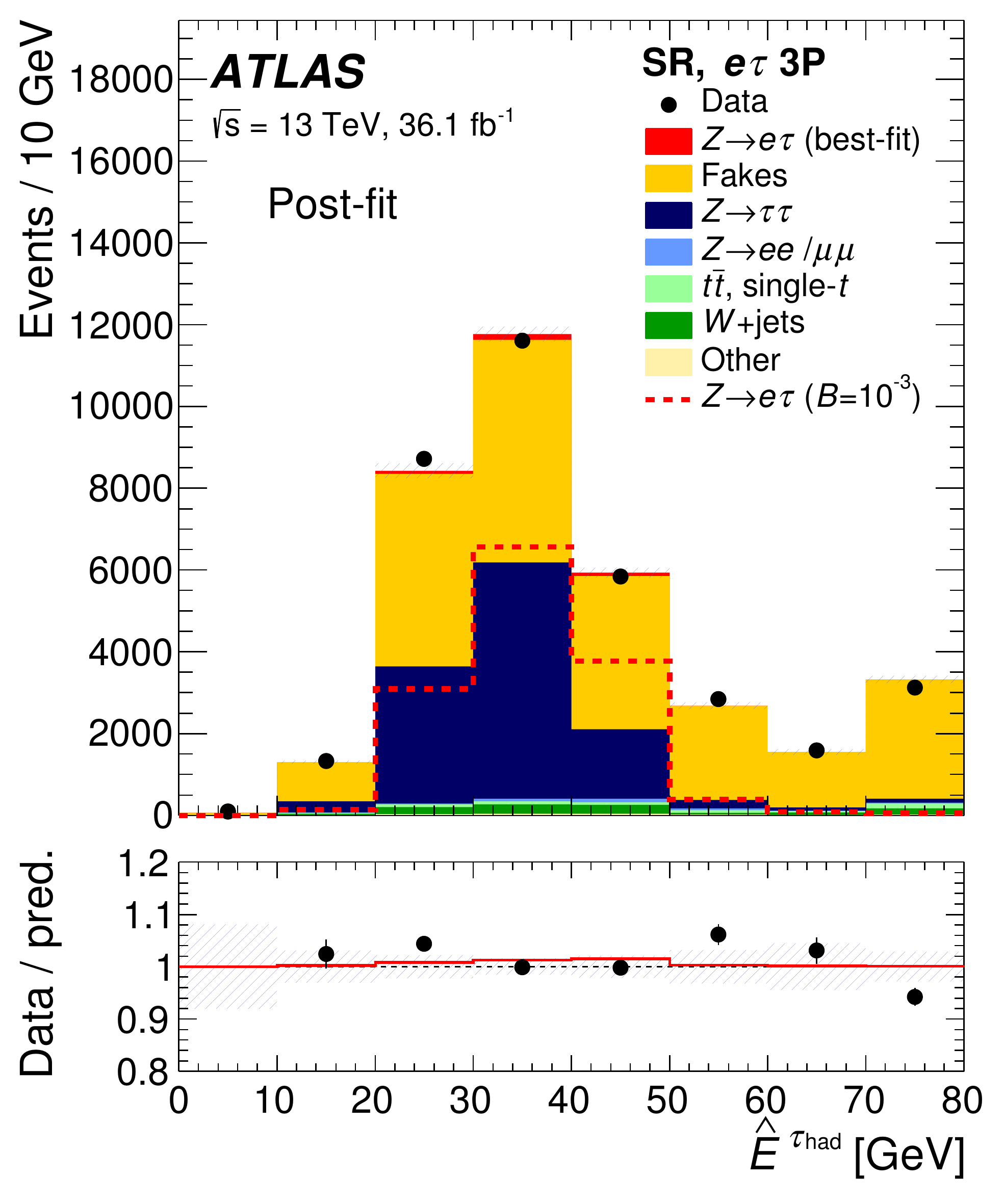}
    \label{}
  }
    \hfill
  \subfloat[$\hat{E}^{\mathrm{miss}}$]{
    \includegraphics[width=0.32\textwidth]{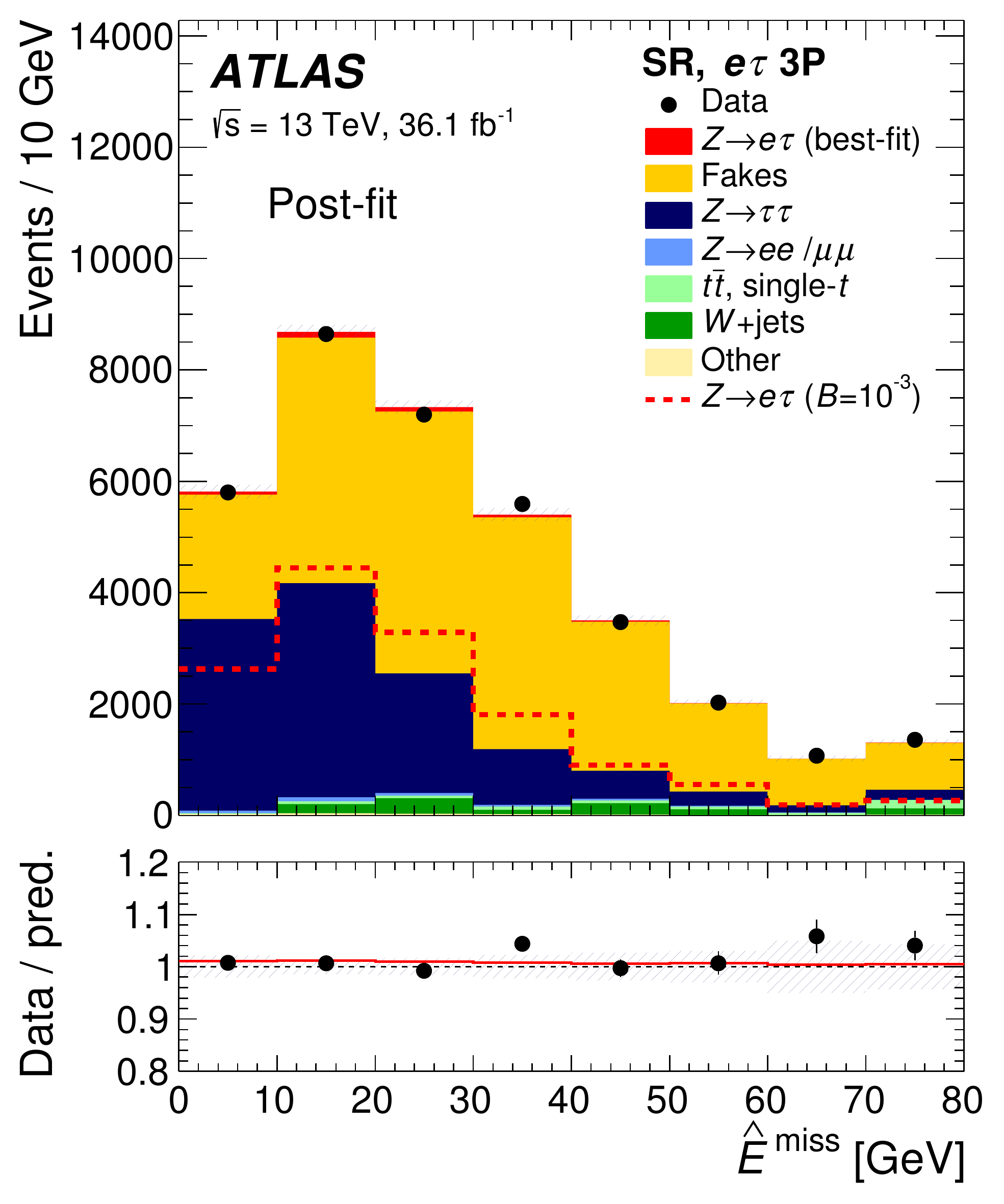}
    \label{}
  }
  \hfill
  \subfloat[$\Delta\alpha$]{
    \includegraphics[width=0.32\textwidth]{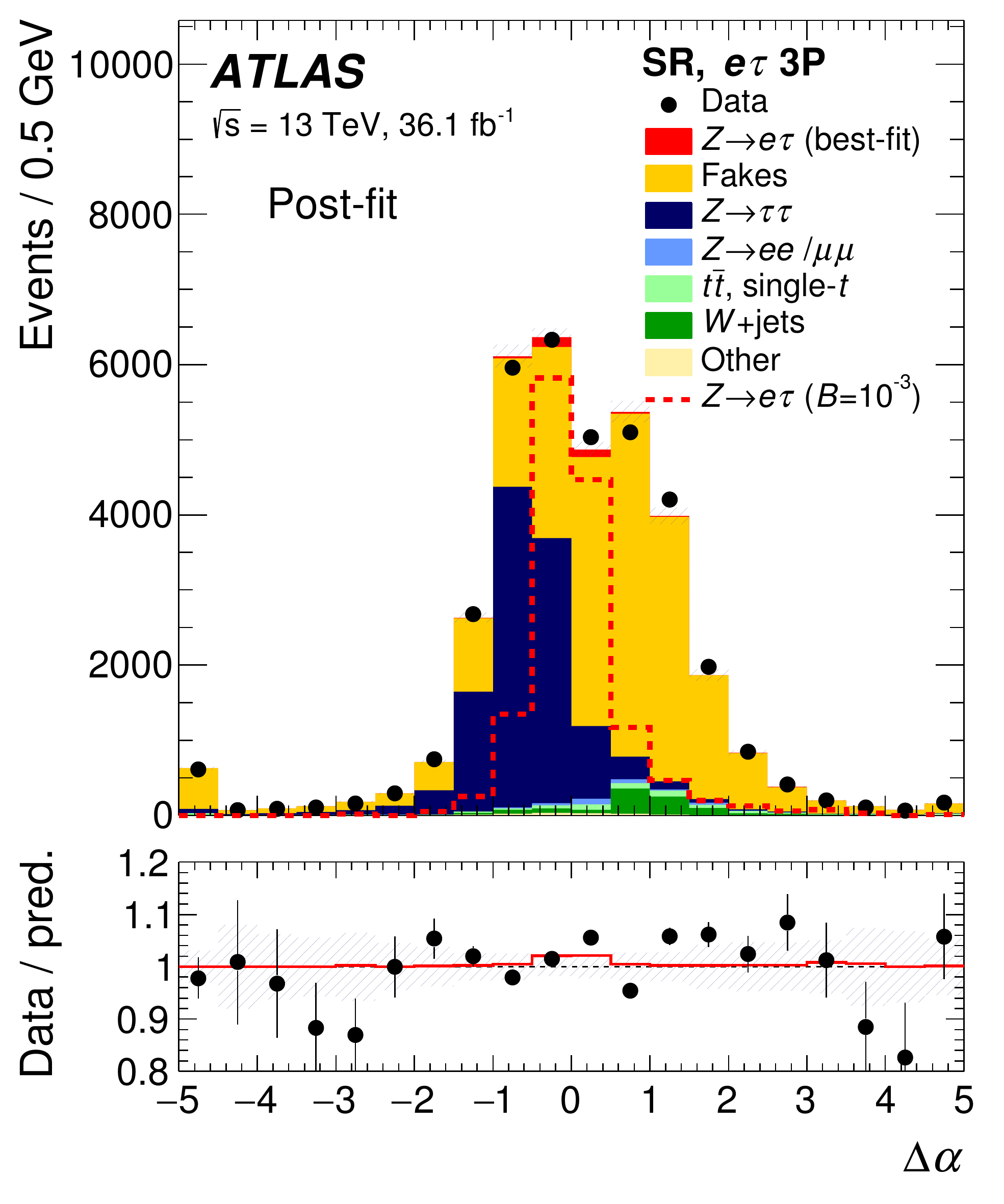}
    \label{}
  }
  \hfill
  \subfloat[$m_{\mathrm{coll}}$]{
    \includegraphics[width=0.32\textwidth]{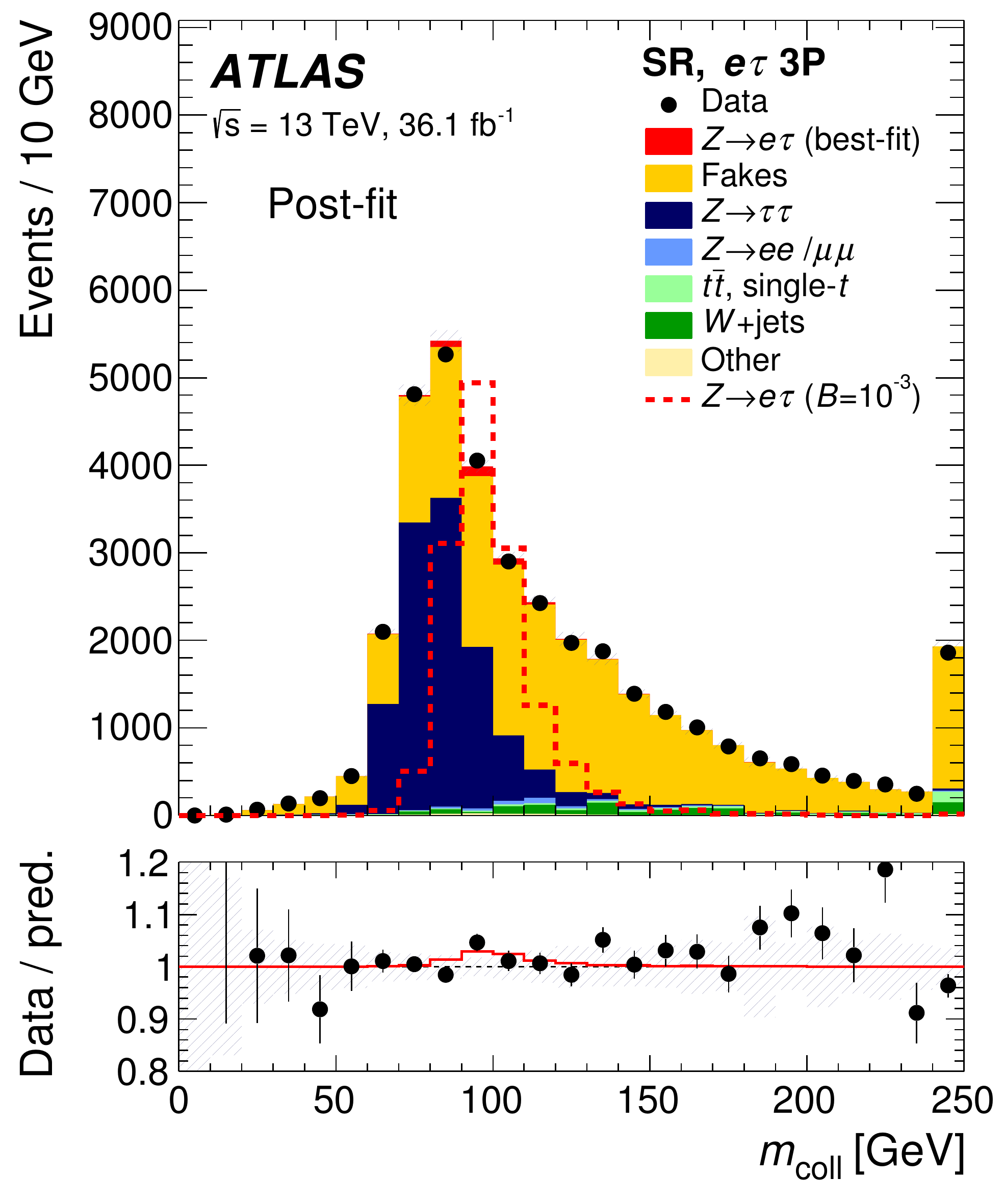}
    \label{}
  }
    \hfill
  \subfloat[$\pt^\mathrm{tot}$]{
    \includegraphics[width=0.32\textwidth]{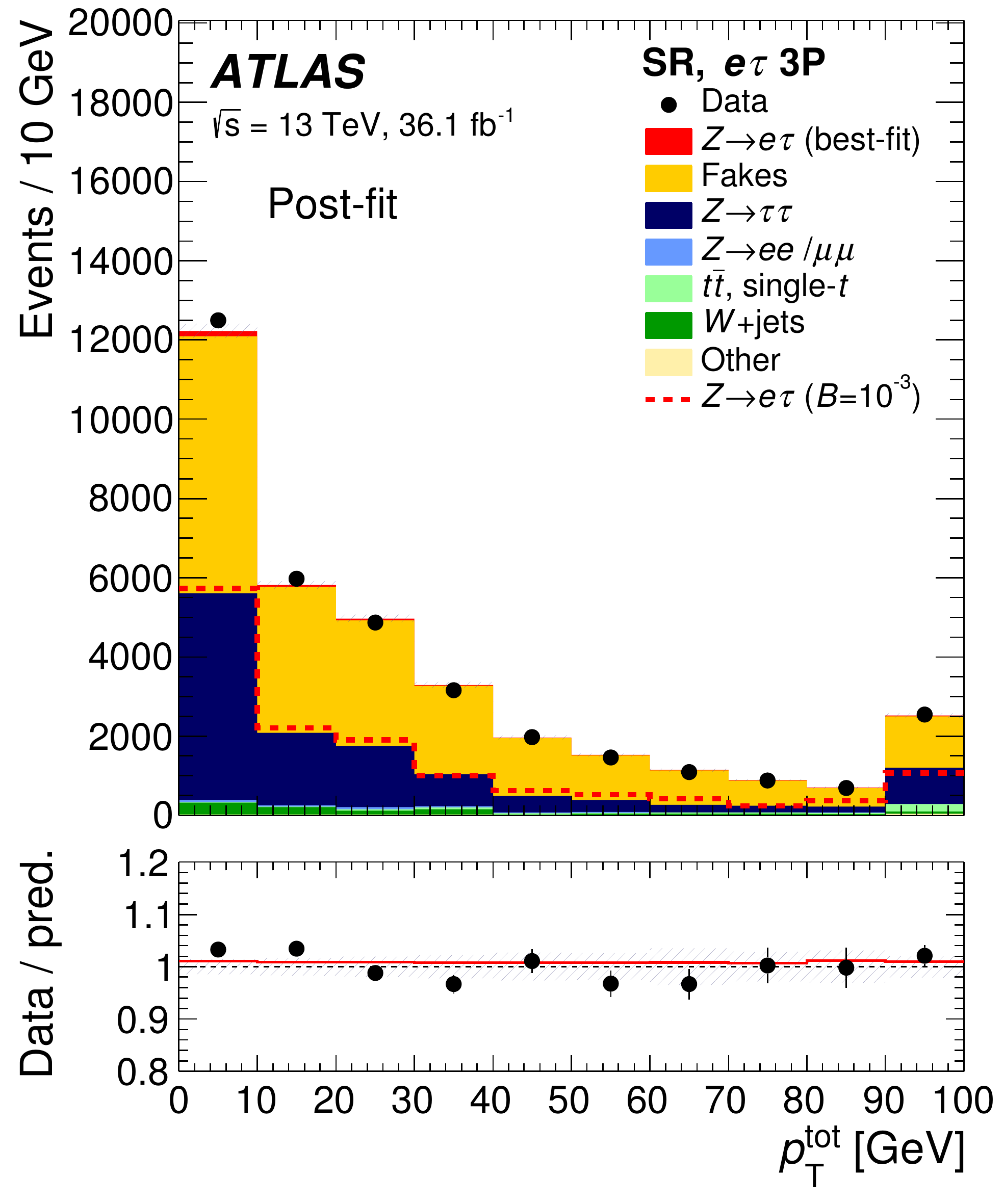}
    \label{}
  }
  \caption{Observed and expected post-fit distributions of unscaled NN inputs in SR for the $e\tau$ channel with 3-prong \tauhadvis candidates. The fit is based on profiling on the combined NN classifier, but not directly on these variables. The filled histogram stacked on top of the backgrounds represents the signal normalized to the best-fit $\BR(Z\to\ell\tau)$. The overlaid dashed line represents the expected distribution for the signal normalized to $\BR(Z\to\ell\tau)=10^{-3}$. In the panels below each plot, the ratios of the observed data (dots) and the post-fit background plus signal (solid line) to the post-fit background are shown. The hatched error bands represent the combined statistical and systematic uncertainties. The first and last bins include underflow and overflow events, respectively.} 
\label{fig:postfit_SR_el_NNInputs_3P}
\end{figure}

\begin{figure}[tbp]
  \centering
  \subfloat[$\hat{E}^{\mathrm{lep}}$]{
    \includegraphics[width=0.32\textwidth]{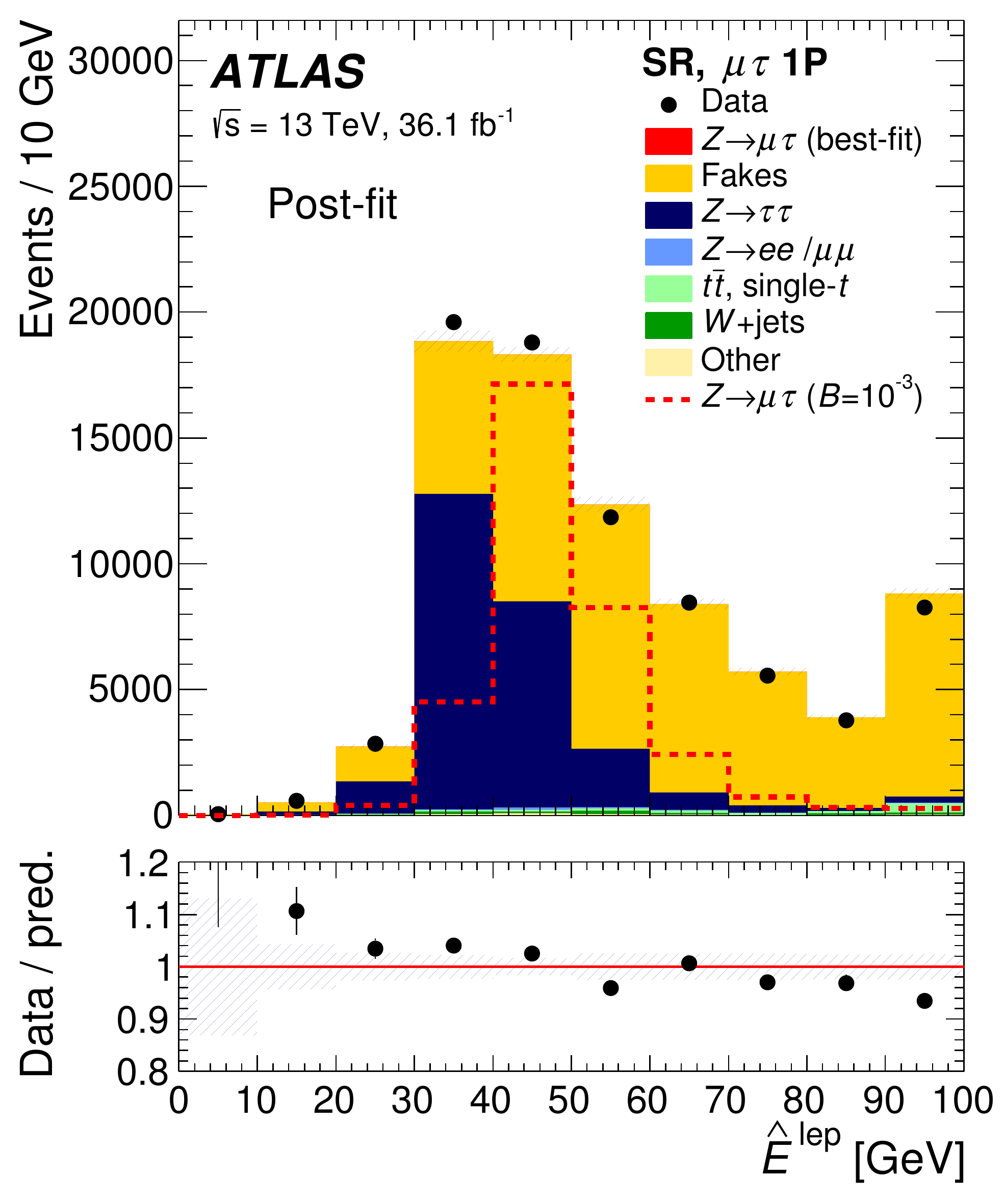}
    \label{}
  }
  \hfill
  \subfloat[$\hat{E}^{\tauh}$]{
    \includegraphics[width=0.32\textwidth]{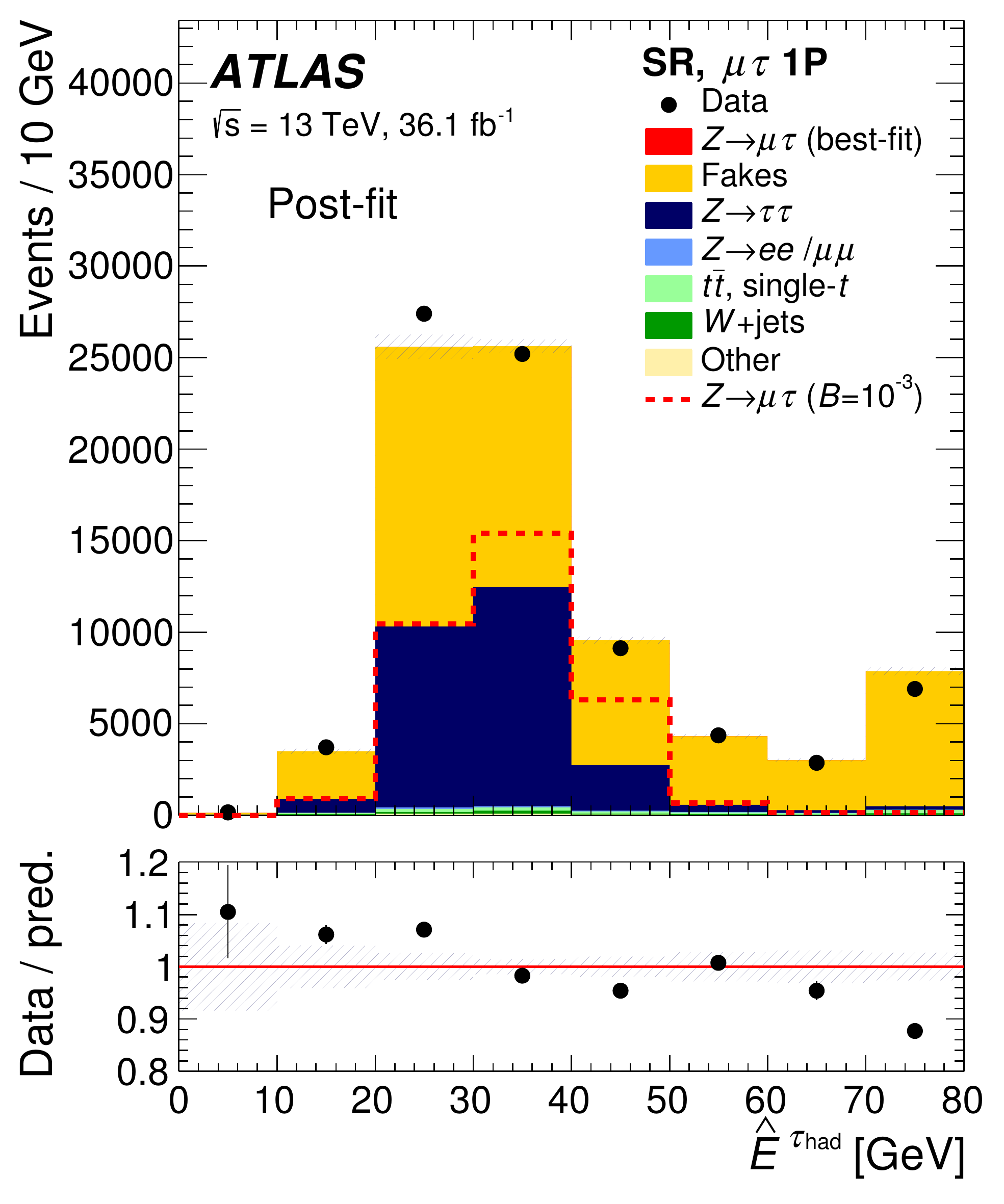}
    \label{}
  }
    \hfill
  \subfloat[$\hat{E}^{\mathrm{miss}}$]{
    \includegraphics[width=0.32\textwidth]{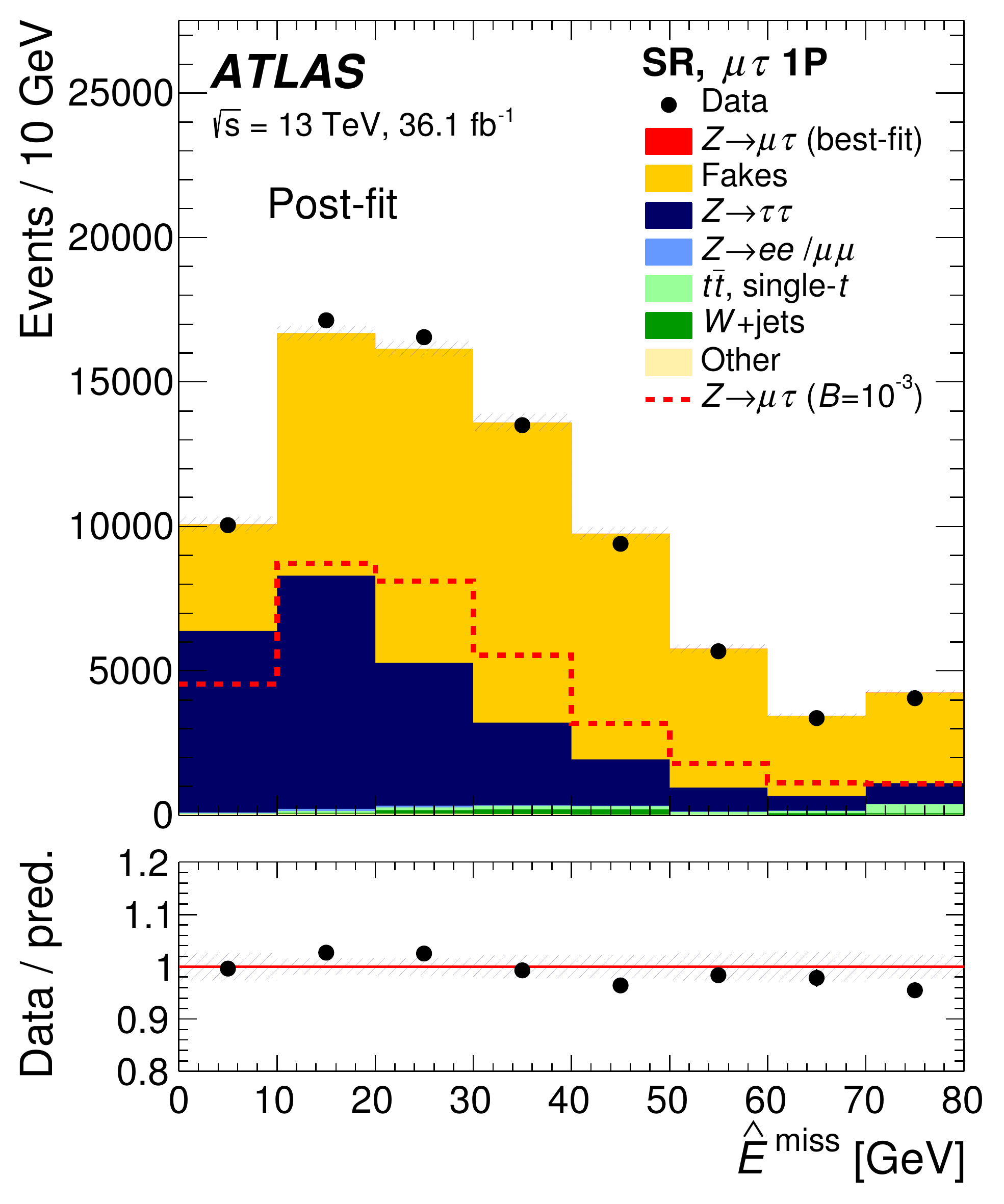}
    \label{}
  }
  \hfill
  \subfloat[$\Delta\alpha$]{
    \includegraphics[width=0.32\textwidth]{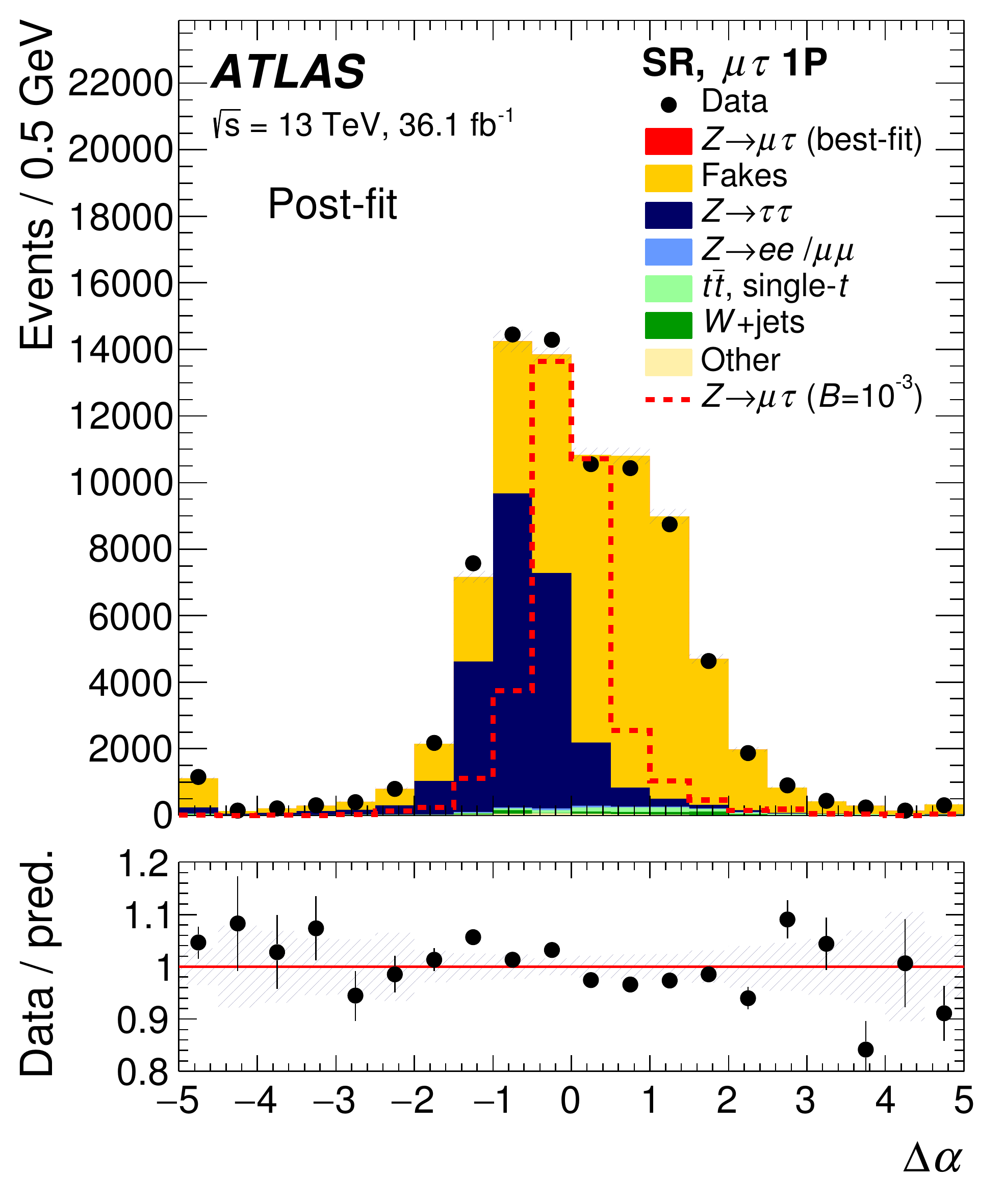}
    \label{}
  }
  \hfill
  \subfloat[$m_{\mathrm{coll}}$]{
    \includegraphics[width=0.32\textwidth]{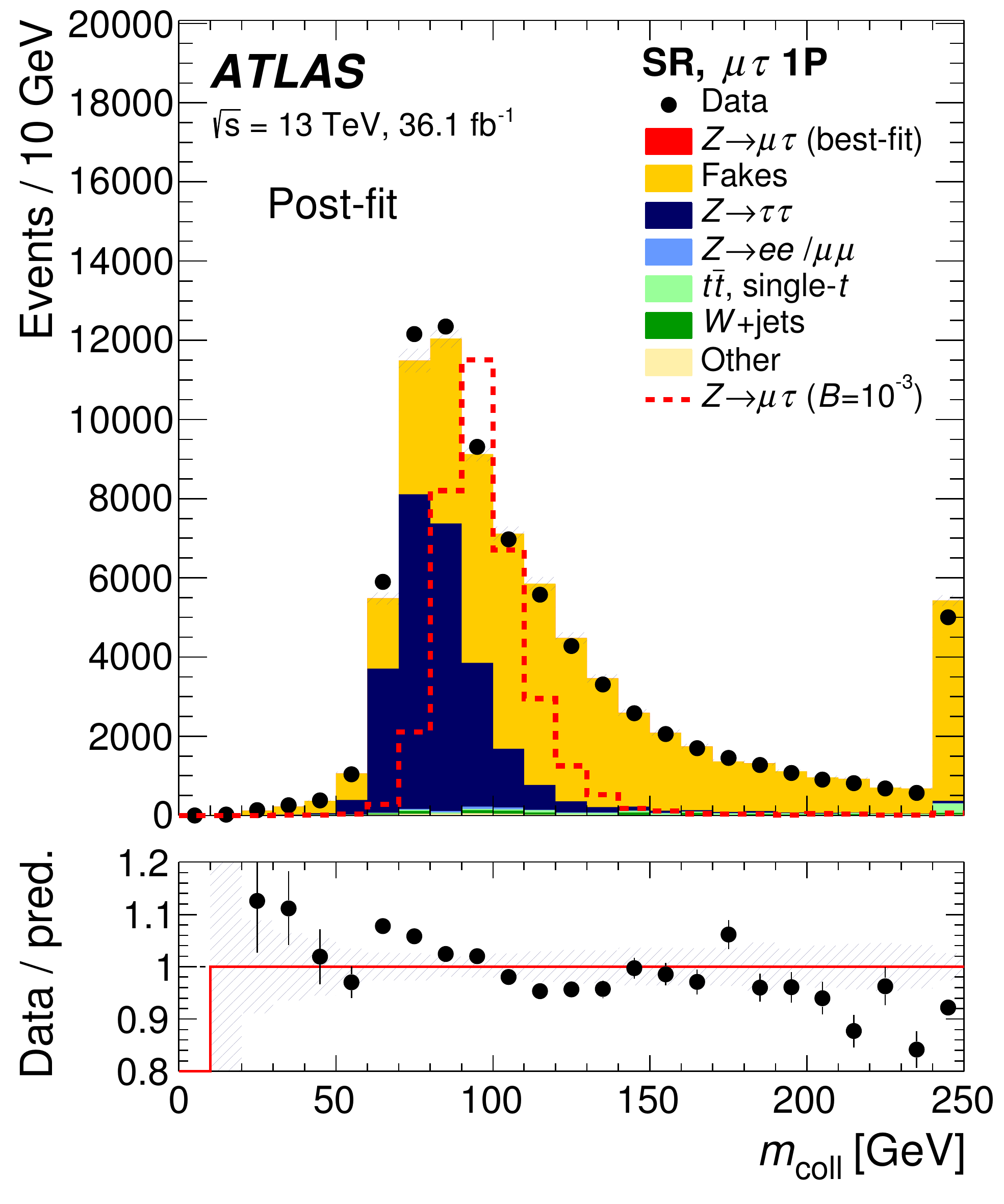}
    \label{}
  }
    \hfill
  \subfloat[$\pt^\mathrm{tot}$]{
    \includegraphics[width=0.32\textwidth]{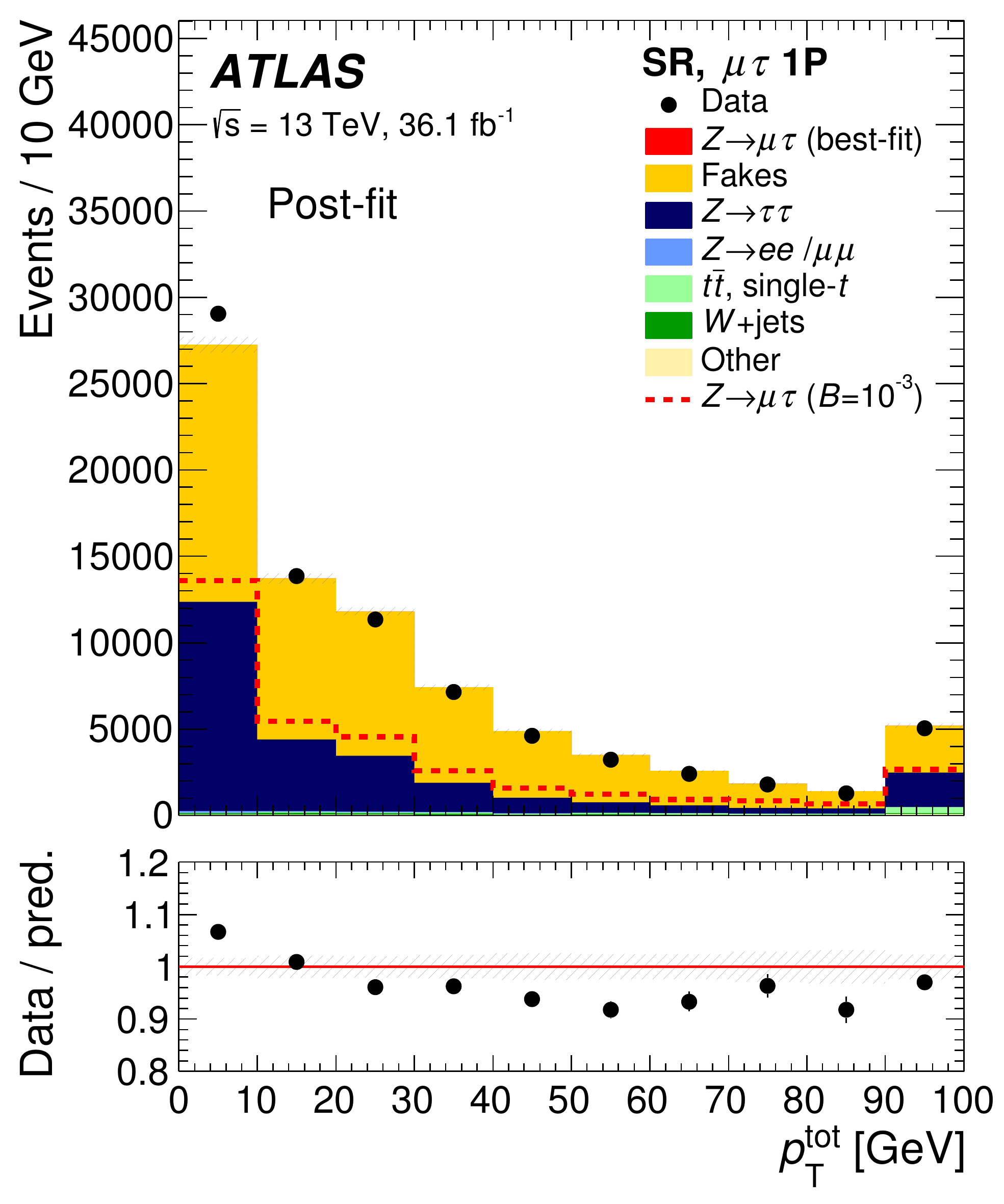}
    \label{}
  }
  \caption{Observed and expected post-fit distributions of unscaled NN inputs in SR for the $\mu\tau$ channel with 1-prong \tauhadvis candidates. The fit is based on profiling on the combined NN classifier, but not directly on these variables. The filled histogram stacked on top of the backgrounds represents the signal normalized to the best-fit $\BR(Z\to\ell\tau)$. The overlaid dashed line represents the expected distribution for the signal normalized to $\BR(Z\to\ell\tau)=10^{-3}$. In the panels below each plot, the ratios of the observed data (dots) and the post-fit background plus signal (solid line) to the post-fit background are shown. The hatched error bands represent the combined statistical and systematic uncertainties. The first and last bins include underflow and overflow events, respectively.} 
\label{fig:postfit_SR_mu_NNInputs_1P}
\end{figure}

\begin{figure}[tbp]
  \centering
  \subfloat[$\hat{E}^{\mathrm{lep}}$]{
    \includegraphics[width=0.32\textwidth]{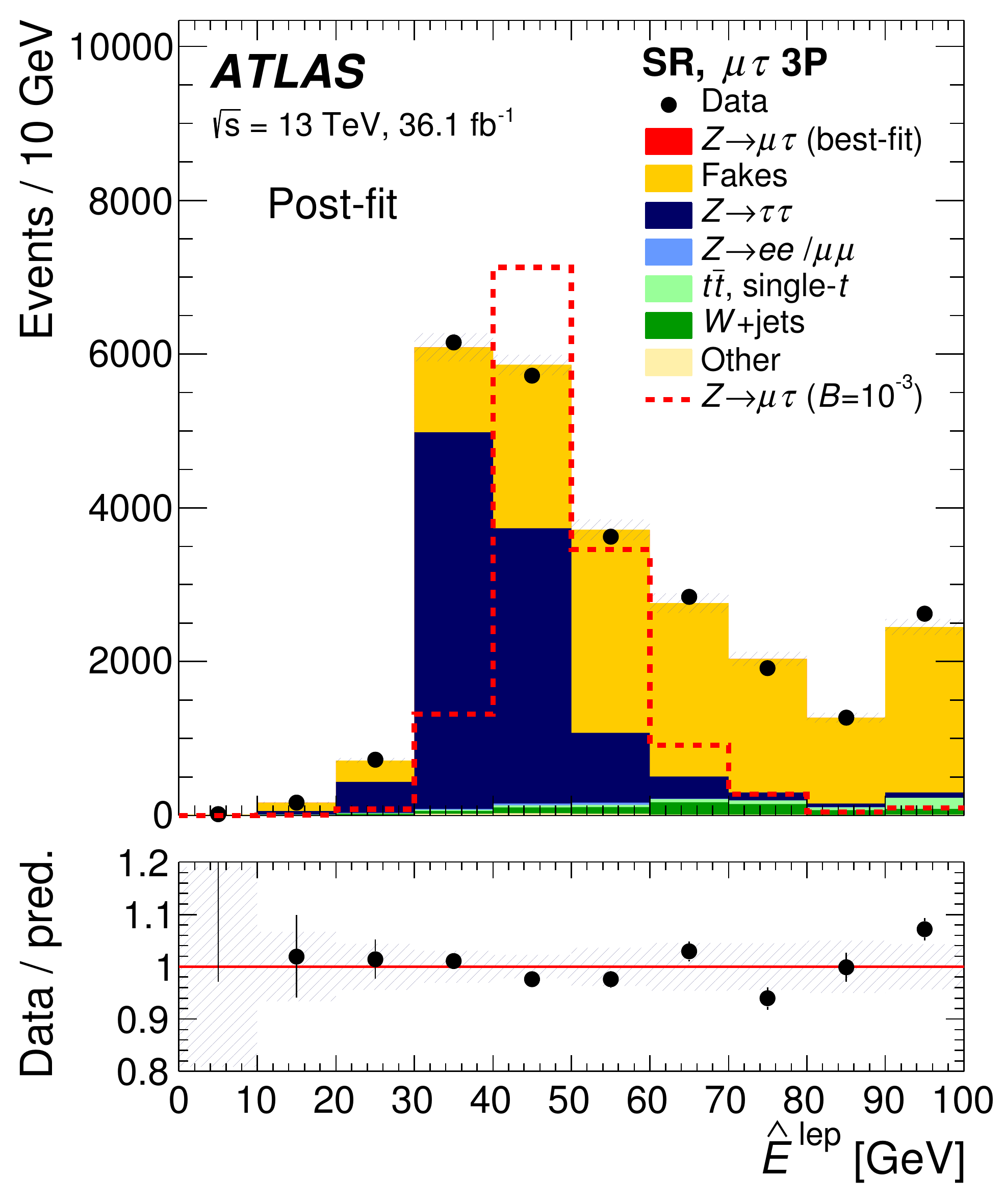}
    \label{}
  }
  \hfill
  \subfloat[$\hat{E}^{\tauh}$]{
    \includegraphics[width=0.32\textwidth]{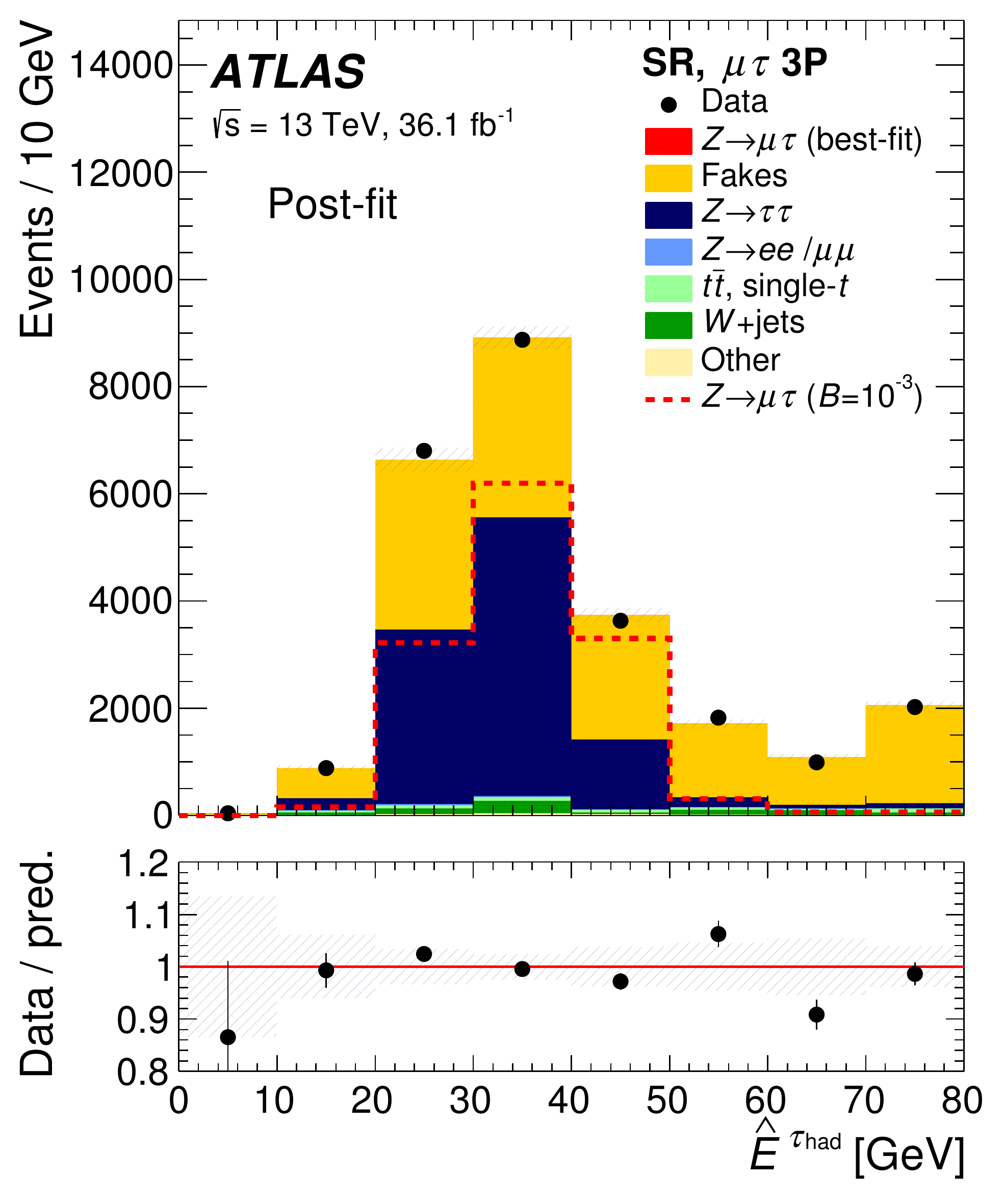}
    \label{}
  }
    \hfill
  \subfloat[$\hat{E}^{\mathrm{miss}}$]{
    \includegraphics[width=0.32\textwidth]{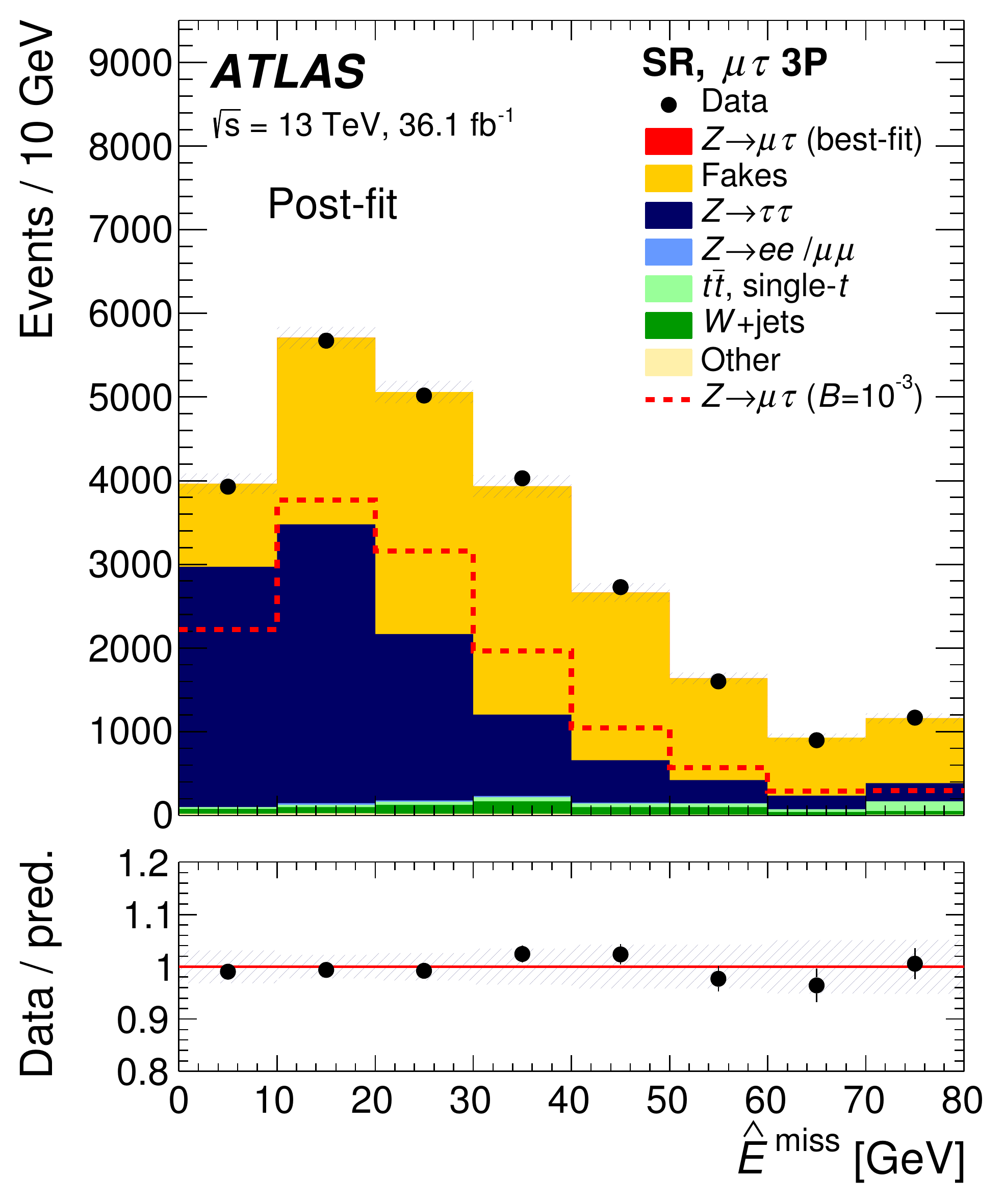}
    \label{}
  }
  \hfill
  \subfloat[$\Delta\alpha$]{
    \includegraphics[width=0.32\textwidth]{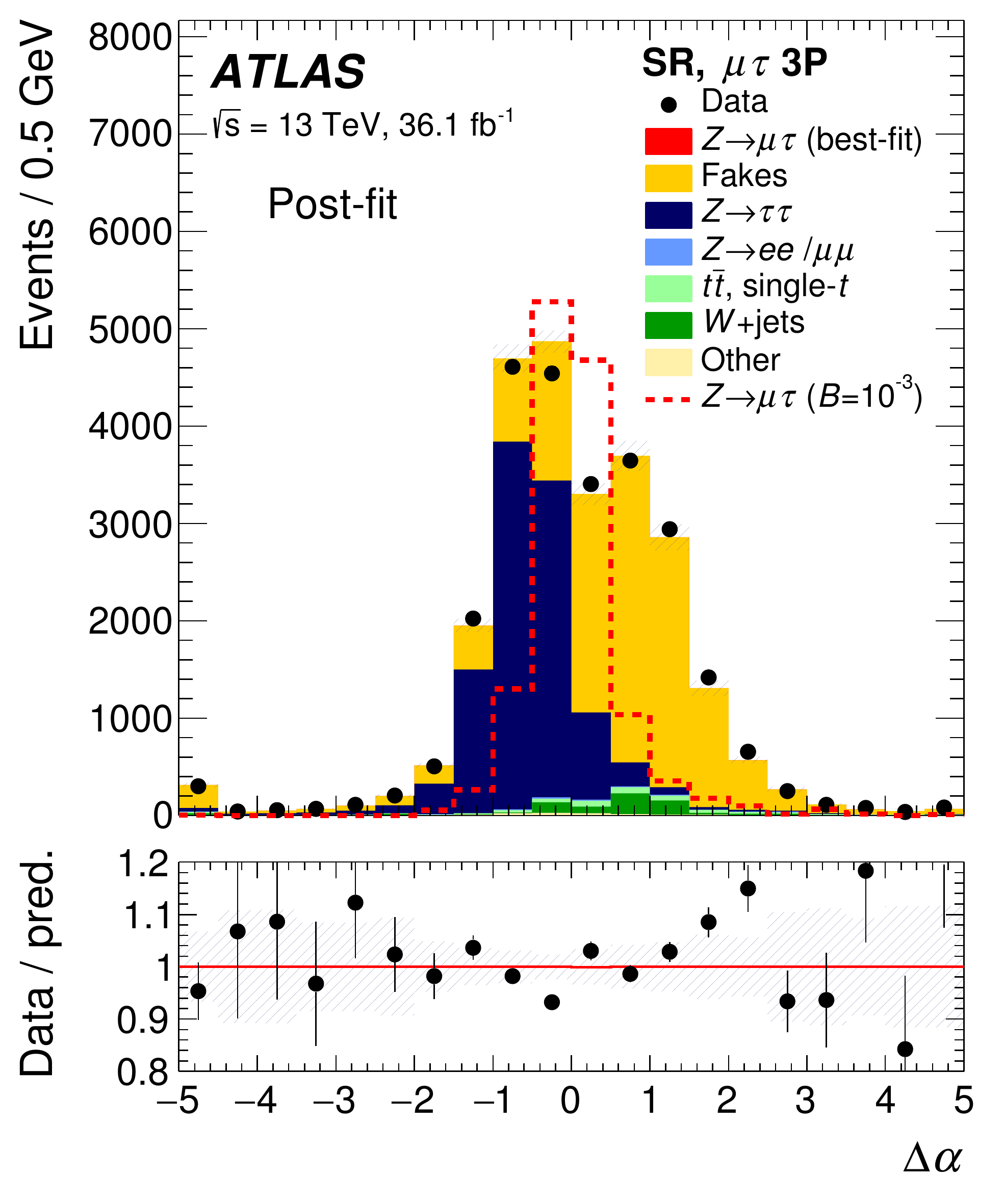}
    \label{}
  }
  \hfill
  \subfloat[$m_{\mathrm{coll}}$]{
    \includegraphics[width=0.32\textwidth]{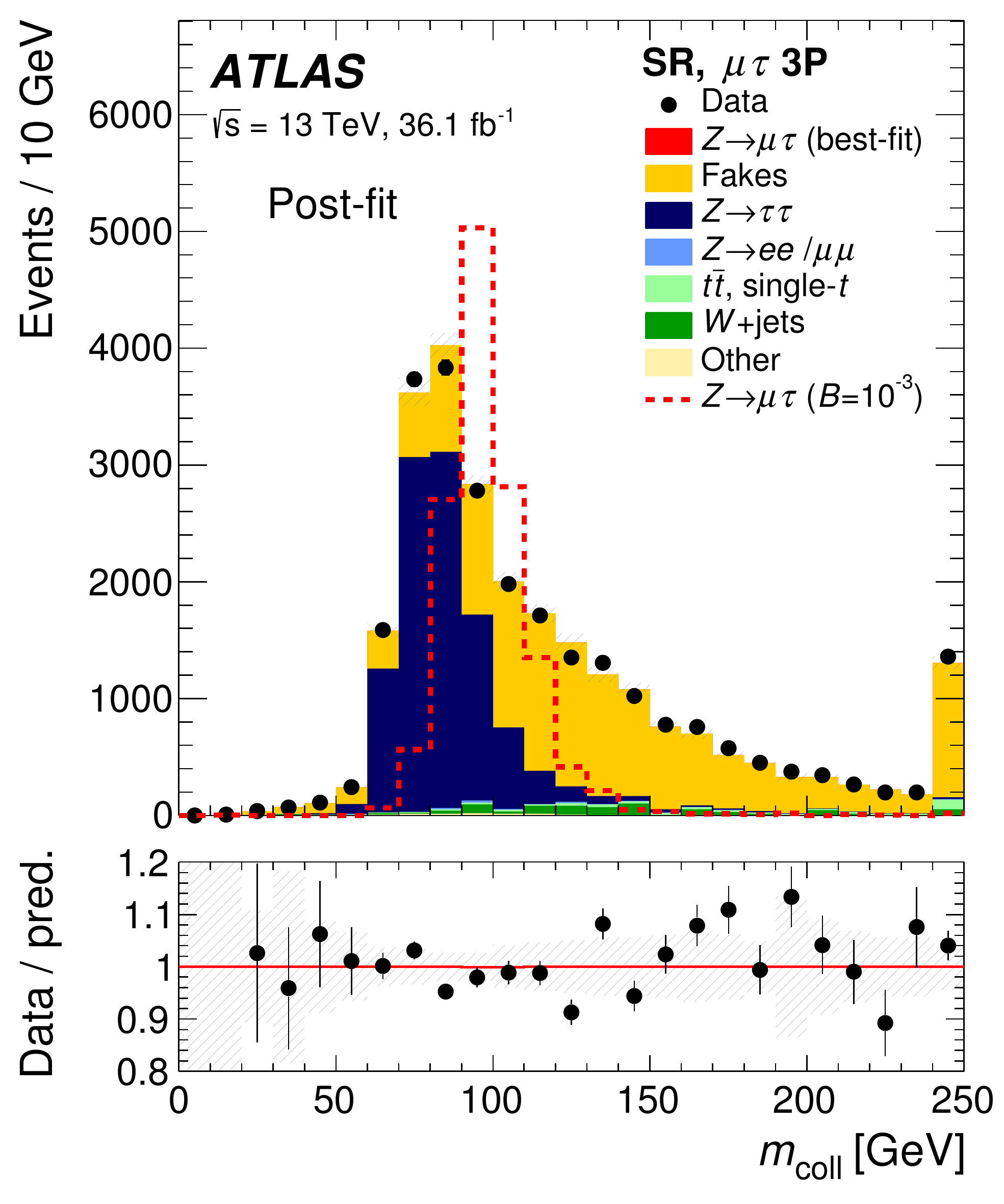}
    \label{}
  }
   \hfill
  \subfloat[$\pt^\mathrm{tot}$]{
    \includegraphics[width=0.32\textwidth]{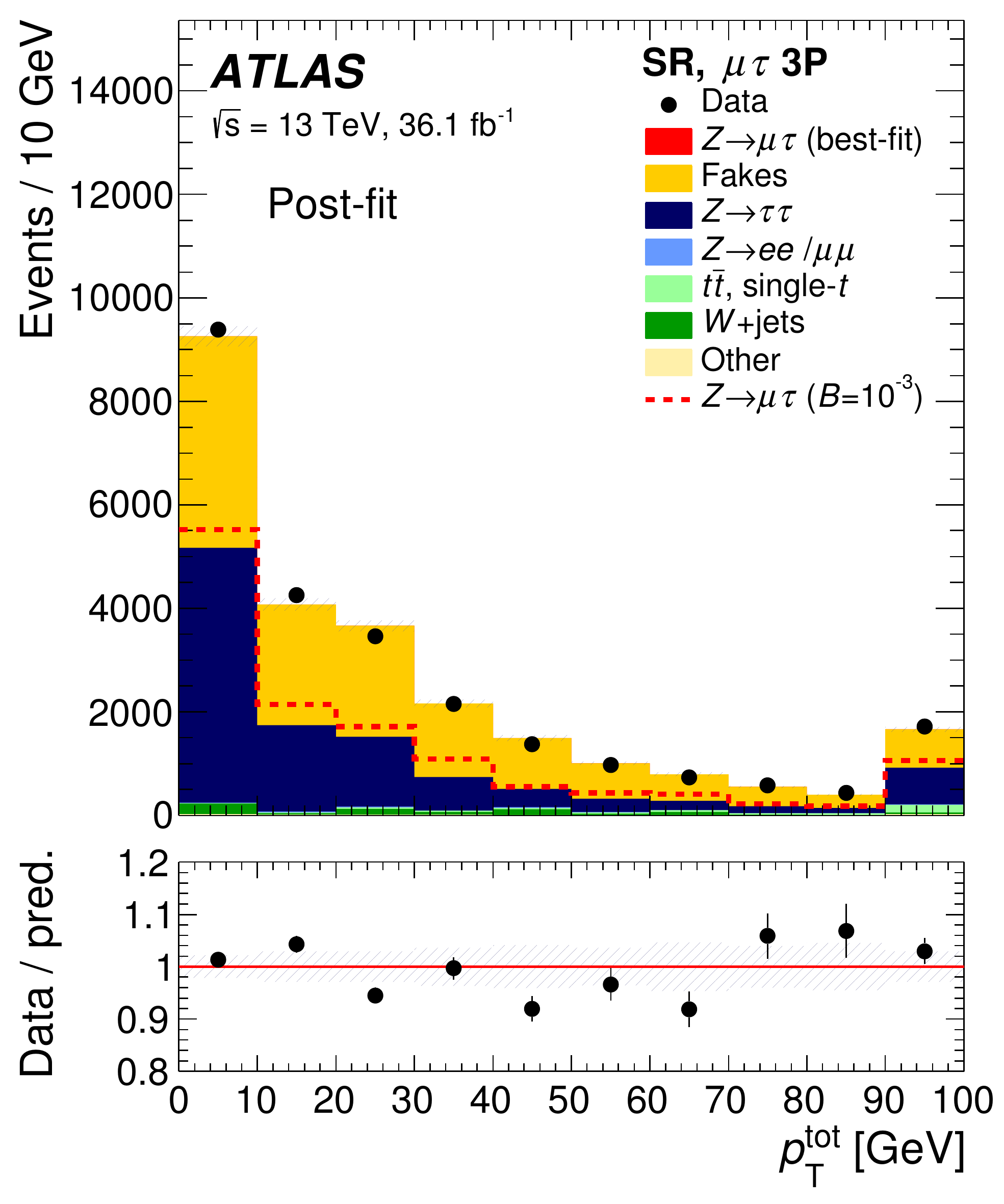}
    \label{}
  }
  \caption{Observed and expected post-fit distributions of unscaled NN inputs in SR for the $\mu\tau$ channel with 3-prong \tauhadvis candidates. The fit is based on profiling on the combined NN classifier, but not directly on these variables. The filled histogram stacked on top of the backgrounds represents the signal normalized to the best-fit $\BR(Z\to\ell\tau)$. The overlaid dashed line represents the expected distribution for the signal normalized to $\BR(Z\to\ell\tau)=10^{-3}$. In the panels below each plot, the ratios of the observed data (dots) and the post-fit background plus signal (solid line) to the post-fit background are shown. The hatched error bands represent the combined statistical and systematic uncertainties. The first and last bins include underflow and overflow events, respectively.} 
\label{fig:postfit_SR_mu_NNInputs_3P}
\end{figure}

Observed and expected post-fit distributions of the unscaled NN inputs of the events in the SR are shown in 
Figures~\ref{fig:postfit_SR_el_NNInputs_1P}--\ref{fig:postfit_SR_mu_NNInputs_3P}. 
The post-fit distributions are compatible with data. An alternative fit combining the $e\tau$ and $\mu\tau$ channels with two independent parameters of interest and the same shared free parameter $\mu(Z)$ yielded the same results as the nominal fit.

\begin{table}[htp]
\caption{The total observed number of events and post-fit event yields in the SR for the $e\tau$ (top) and $\mu\tau$ (bottom) channels after an fit to data. The
uncertainties include both the statistical and systematic contributions.
The correlations between the uncertainties in individual contributions are accounted for in the quoted uncertainties in the total post-fit event yields.
}
\label{tab:yields_unblindNNfit_combined}
\begin{center}
\begin{tabular}{lr@{$\pm$}lr@{$\pm$}l}
\toprule
{\bf }           & \multicolumn{2}{c}{1-prong}            & \multicolumn{2}{c}{3-prong}              \\[-0.05cm]
\midrule
Total observed $e\tau$ events          & \multicolumn{2}{c}{$89\,294$}              & \multicolumn{2}{c}{$35\,148$}                    \\
\noalign{\smallskip}\hline\noalign{\smallskip}
Total post-fit $e\tau$ event yield         & 89\,300 & 300           &  35\,200 & 200               \\
\noalign{\smallskip}\hline\noalign{\smallskip}
        Fakes         & 57\,000 & 1000           & 21\,500 & 500               \\
        $Z\to\tau\tau$         & 26\,000 & 1000           & 11\,500 & 500              \\
        $Z\to\ell\ell$         & 3200 & 100          & 250 & 150              \\
        Top         & 770 & 120           & 440 & 70              \\
        $W$+jets         & 540 & 100         & 950 & 180              \\
        Other         & 340 & 70          & 150 & 30              \\
        $Z\to\tau e$ signal         & 900 & 400           & 390 & 160              \\
\midrule
\midrule
Total observed $\mu\tau$ events          & \multicolumn{2}{c}{$79\,744$}              & \multicolumn{2}{c}{$25\,050$}                    \\
\noalign{\smallskip}\hline\noalign{\smallskip}
Total post-fit $\mu\tau$ event yield         & 79\,700 & 500           & 25\,100 & 700               \\
\noalign{\smallskip}\hline\noalign{\smallskip}
        Fakes         & 52\,000 & 1000           & 13\,600 & 800               \\
        $Z\to\tau\tau$         & 26\,000 & 1000           & 10\,300 & 300              \\
        $Z\to\ell\ell$         & 240 & 110          & 80 & 40              \\
        Top         & 890 & 140           & 360 & 60               \\
        $W$+jets         & 610 & 120          & 680 & 130               \\
        Other         & 290 & 70           & 110 & 20               \\
        $Z\to\tau\mu$ signal         & $-20$ & 360          & $-10$ & 140             \\
\bottomrule
\end{tabular}
\end{center}
\end{table}

\begin{table}[htp]
\caption{Best-fit values for $\BR (Z\to\ell\tau)$ and the other free parameters, and exclusion upper limits in the $e\tau$ and $\mu\tau$ channels. The uncertainties include both the statistical and systematic contributions.}
\begin{center}
\begin{tabular}{l|c||c}
\toprule
 & $e\tau$ & $\mu\tau$ \\
\midrule
$\BR (Z\to\ell\tau)$ & $(3.3^{+1.5}_{-1.4})\times10^{-5}$ & $(-0.1^{+1.2}_{-1.2})\times10^{-5}$ \\[2.0pt]
$\mu(Z)$ & $0.83^{+0.09}_{-0.07}$ & $0.87^{+0.09}_{-0.08}$ \\[2.0pt]
$\mu(\text{fakes\_1P})$ & $1.18^{+0.06}_{-0.06}$ & $1.12^{+0.09}_{-0.08}$ \\[2.0pt]
$\mu(\text{fakes\_3P})$ & $1.01^{+0.06}_{-0.05}$ & $1.09^{+0.13}_{-0.14}$ \\[2.0pt]
\midrule
Observed (expected) upper limit at 95\% CL & $5.8(2.8)\times10^{-5}$ & $2.4(2.4)\times10^{-5}$ \\
\bottomrule
\end{tabular}
\end{center}
\label{tab:post-fit-mu}
\end{table}%

After the fit, the probabilities of compatibility between the data and the background-plus-signal and background-only hypotheses are assessed using the profile log-likelihood ratio method \cite{Cowan:2010js}, where the nuisance parameters are profiled as a function of the parameter of interest. The normalization parameters are not profiled. As no significant deviation from the background-only hypothesis is observed, exclusion upper limits on $\BR (Z\to\ell\tau)$ are set using the CL$_\text{s}$ method \cite{CLs}.
The resulting observed (expected) upper limits at 95\% CL are $\BR (Z\to e\tau)<5.8\times10^{-5}$ ($2.8\times10^{-5}$) and $\BR (Z\to\mu\tau)<2.4\times10^{-5}$ ($2.4\times10^{-5}$). 
The significance of the excess in the $e\tau$ channel is 2.3$\sigma$.

The result of the search for $Z\to\mu\tau$ decays presented here is combined with the result published by ATLAS with \SI{20.3}{\ifb} of data at a center-of-mass energy of $\sqrt{s} = \SI{8}{\TeV}$ \cite{HIGG-2015-09}. In this previous analysis, a 95\% CL 
upper limit was set at $\BR (Z\to \mu\tau) < 1.7\times 10^{-5}$. The expected upper limit was $2.6\times 10^{-5}$. 

The analysis of the $\SI{8}{\TeV}$ data was 
based on a template fit to the observed distributions in data of the $m^\mathrm{MMC}_{\tau\mu}$ mass, as reconstructed by using the Missing Mass Calculator~\cite{MMC}. This is a likelihood-based mass estimator optimized for \Ztt events. The dominant irreducible \Ztt background was estimated using so-called embedded events~\cite{HIGG-2014-09} and was normalized to data. The reducible background of events with \tauhadvis candidates originating from misidentified jets was also estimated from data using events with $\mu\tau$ pairs with the same electric charges. The other smaller background contributions were estimated from simulation. The $Z\to\mu\tau$ signal was simulated and was normalized using the predicted $Z$ production cross section at $\SI{8}{\TeV}$.

The $\SI{8}{\TeV}$ and $\SI{13}{\TeV}$ analyses are combined using the same parameter of interest,
but assuming no other correlation. Indeed, the estimates of the two dominant sources of background, \Ztt and fakes, are based on different data and different methods. The signal predictions are also uncorrelated since the $Z$ production cross section is either predicted, in the $\SI{8}{\TeV}$ analysis, or determined from data, in the $\SI{13}{\TeV}$ analysis. Furthermore, the systematic uncertainties related to the detector modeling in simulated data are typically based on auxiliary measurements performed on different data. 
If these modeling uncertainties are set to zero, the combined upper limit changes by only 3\%.
This 3\% represents an upper bound on how much the combined limit can change if different assumptions are made about correlations in systematic uncertainties related to detector modeling.

The combined best-fit value of $\BR (Z\to \mu\tau)$ is $(-0.8^{+0.9}_{-0.8})\times 10^{-5}$ and the combined observed (expected) 95\% CL 
upper limit is $\BR (Z\to \mu\tau) < $1.3 (1.8) $\times 10^{-5}$.

\FloatBarrier

\section{Conclusions}
\label{sec:summary}

Direct searches for lepton flavor violation in decays of the $Z$ boson are performed using a data sample of proton--proton collisions recorded by the ATLAS detector at the LHC corresponding to an integrated luminosity of \SI{36.1}{\ifb} at a center-of-mass energy of $\sqrt{s} = \SI{13}{\TeV}$. 
The analysis selects events consistent with the decay of a $Z$ boson into an electron or muon and a hadronically decaying $\tau$ lepton. 
In these decays the $\tau$ lepton is assumed to be unpolarized. Neural network 
classifiers are used to discriminate signal from backgrounds, and the NN output distributions are analyzed in a template fit to data. 

No significant excess of events above the expected background is observed and upper limits on the lepton-flavor-violating branching ratios are set at the 95\% confidence level using the CL$_\text{s}$ method: $\BR (Z\to \mu\tau) < 2.4 \times 10^{-5}$ and $\BR (Z\to e\tau) < 5.8 \times 10^{-5}$. 
The corresponding expected upper limits are $2.4 \times 10^{-5}$ and $2.8 \times 10^{-5}$, respectively.
An excess of data over the expected backgrounds is observed in the $e\tau$ final state with a significance of 2.3$\sigma$. 

No upper limits on $\BR  (Z\to e\tau)$ from ATLAS data have been published previously. The current best upper limit is from LEP at $\BR (Z\to e\tau) < 0.98 \times 10^{-5}$.

The result on $\BR (Z\to \mu\tau)$ presented here is combined with the previous ATLAS result based on \SI{20.3}{\ifb} of data at a center-of-mass energy of $\sqrt{s} = \SI{8}{\TeV}$. The combined 95\% CL 
upper limit is $\BR (Z\to \mu\tau) < $1.3 $\times 10^{-5}$, to be compared with LEP upper limit of $\BR (Z\to \mu\tau) < $1.2 $\times 10^{-5}$.

\section*{Acknowledgments}


We thank CERN for the very successful operation of the LHC, as well as the
support staff from our institutions without whom ATLAS could not be
operated efficiently.

We acknowledge the support of ANPCyT, Argentina; YerPhI, Armenia; ARC, Australia; BMWFW and FWF, Austria; ANAS, Azerbaijan; SSTC, Belarus; CNPq and FAPESP, Brazil; NSERC, NRC and CFI, Canada; CERN; CONICYT, Chile; CAS, MOST and NSFC, China; COLCIENCIAS, Colombia; MSMT CR, MPO CR and VSC CR, Czech Republic; DNRF and DNSRC, Denmark; IN2P3-CNRS, CEA-DRF/IRFU, France; SRNSFG, Georgia; BMBF, HGF, and MPG, Germany; GSRT, Greece; RGC, Hong Kong SAR, China; ISF, I-CORE and Benoziyo Center, Israel; INFN, Italy; MEXT and JSPS, Japan; CNRST, Morocco; NWO, Netherlands; RCN, Norway; MNiSW and NCN, Poland; FCT, Portugal; MNE/IFA, Romania; MES of Russia and NRC KI, Russian Federation; JINR; MESTD, Serbia; MSSR, Slovakia; ARRS and MIZ\v{S}, Slovenia; DST/NRF, South Africa; MINECO, Spain; SRC and Wallenberg Foundation, Sweden; SERI, SNSF and Cantons of Bern and Geneva, Switzerland; MOST, Taiwan; TAEK, Turkey; STFC, United Kingdom; DOE and NSF, United States of America. In addition, individual groups and members have received support from BCKDF, the Canada Council, CANARIE, CRC, Compute Canada, FQRNT, and the Ontario Innovation Trust, Canada; EPLANET, ERC, ERDF, FP7, Horizon 2020 and Marie Sk{\l}odowska-Curie Actions, European Union; Investissements d'Avenir Labex and Idex, ANR, R{\'e}gion Auvergne and Fondation Partager le Savoir, France; DFG and AvH Foundation, Germany; Herakleitos, Thales and Aristeia programmes co-financed by EU-ESF and the Greek NSRF; BSF, GIF and Minerva, Israel; BRF, Norway; CERCA Programme Generalitat de Catalunya, Generalitat Valenciana, Spain; the Royal Society and Leverhulme Trust, United Kingdom.

The crucial computing support from all WLCG partners is acknowledged gratefully, in particular from CERN, the ATLAS Tier-1 facilities at TRIUMF (Canada), NDGF (Denmark, Norway, Sweden), CC-IN2P3 (France), KIT/GridKA (Germany), INFN-CNAF (Italy), NL-T1 (Netherlands), PIC (Spain), ASGC (Taiwan), RAL (UK) and BNL (USA), the Tier-2 facilities worldwide and large non-WLCG resource providers. Major contributors of computing resources are listed in Ref.~\cite{ATL-GEN-PUB-2016-002}.

\printbibliography

\clearpage 
 
\begin{flushleft}
{\Large The ATLAS Collaboration}

\bigskip

M.~Aaboud$^\textrm{\scriptsize 34d}$,    
G.~Aad$^\textrm{\scriptsize 99}$,    
B.~Abbott$^\textrm{\scriptsize 124}$,    
O.~Abdinov$^\textrm{\scriptsize 13,*}$,    
B.~Abeloos$^\textrm{\scriptsize 128}$,    
D.K.~Abhayasinghe$^\textrm{\scriptsize 91}$,    
S.H.~Abidi$^\textrm{\scriptsize 164}$,    
O.S.~AbouZeid$^\textrm{\scriptsize 143}$,    
N.L.~Abraham$^\textrm{\scriptsize 153}$,    
H.~Abramowicz$^\textrm{\scriptsize 158}$
% ...
\bigskip
\\

% ...
$^{13}$Institute of Physics, Azerbaijan Academy of Sciences, Baku; Azerbaijan.\\
% ...
$^{34}$$^{(a)}$Facult\'e des Sciences Ain Chock, R\'eseau Universitaire de Physique des Hautes Energies - Universit\'e Hassan II, Casablanca;$^{(b)}$Centre National de l'Energie des Sciences Techniques Nucleaires (CNESTEN), Rabat;$^{(c)}$Facult\'e des Sciences Semlalia, Universit\'e Cadi Ayyad, LPHEA-Marrakech;$^{(d)}$Facult\'e des Sciences, Universit\'e Mohamed Premier and LPTPM, Oujda;$^{(e)}$Facult\'e des sciences, Universit\'e Mohammed V, Rabat; Morocco.\\
% ...
$^{91}$Department of Physics, Royal Holloway University of London, Egham; United Kingdom.\\
% ...
$^{99}$CPPM, Aix-Marseille Universit\'e, CNRS/IN2P3, Marseille; France.\\
% ...
$^{124}$Homer L. Dodge Department of Physics and Astronomy, University of Oklahoma, Norman OK; United States of America.\\
% ...
$^{128}$LAL, Universit\'e Paris-Sud, CNRS/IN2P3, Universit\'e Paris-Saclay, Orsay; France.\\
% ...
$^{143}$Santa Cruz Institute for Particle Physics, University of California Santa Cruz, Santa Cruz CA; United States of America.\\
% ...
$^{153}$Department of Physics and Astronomy, University of Sussex, Brighton; United Kingdom.\\
% ...
$^{158}$Raymond and Beverly Sackler School of Physics and Astronomy, Tel Aviv University, Tel Aviv; Israel.\\
% ...
$^{164}$Department of Physics, University of Toronto, Toronto ON; Canada.\\
% ...
$^{a}$ Also at  Department of Physics, University of Malaya, Kuala Lumpur; Malaysia.\\
$^{b}$ Also at Borough of Manhattan Community College, City University of New York, NY; United States of America.\\
$^{c}$ Also at Centre for High Performance Computing, CSIR Campus, Rosebank, Cape Town; South Africa.\\
$^{d}$ Also at CERN, Geneva; Switzerland.\\
$^{e}$ Also at CPPM, Aix-Marseille Universit\'e, CNRS/IN2P3, Marseille; France.\\
% ...
$^{*}$ Deceased

\end{flushleft}


\end{document}